 \documentclass[final,1p,times]{elsarticle}
\usepackage{amssymb}
\usepackage{graphicx}
\usepackage{amsmath}
\usepackage{amssymb}
\usepackage{psfrag}
\usepackage{setspace}
\usepackage{subfig}
\usepackage{wrapfig}
\usepackage{sidecap}

\newcommand{\bs}[1]{\boldsymbol{#1}}
\def\RR{ \mathbb R}

\def\bt{\bs{\Theta}}

\usepackage{epsf}
\newcommand{\ee}{\end{equation}}
\newcommand{\be}{\begin{equation}}
\newcommand{\ec}{\end{center}}
\newcommand{\bc}{\begin{center}}
\newcommand{\eea}{\end{eqnarray}}
\newcommand{\bea}{\begin{eqnarray}}
\newcommand{\bd}{\begin{description}}
\newcommand{\ed}{\end{description}}
\newcommand{\bi}{\begin{itemize}}
\newcommand{\ei}{\end{itemize}}
\newcommand{\pa}{\partial}

\newcommand{\mb}[1]{\mathbf{#1}}

\newcommand{\bpsi}{\bs{\Psi}}
\newcommand{\dpsi}{d_{\Psi}}
\newcommand{\dth}{d_{\Theta}}
\newcommand{\tcb}{\textcolor{blue}}
\newcommand{\tcc}{\textcolor{cyan}}

\newcommand{\tcr}{\textcolor{red}}
\newcommand{\tcreview}{\textcolor{black}} % added changes based on first review
\newcommand{\tcrereview}{\textcolor{black}} % added changes based on second review

\newcommand{\refeq}[1]{Equation (\ref{#1})}
\newcommand{\reffig}[1]{Figure \ref{#1}}

\journal{Computer Methods in Applied Mechanics and Engineering}

\usepackage{amsmath}
\usepackage{algorithm} % for algorithm
\usepackage{algpseudocode} % add [noend] if in the algorithm the 'end' part should not be shown

\usepackage{rotating} % for rotating tables

\usepackage{tikz} % for using a flow chart
\usetikzlibrary{shapes,arrows}  % for using a flow chart

\usepackage{url}%for urls

\begin{document}

\begin{frontmatter}

\title{Sparse Variational Bayesian Approximations for Nonlinear Inverse Problems: applications in nonlinear elastography}



\author[rvt]{Isabell M. Franck}
\ead{franck@tum.de}
\author[rvt]{P.S. Koutsourelakis\corref{cor1}}
\ead{p.s.koutsourelakis@tum.de}
\cortext[cor1]{Corresponding Author. Tel: +49-89-289-16690}
\address[rvt]{Professur  f\"ur Kontinuumsmechanik, Technische Universit\"at M\"unchen, Boltzmannstrasse 15, 85747 Garching (b. M\"unchen), Germany}
\ead[url]{http://www.contmech.mw.tum.de}

\begin{abstract}

This paper presents an efficient Bayesian framework for solving nonlinear, high-dimensional model calibration problems.  It is based on  a Variational Bayesian formulation that aims at approximating the exact  posterior by means of solving an optimization problem over  an appropriately selected family of distributions. The goal is two-fold. Firstly, to find lower-dimensional representations of the unknown parameter vector  that capture as much as possible of the associated posterior density, and secondly to  enable the computation of the  approximate posterior density with as few forward calls as possible. We discuss how these objectives can be achieved by using a fully Bayesian argumentation and employing the
marginal likelihood or evidence  as the ultimate model validation metric for
any proposed dimensionality reduction. We demonstrate the performance of the proposed methodology for problems in nonlinear elastography where  the identification of the mechanical
properties of biological materials can inform non-invasive, medical diagnosis.
\tcreview{An Importance Sampling scheme is finally employed in order to validate the results and assess the efficacy of the approximations provided.}

% 
%  novel and practical Bayesian framework for nonlinear inverse models with a high number of dimensions. We work with a Variational Bayes algorithm which reduces the number of forward calls compared to standard MCMC methods dramatically. In this context we also enforce sparsity by transposing the large sum of unknowns to a lowerdimensional manifold, reducing the number of dimensions and making uncertainty quantification for larger applications applicable. We attempt to provide probabilistic estimates of material properties which is especially important in biomedical settings. The inferred material properties and their uncertainties can be used for medical diagnosis.

\end{abstract}


\end{frontmatter}
Copyright \copyright\, 2015 I. M. Franck and P.S. Koutsourelakis. This manuscript version is made available under the CC-BY-NC-ND 4.0 license \url{http://creativecommons.org/licenses/by-nc-nd/4.0/}


\section{Introduction}\label{sec:Introduction}
The extensive use of large-scale computational models  poses several challenges in model calibration as the  accuracy of the predictions provided depends strongly on assigning proper values to the various model parameters.
 In mechanics of materials, accurate mechanical property identification can guide damage  detection  and an informed assessment of the system's reliability \cite{johannesson_multi-resolution_2005}. Identifying property-cross correlations can lead to the design of multi-functional materials \cite{torquato_random_2002}. Permeability estimation for soil transport processes can assist in detection of contaminants, oil exploration \cite{wang_markov_2006}.

Deterministic optimization techniques which have been developed to address these problems \cite{vogel_computational_2002},
%tikhonov_solutions_1977,groetsch_inverse_1993,engl_regularization_1996,calvetti_tikhonov_2003,kaipio_statistical_2006}
 lead to point estimates for the unknowns without rigorously considering the statistical nature of the problem   and without providing quantification of the uncertainty in the inverse solution.
 Statistical approaches based on the Bayesian paradigm \cite{kaipio_statistical_2006} on the other hand, aim at  computing  a (posterior) probability distribution  on the parameters of interest. Bayesian formulations  offer several advantages as they  provide a unified framework for dealing with the uncertainty introduced by the incomplete and noisy measurements.  Significant successes have been noted in applications  such as geological tomography \cite{glaser_stochastic_2004}
%,andersen_bayesian_2003}
, medical tomography \cite{weir_fully_1997},  petroleum engineering \cite{hegstad_uncertainty_2001}
%, craig_bayesian_2001}
, as well as a host of other physical, biological, or social systems \cite{wang_hierarchical_2005,liu_bayesian_2008}.
%\cite{kitanidis_parameter_1986,schmidt_bayesian_1998,wang_hierarchical_2005,liu_bayesian_2008}
  Representations of the parametric fields in existing deterministic and Bayesian approaches (artificially) impose a minimum length scale of variability usually determined by the discretization size of the governing PDEs \cite{lee_markov_2002}. As a result  they give rise to  a very large  vector of unknowns. Inference in high-dimensional spaces using standard Markov Chain Monte Carlo (MCMC) schemes  is generally impractical as it requires an  exuberant number of calls to the forward simulator in order to achieve convergence. Advanced schemes such as those employing  Sequential Monte Carlo samplers \cite{moral_sequential_2006, koutsourelakis_multi-resolution_2009}, adaptive MCMC \cite{chopin_free_2012}\tcreview{, accelerated MCMC methods \cite{hoang_complexity_2013}} or spectral methods \cite{marzouk_stochastic_2007} can alleviate some of these difficulties particularly when the posterior is multi-modal but still pose significant challenges in terms of the computational cost \cite{bilionis_solution_2014}.
  %These problems are amplified if the posterior distribution is multi-modal i.e. several significantly different hypotheses are likely given the available data.  %While it is apparent that, computationally inexpensive surrogates \cite{Marzouk2007} and   coarser scale simulations can assist the identification process \cite{Dostert:2006}, the critical task of efficiently transferring the information across resolutions still remains \cite{liu00gen,hig03mar}.

This work is particularly concerned with the identification  of the mechanical properties of  biological materials, in the context non-invasive medical diagnosis.
While in certain cases mechanical properties can also be measured directly by excising multiple tissue samples, non-invasive procedures offer obvious advantages in terms of ease, cost and reducing risk of complications to the patient.  Rather than x-ray techniques which capture variations in density, the identification of stiffness or mechanical properties in general, can potentially  lead to earlier and more accurate diagnosis \cite{oberai_linear_2009,ganne-carrie_accuracy_2006},  provide valuable insights that  differentiate between modalities of the same pathology \cite{curtis_genomic_2012} and monitor the progress of treatments. 
In this paper we  do not propose   new imaging techniques but rather aim at  developing  rigorous statistical models and efficient computational tools that can make use of the data/observables (i.e. noisy displacements of deformed tissue) from existing imaging modalities (such as magnetic 
resonance \cite{muthupillai_magnetic_1995}, ultrasonic) in order to produce certifiable estimates of mechanical properties. 
The primary imaging modality considered in this project is ultrasound elasticity imaging (elastography \cite{sarvazyan_editorial_2011,doyley_model-based_2012}).
It is based on ultrasound tracking of pre- and post-compression images to obtain a map of position changes and  deformations of the specimen due to an external pressure/load. 
The pioneering work of   Ophir and coworkers \cite{ophir_elastography:_1991}
 followed by several clinical studies \cite{bamber_progress_2002,thomas_real-time_2006,parker_imaging_2011}
% \cite{garra_elastography_1997,bamber_progress_2002,thomas_real-time_2006,parker_imaging_2011}
 have demonstrated that the resulting strain images typically improve the diagnostic accuracy over ultrasound alone.

Beyond a mere strain imaging there are two approaches for inferring the
constitutive material parameters. In the {\em direct approach}, the equations
of equilibrium  are interpreted as  equations for the material parameters of interest, where the inferred strains and their derivatives appear as coefficients \cite{barbone_adjoint-weighted_2010}. %,ISI:000237030900021}.
While such an approach provides a computationally efficient strategy,  it does not use the raw data (i.e. noisy displacements) but transformed versions i.e. strain fields (or even-worse, strain derivatives) which arise by applying sometimes ad hoc filtering and  smoothing operators. As a result  the informational content of the data is compromised and the quantification of the effect of observation noise is cumbersome. 
Furthermore, the smoothing employed can smear regions with sharply varying properties and hinder proper identification.

The alternative to direct methods, i.e. {\em indirect or iterative} procedures admit an inverse problem formulation where the discrepancy %(in various norms,  %\cite{ISI:000240849100006,ISI:000255220100005}   \cite{ISI:000240849100006}) 
 between observed and model-predicted displacements is minimized with respect to the material fields of interest \cite{oberai_evaluation_2004,doyley_enhancing_2006,arnold_efficient_2010,olson_numerical_2010}. %\cite{ISI:000237981300009,ISI:000223500200013,ISI:000238422400023,lakisorig,ISI:000232236800008,ISI:000228126000006,ISI:000262358700008,ISI:000257837900010,ISI:000275756200016,ISI:000280774700004,bon05inv}.
 While these approaches utilize directly the raw data, they generally imply an increased computational cost as the forward problem and potentially derivatives have to be solved/computed several times.
This effort is amplified when stochastic/statistical formulations are employed as those arising in the Bayesian paradigm. Technological advances have led to the development of hand-carried ultrasound systems in the size of a smartphone \cite{schleder_diagnostic_2013}. Naturally their accuracy and resolution does not compare with the more expensive traditional ultrasound machines or even more so MRI systems. If however computational tools are available that can distill the informational content from noisy and incomplete data then this would constitute a major advance. Furthermore, significant progress is needed in improving the computational efficiency of these tools if they are to be made applicable on a patient-specific basis.
%, whose cost is comparable to that of a deterministic global optimization technique \cite{kai05com}.

In this work we advocate a Variational Bayesian (VB) perspective 
\cite{beal_variational_2003, bishop_pattern_2006}. Such methods have risen into prominence for probabilistic inference tasks in the machine learning community    \cite{jordan_introduction_1999, attias_variational_2000,wainwright_graphical_2008} but have  recently been   employed also in the context of inverse problems  \cite{chappell_variational_2009,jin_hierarchical_2010}.
They provide {\em approximate} inference results   by solving an optimization problem over a family of appropriately selected probability densities  with the objective of minimizing the Kullback-Leibler divergence \cite{cover_elements_1991} with  the exact posterior. The success of such an approach  hinges upon the selection of appropriate densities that have the capacity of providing good approximations while enabling efficient (and preferably) closed-form optimization with regards to their parameters.  We note that an alternative  optimization strategy originating from a different perspective and  founded on  map-based representations of the posterior has been proposed in \cite{el_moselhy_bayesian_2012}.

A pivotal role in Variational Bayesian strategies  or any other inference method, is dimensionality reduction i.e. the identification of lower-dimensional features that provide the strongest signature to the unknowns and the corresponding posterior.
Discovering a sparse set of features  has attracted great interest in many applications as in  the representation of natural images \cite{olshausen_sparse_1997} and a host of algorithms have been developed not only for finding such representations but also an appropriate dictionary for achieving this goal \cite{lewicki_learning_2000}.
% \cite{lewicki_learning_2000, lee_efficient_2006,mairal_online_2009,dobigeon_bayesian_2010}. 
   While all these tools are pertinent to the present problem they differ in a fundamental way. They are based  on several data/observations/instantiations of the vector that we seek to represent.  In our problem however we do not have such direct observations i.e. the data available pertains to the output of a model  which is nonlinearly and implicitly dependent on the vector of unknowns. Furthermore we are primarily interested in  approximating the posterior of  this vector rather than the dimensionality reduction itself.
{\em  We demonstrate how this can be done by using a fully Bayesian argumentation and employing the marginal likelihood or evidence   as the ultimate model validation metric for any proposed dimensionality reduction.}

The paper is organized as follows: The next section  (Section \ref{sec:method}) presents the essential ingredients of the forward model (Section \ref{sec:forward})  which are common with a wide range of nonlinear, high-dimensional problems encountered in several simulation contexts. 
We also discuss  the VB framework advocated, the dimensionality reduction scheme proposed, the prior densities for all model parameters, an iterative, coordinate-ascent  algorithm that enables the identification of all the unknowns (Section \ref{sec:BayesianModel}) as well as an information-theoretic criterion for determining the number of dimensions needed (Section \ref{sec:card}). We finally describe a Monte Carlo scheme based on Importance Sampling that can provide statistics of the exact posterior as well as a quantitative assessment of the VB approximation (Section \ref{sec:is}).  Section \ref{sec:NumericalExamples} demonstrates the performance and features of the proposed methodology in two problems from solid mechanics that are of relevance to the elastography settings.  Various signal-to-noise ratios are considered and the performance of the method, in terms of forward calls and accuracy, is assessed.

\section{Methodology} 
\label{sec:method}
The motivating application is related to continuum mechanics in the nonlinear elasticity regime. We describe below the governing equations in terms of conservation laws and the constitutive equations. 
The proposed model calibration  process can be readily adapted to other forward models. As it will be shown,  the only information utilized by the Bayesian inference engine proposed is a) the response quantities at the locations where measurements are available, and b) their derivatives with respect to the unknown  model parameters.

\subsection{Forward model - Governing equations}
 \label{sec:forward}

The following  expressions are formulated in the general case which includes nonlinear material behavior and  large deformations. 
\tcrereview{Our physical domain is described by $\Omega_0$ in $\mathcal{R}^3$ in the reference configuration.}
Let $\mathbf{X}$ denote the coordinates of the continuum particles in the undeformed configuration and $\bs{x}$ in the deformed. Their relation (in the static case) is provided by the deformation map $\bs{\phi}$ such that:  $\bs{x}=\bs{\phi}(\bs{X})$. The displacement field is defined as $\bs{u}(\bs{X})=\bs{x}-\bs{X}= \phi(\bs{X}) - \bs{X}$. The gradient of the deformation map is denoted by $\bs{F}=\nabla \bs{\phi}$ and $\bs{E}=\frac{1}{2} (\textbf{F}^T\textbf{F}-\bs{I})$ is then the  Lagrangian (finite)  strain tensor used as the primary kinematic state variable in our constitutive law \cite{holzapfel_nonlinear_2000,mase_continuum_2009,bonet_worked_2012}. 
 The governing equations consist of the  conservation of linear momentum:
\be
   \bigtriangledown \cdot (\textbf{FS}) + \rho_0 \textbf{b} = 0	\quad \tcrereview{in \quad \Omega_0}
   \label{eq:equilibrium}
\ee
\tcrereview{
and the Dirichlet and Neumann boundary conditions as
\be
  \bs{u} = \bs{\hat{u}}   \quad on \quad \Gamma_u
\ee
\be
  \bs{FS}\cdot \bs{N} = \bs{\hat{T}}   \quad on \quad \Gamma_S.
\ee
}
$\textbf{b}$ is  body force vector (per unit mass), $\rho_0$ is the initial density,  $\textbf{S}$ is the second Piola-Kirchhoff stress tensor 
\tcrereview{and $\bs{N}$ is the outward  normal at $\Gamma_S$.}
\tcrereview{$\Gamma_u$ and $\Gamma_S$ are subsets of the boundary, {$\Gamma_0 = \pa \Omega_0$}, on which displacement and traction boundary data, $\bs{\hat{u}}$ and $ \bs{\hat{T}}$, respectively, are specified.}
For a hyperelastic material, it is assumed that the strain energy density function $w(\bs{E}; \bs{\psi})$  exists and depends  on the invariants of the  Lagrangian strain tensor $\textbf{E}$ and the constitutive material parameters $\bs{\psi}(\bs{X})$. We note that the latter in general exhibit spatial variability which we intend to estimate using the methods discussed. The conjugate stress variables described by the second Piola-Kirchhoff stress tensor can be found as:
\be
  \textbf{S} = \frac{\pa w}{\pa \textbf{E}}=\bs{S}(\bs{E}; \bs{\psi}).
  \label{eq:const}
\ee
The aforementioned governing equations should  be complemented with any other information about the problem or the material, such as incompressibility. In fact incompressibility is frequently encountered in bio-materials and corresponds to the condition $det(\bs{F})=1$ at all points in the problem problem domain.

The governing equations presented thus far cannot be solved analytically for the vast majority of problems and one must resort to numerical techniques that discretize these equations and the associated fields. The most prominent such approach is the Finite Element Method (FEM) which is employed in this study as well. In the first step, the \textit{weak} form of the PDEs needs to be derived.  
\tcrereview{To that end, we define the usual  function spaces $\mathcal{S}$ and $\mathcal{V}$ for the set of admissible solutions and weighting functions respectively, as follows \cite{gokhale_solution_2008}:
\be
\begin{array}{l}
\mathcal{S} = \{\bs{u} |u_i \in H^1(\Omega_0): u_i = \hat{u}_i \, on \, \Gamma_u \}, \mathcal{V} = \{\bs{v} |v_i \in H^1(\Omega_0): v_i = 0 \, on \, \Gamma_u \}
\end{array}
\ee
where $H^1(\Omega_0)$ denotes the Sobolev space of square integrable functions with square integrable derivatives in $\Omega_0$ \cite{adams_sobolev_2003}. 
By multiplying \refeq{eq:equilibrium} with a weighting function $\bs{v} \in \mathcal{V}$, integrating by parts and  exploiting the essential and non-essential boundary conditions  above, we obtain:
\be
\int_{\Omega_0} F_{iK}S_{KL} v_{i,L} ~d\Omega_0=\int_{\Gamma_S} \hat{T}_i v_i~d\Gamma_S +\int_{\Omega_0} \rho_0 b_i v_i ~ d\Omega_0
\ee
In the incompressible case, pressure must be taken into account and for that purpose the pressure trial solutions $p \in  L_2(\Omega_0)$  and weighting functions $q \in   L_2(\Omega_0)$ should also be introduced \cite{goenezen_solution_2011}. 
 }

Subsequently the problem domain is discretized into finite elements and shape functions are used for interpolating the unknown fields. 
As this is a very mature subject, from a theoretical and computational point of view, we do not provide further details here but refer the interested reader  to one of many books available \cite{zienkiewicz_finite_1977, hughes_finite_2000} or more specifically in the context of inverse problems for (in)compressible elasticity in \cite{gokhale_solution_2008,goenezen_solution_2011}. Most often all unknowns are expressed in terms of the discretized displacement field denoted here by a vector $\bs{U} \in \RR^n$. An approximate solution can be found by solving an $n-$dimensional  system of nonlinear algebraic equations which in 
residual form can be written as:
\be
\bs{r}(\bs{U} ; \bs{\Psi})=\bs{0}.
\label{eq:discr}
\ee
We denote here by $\bs{r}: \RR^n \times \RR^{\dpsi} \to \RR^n$ the residuals and by  $\bpsi \in \RR^{ \dpsi}$, the {\em discretized} vector of constitutive material parameters $\psi(\bs{X})$.

The discretizations can be done in many different ways. For example if the same mesh and shape functions as for the discretization of the displacements are adopted, then each entry of the vector $\bpsi$ corresponds to the value of the material parameter of interest at each nodal point. Frequently it is assumed that the value of the constitutive parameters are constant within each finite element in which case $\dpsi$ coincides with the number of elements in the FE mesh.
While the representation of $\bs{\Psi}$ is discussed in detail in the sequence, we point out that the discretization of $\bs{\psi}(\bs{X})$  does not need to be associated with the discretization used for the governing equations. Usually in practice the two are related, but if one  aims at inferring  as many details about the variability of  $\psi(\bs{X})$ that the discretized equations $\refeq{eq:discr}$ can provide, a finer discretization might be employed for $\bs{\psi}(\bs{X})$. We note however that if the material properties exhibit significant variability within each finite element i.e. if $\dpsi \gg n$ , special care has to be taken in formulating the finite element solution and  multiscale schemes might need to be employed \cite{e_principles_2011}.%\cite{hughes_variational_1998,hou_multiscale_1997,dorobantu_wavelet-based_1998,abdulle_analysis_2006}.

We note here:
\bi
\item Frequently the size $n$ of the system of the equations that need to be solved is large. This is necessitated by accuracy requirements  in capturing the underlying physics and mathematics. It  can impose a significant computational burden as in general repeated solutions of this system, under different values of $\bs{\Psi}$, are needed.
If for  example a Newton-Raphson method is employed then repeated solutions of the linearized \refeq{eq:discr} will need to be performed:
\be
\left\{ 
\begin{array}{l}
 0 = \bs{r}(\bs{U}^{(t)})+ \bs{J}(\bs{U}^{(t)}) \delta \bs{U}^{(t)}  \\ \bs{U}^{(t+1)}=\bs{U}^{(t)}+\delta \bs{U}^{(t)}
\end{array}
\right.
\label{eq:nr}
\ee
where $t$ is the iteration number and $\bs{J}=\frac{\pa \bs{r}}{\pa \bs{U} }$ is the Jacobian matrix. 
Hence for large $n$ as in applications of interest,  the number of such forward solutions is usually what dictates the overall computational cost and this is what we report in subsequent numerical experiments. 
Depending on the particular solution method employed, converged solutions $\bs{U}(\bs{\Psi})$ at a certain stage of the inversion procedure can be used as initial guesses for subsequent solutions under different $\bs{\Psi}$ reducing as a result the overall cost. In this work we do not make use of such techniques.

\item The data available generally concerns a subset or more generally a lower-dimensional function of $\bs{U}$. In this work, the experimental measurements/ observations are (noisy) displacements at specific locations in the \tcreview{physical} domain. 
%These are extracted from appropriate processing of the ultrasound/MRI images \cite{Rivaz2011}. This processing naturally introduces model errors which are considered negligible in this study as the emphasis is on the inversion of the continuum mechanics model. In reality however, the model calibration process should consider directly the image data. 
We denote these displacements by $\bs{y} \in \RR^{d_y}$ and they can be formally expressed as $\bs{y} = \bs{Q} \bs{U}$ where $\bs{Q}$ is a Boolean matrix which picks out the entries of interest from $\bs{U}$.
Naturally, since $\bs{U}$ depends on $\bs{\Psi}$, $\bs{y}$ is also a function of $\bs{\Psi}$ i.e. $\bs{y}(\bs{\Psi})$. We emphasize that this function is generally {\em highly nonlinear} and most often than not, many-to-one  \cite{schillings_sparse_2013}. The unavailability of the inverse as well as the high nonlinearity constitute two of the basic difficulties of the associated inverse problem.

\item In addition to the solution vector $\bs{U}(\bs{\Psi})$, the proposed inference scheme will  
make use of the derivatives $\frac{\pa \bs{y}(\bs{\Psi})}{\pa \bs{\Psi}}$. The computation of derivatives of the response with respect to model parameters is a well-studied subject in the context of PDE-constrained optimization \cite{giles_introduction_2000,hinze_optimization_2009,papadimitriou_direct_2008} and we make use of it in this work.
For any scalar function $f(\bs{U})$, one can employ the adjoint form  of \refeq {eq:discr} according to which:
\be
\frac{d f}{d{\Psi}_k}= -\nu_j \frac{\pa r_i}{\pa \Psi_k}
\label{eq:der}
\ee
where $\bs{\nu}\in \RR^n$ is defined such as:
\be
\nu_j \frac{\pa r_j}{\pa U_i}=\frac{\pa f}{\pa U_i} \quad or \quad \bs{J}^T \bs{\nu} = \frac{\pa f}{\pa \bs{U}}.
\label{eq:adj}
\ee
We note that $\frac{\pa r_j}{\pa U_i}$ is the Jacobian of the residuals in \refeq{eq:discr} evaluated at the solution $\bs{U}(\bs{\Psi})$. We point out that if a direct solver  
for the solution of the linear system in $\refeq{eq:nr}$ is employed, then the additional cost of evaluating $\frac{d f}{d  \bs{\Psi} }$ is minimal as the Jacobian would not need to be re-factorized for solving \refeq{eq:adj} \footnote{The cost of evaluating  $\frac{\pa r_i}{\pa \Psi_k}$ is negligible compared to other terms as it scales linearly with the number of elements/nodes.}. 
\tcreview{ In the context of the problems considered in this paper (see Section 3), repeated use of \refeq{eq:adj} is made where $f$ is a different component of the observables and as such the overall cost increases proportionally with the number of observables (displacements in our problems) that are available. In problems where $n$ is so large that it precludes the use of direct solvers, then the cost of its solution of the adjoint equations can be augmented but nevertheless comparable to the cost of a forward solution.
}
\tcrereview{In cases where both $n$ as well as the dimension of $\bs{\Psi}$ are high, then advanced iterative solvers, suitable for multiple right-hand sides must be employed \cite{Orginos:1118470,gutknecht_block_2009}. These imply an added computational burden which nevertheless scales sublinearly with the dimension of   $\bs{\Psi}$.}
%Especially when the dimension $d_y$ of $\bs{y}$ is much smaller than that of $\bs{U}$, repeated use of \refeq{eq:der} and \refeq{eq:adj} with $f$ being each of the $y_i, i=1, \dots d_y$ can readily yield the desired matrix $\frac{\pa \bs{y}}{ \pa \bs{\Psi}}$.

%sIn contrast to determinstic inversion techniques where a point eastimate of $\bs{\Psi}$ is sought, Bayesian formulation attempt to infer the (posterior) distribution on $\bs{\Psi}$ incurring potentially a greater  cost as it will be discussed later/earlier. 

\ei

\subsection{Bayesian Model} \label{sec:BayesianModel}

The following discussion is formulated in general terms and can be applied for the calibration of any model with parameters represented by the vector $\bs{\Psi} \in \RR^{\dpsi}$ when output $\bs{y}(\bs{\Psi}) \in \RR^{d_y}$ is available. We also presuppose the availability of the derivatives $\frac{\pa \bs{y}}{ \pa \bs{\Psi}}$. For problems of practical interest, it is assumed that the dimension $\dpsi$ of the unknowns is very large which poses a significant hindrance in the solution of the associated inverse problem as well as in finding proper regularization (in deterministic settings \cite{calvetti_tikhonov_2003}) or in specifying appropriate priors (in probabilistic settings \cite{bardsley_gaussian_2013, schwab_sparse_2012}). The primary focus of the Bayesian model developed is two-fold:
\bi
\item find lower-dimensional representations of the unknown parameter vector $\bs{\Psi}$ that capture as much as possible of the associated posterior density
\item enable the computation of the  posterior density with as few forward calls (i.e. evaluations of $\bs{y}(\bs{\Psi}), \frac{\pa \bs{y}}{ \pa \bs{\Psi}}$) as possible.
\ei

We denote $\mb{\hat{y}} \in  \mathbb{R}^{d_y}$ the vector of observations/measurements. In the context of elastography the observations are displacements (in the static case) and/or velocities (in the dynamics). The extraction of this data from images (ultrasound or MRI) is a challenging topic that requires sophisticated image registration techniques \cite{richards_quantitative_2007,rivaz_real-time_2011}.                                                   %\cite{richards_quantitative_2007,lindop_2d_2008, rivaz_real-time_2011}. 
 Naturally, these compromise the informational content of the raw data (i.e. the images). In this study we ignore the error introduced by the image registration  process, as the emphasis is on the inversion of the continuum mechanics, PDE-based,  model, and assume that the displacement data are contaminated with noise. We postulate the presence of i.i.d. Gaussian noise denoted here by the random vector $\bs{z} \in \RR^{d_y}$ such that:
\be
 \mb{\hat{y}} = \mb{y(\Psi)} + \mb{\mathbf{z}} , \quad \mb{\mathbf{z}} \tcreview{\sim} \mathcal{N}(\mb{0},\tau^{-1} \mb{I}_{d_y}). \label{eq:objective}
\ee
We  assume that each entry of $\bs{z}$ has zero mean and an unknown variance $\tau^{-1}$ which will also be inferred from the data.
We note that other  models can also be employed as for example  impulsive noise  to account for  outliers  due to instrument calibration or experimental conditions  \cite{jin_variational_2012}.
Generally the difference between observed and model-predicted outputs can be attributed not only to observation errors (noise) but also to model discrepancies \cite{arridge_approximation_2006, kaipio_statistical_2007,koutsourelakis_novel_2012}. In this work such model errors are lumped with observation errors in the $\mb{\mathbf{z}}$-term.  %In this work we presuppose that the magnitude of this error is much smaller than that of the noise.

The likelihood function of the observed data $\mb{\hat{y}}$ i.e. its conditional probability density given the model parameters $\bs{\Psi}$ (and implicitly the model $\mathcal{M}$ itself as described by the aforementioned governing Equations (\ref{eq:discr})) and $\tau$ is:
\be
p(\mb{\hat{y}} |\mb{\Psi},\tau) = \left( \frac{\tau}{2\pi} \right)^{d_y/2} e^{-\frac{\tau}{2} |\hat{\mb{y}}- \mb{y}(\mb{\Psi})|^2}.
\label{eq:like}
\ee
% with
% \be
%  A(\mb{\Psi}) = [\hat{\mb{y}}- \mb{y}(\mb{\Psi})]^T [\hat{\mb{y}}- \mb{y}(\mb{\Psi})].
% \ee

In the Bayesian framework advocated one would also  need to specify priors on the unknown parameters. We defer a detailed discussion of the priors associated with $\bs{\Psi}$ for the next section where the dimensionality reduction aspects are discussed.   With regards to the noise precision  $\tau$ we employ a \tcreview{(conditionally) conjugate} Gamma prior i.e. 
\be
 \tau   \sim Gamma(a_0,b_0).  \label{eq:tauPrior}
\ee
The values of the parameters are taken $a_0=b_0=0$ in the following examples. This corresponds to a limiting case where the density degenerates to an improper, non-informative Jeffreys  prior i.e. $p(\tau) \propto \frac{1}{\tau}$ that is scale invariant \cite{gelman_bayesian_2003}. Naturally more informative choices can be made if such information is available a priori.

%  According to Bayes rule, for a chosen model $\mathcal{M}$, with
% \be
% p(\mb{\Psi}|\mb{\hat{y}}, \mathcal{M}) = \frac{  p(\mb{\hat{y}}|\mb{\Psi},\mathcal{M})  p(\mb{\Psi}| \mathcal{M})}  { p(\mb{\hat{y}} |\mathcal{M})}.
% \ee
% $p(\mb{\Psi}|\mb{\hat{y}}, \mathcal{M})$ is the \textit{posterior}, $p(\mb{\hat{y}}|\mb{\Psi},\mathcal{M})$ is the \textit{likelihood} probability distribution and includes the model and the measured displacements.  $p(\mb{\Psi}|\mathcal{M})$ is the \textit{prior} which includes prior information/knowledge about the material parameter which we have before making experiments. $ p(\mb{\hat{y}} |\mathcal{M})$ is the \textit{evidence} of the measurements given the chosen model.
% \\The likelihood distribution is modeled by a multivariate Gaussian distribution capturing the displacement relationship from equation (\ref{eq:objective}): 
% \be
% p(\mb{\hat{y}} |\mb{\Psi},\tau) = \mathcal{N}( \mb{y}(\mb{\Psi}), \frac{1}{\tau}) = [\frac{\tau}{2\pi}]^{N/2} e^{-\frac{\tau}{2}A(\mb{\Psi})}
% \ee
% with
% \be
%  A(\mb{\Psi}) = [\hat{\mb{y}}- \mb{y}(\mb{\Psi})]^T [\hat{\mb{y}}- \mb{y}(\mb{\Psi})].
% \ee
% When $\mb{y}(\mb{\Psi})$ becomes closer to $\hat{\mb{y}}$ the probability  $p(\mb{\hat{y}} |\mb{\Psi},\tau)$  is increases.
% We use a conjugate Gamma prior for the inverse of the noise variance:
% \be
%  p(\tau)  \approx Gamma(a_0,b_0)  \label{eq:tauPrior}
% \ee
% We assume no model error for the time being. The prior on the unknown material parameters follows within the next subsection.

\subsubsection{Dimensionality Reduction  for $\bs{\Psi}$}

%\subsection{Subspaces}
\label{sec:ARDSubspaces}

As mentioned earlier one of the primary goals of the present work to is identify, with the least number of forward calls,  a lower-dimensional subspace in $\RR^{\dpsi}$ on which the posterior probability density  can be sufficiently-well approximated. 
Dimensionality reduction could be enforced directly by appropriate prior specification. For example in \cite{honarvar_sparsity_2012} the Fourier transform coefficients of $\bpsi$ corresponding to small-wavelength fluctuations were turned-off by assigning zero prior probability to non-zero values.
While such an approach achieves the goal of dimensionality reduction it does not take into account the forward model in doing so. The nonlinear map $\bs{y}(\bs{\Psi})$ as well as the available data $\hat{\bs{y}}$ provide \tcreview{varying} amounts of information for  identifying different features of $\bs{\Psi}$. One would expect the likelihood (which measures the degree of fit of model predictions with the data) to exhibit different levels of sensitivity  along different  directions in the $\bs{\Psi}$-space. 
\tcreview{
Consider for example Laplace's method which is based on a semi-analytic Gaussian approximation around the Maximum-A-Posteriori estimate $\bs{\Psi}_{MAP}$ \cite{mackay_choice_1998, bishop_pattern_2006}.
 The negative of the Hessian of the log-posterior (assuming this is positive-definite) serves as the covariance matrix. As it was shown 
in \cite{bui-thanh_extreme-scale_2012} in many inverse problems this covariance matrix exhibits a significant discrepancy in its eigenvalues which was exploited in constructing low-rank approximations. At one extreme, there would be principal directions (with small variance) along which the slightest change  from  from $\bs{\Psi}_{MAP}$ would cause a huge decrease in the posterior and on the other, there would principal directions (with large variance)  along which the posterior  would remain almost constant.  Such principal directions  will naturally encapsulate the effect of the log-prior. 
%It is obvious that along the latter directions the posterior in the vicinity of $\bs{\Psi}_{MLE}$ would be largely determined by the prior whereas in the former directions, the likelihood term would dominate the posterior.
In the proposed scheme however, {\em only} the data log-likelihood affects the directions with the maximal posterior variance \cite{tiangang_cui_likelihood-informed_2014}.
 More importantly perhaps we propose a unified framework where the identification of the subspace with the largest posterior variance is performed {\em simultaneously} with the inference of the posterior under the same Variational Bayesian objective. This yields not  only  a highly efficient algorithm (in terms of the number of forward solves) but also a highly extendable framework as discussed in the conclusions.
}

%Consider for example the Maximum-Likelihood-Estimate (MLE) $\bs{\Psi}_{MLE}$ (or in general a local maximum) in \refeq{eq:like} \footnote{A similar argument can be made if one considers the Maximum-A-Posteriori estimate $\bs{\Psi}_{MAP}$}. At one extreme, there would be variations from $\bs{\Psi}_{MLE}$ along directions where even the slightest change would cause a huge drop in the likelihood, and on the other, there would be other variations from $\bs{\Psi}_{MLE}$ along which the likelihood would remain constant. 
%It is obvious that along the latter directions the posterior in the vicinity of $\bs{\Psi}_{MLE}$ would be largely determined by the prior whereas in the former directions, the likelihood term would dominate the posterior.

%The goal of this work is to identify such directions. \tcr{ We approach this problem by employing  probabilistic arguments and the invoking the quality of the approximation to the posterior as our guiding objective. We note that a similar effort based on  linear-algebraic objectives with regards to the Hessian of the map $\bs{y}(\bs{\Psi})$ around $\bs{\Psi}_{MLE}$ or $\bs{\Psi}_{MAP}$  %(which might not be unique anyway)
% was implemented  in \cite{bui-thanh_extreme-scale_2012}. Differences with this approach are discussed in more detail at the end of section  \ref{sec:vba}. {\em  We demonstrate how this can be done by using a fully Bayesian argumentation and employing the marginal likelihood or evidence  \cite{bishop_pattern_2006} as the ultimate model validation metric for any proposed dimensionality reduction.} }

\tcreview{The inference and dimensionality reduction problems  are approached by employing fully Bayesian argumentation and invoking the quality of the approximation to the posterior as our guiding objective.} To that end we postulate the following representation for the high-dimensional vector of unknowns $\bs{\Psi}$:
\be
\underbrace{\mb{\Psi}}_{\dpsi \times 1} =\underbrace{ \bs{\mu}}_{\dpsi \times 1}  + \underbrace{ \mb{W}}_{\dpsi \times \dth} \underbrace{\mb{\Theta}}_{\dth \times 1}. 
\label{eq:red}
\ee
 The motivation behind such a decomposition is quite intuitive as it resembles a Principal Component Analysis (PCA) model \cite{tipping_probabilistic_1999}. The vector  $\bs{\mu}$ represents the mean value of the representation of $\bpsi$ whereas $\bs{\Theta}$ the reduced (and latent) coordinates 
 of $\mb{\Psi}$ along the linear subspace spanned by the $\dth$ columns of the matrix $\bs{W}$. 
 The linear decomposition of a high-dimensional vector such as $\bs{\Psi}$ has received a lot of attention in several different fields. Most commonly $\bpsi$ represents a high-dimensional signal  (e.g. an image, an audio/video recording)  and $\bs{W}$ consists of  an over- or under-complete basis set   \cite{olshausen_sparse_1997,dobigeon_bayesian_2010} % baraniuk_compressive_2007,dobigeon_bayesian_2010} 
 which attempts to encode the signal as {\em sparsely} as possible. Significant advances in Compressed Sensing \cite{candes_robust_2006} or Sparse Bayesian Learning \cite{wipf_sparse_2004} have been achieved in recent years along these lines. A host of deterministic \cite{lee_efficient_2006} or probabilistic \cite{seeger_variational_2010} algorithms have been developed for identifying the reduced-coordinates $\bs{\Theta}$ (or their posterior)  as well as techniques for learning the most appropriate set of basis $\bs{W}$  (dictionary learning) i.e. the one that can lead to the sparsest possible 
representation.  While all these 
tools are pertinent to the present problem they differ in a fundamental way. They are based  on several data/
observations/instantiations of $\bs{\Psi}$ whereas in our problem we do not have such direct observations i.e. the data available pertains to $\bs{y}$ which is nonlinearly and implicitly dependent on $\bs{\Psi}$. Furthermore we are primarily interested in  approximating the posterior on $\bpsi$ rather than the dimensionality reduction itself.

% Before we embark in the discussion of how these objectives  can be achieved we note that an interesting possibility which is not explored here is to use a reduced but continuous representation of the material parameter field(s) $\psi(\bs{X})$ directly instead of its discretized representation. In this case one would express $\psi(\bs{X})$ as:
% \be
% \psi(\bs{X}) = \mu(\bs{X}) + \sum_{i = 1}^{\dth} \Theta_i w_i(\bs{X}) 
%\ee
%where $\Theta_i$ play again the role of reduced coordinates as in \refeq{eq:red} and $ \mu(\bs{X})$, $w_i(\bs{X})$ are the continuous counterparts of $\bs{\mu}$ and $\bs{W}$.

We focus now on the representation of \refeq{eq:red} and proceed to discuss the identification of $\bs{\mu}$, $\bs{W}$ and $\bs{\Theta}$. In a fully Bayesian setting these parameters would be equipped with priors,  say $p(\bs{\mu}), p(\bs{W}), p(\bt)$ respectively\footnote{We use the same symbol \textbf{$p( )$} for various densities without super/subscripts for economy of notation. The parameters each density pertains to, can be identified from its arguments.}, and their {\em joint} posterior would be sought:
\be
p(\bs{\mu}, \bs{W}, \bs{\Theta}, \bs{\tau}| \hat{\bs{y}}) \propto p( \hat{\bs{y}} |\bs{\mu}, \bs{W}, \bs{\Theta}, \bs{\tau})~ p(\bs{\mu})p(\bs{W})p(\bt) p(\tau)
\label{eq:jpost}
\ee
where   $p(\tau)$ represents  the Gamma prior for $\tau$ discussed in \refeq{eq:tauPrior}. Such an inference problem would in general be formidable particularly with regards to $\bs{\mu}$ and $\bs{W}$ whose dimension is dominated by $\dpsi >>1$. To \tcreview{address} this difficulty we propose computing  point estimates for $\bs{\mu}$ and $\bs{W}$ while inferring the whole posterior of $\bs{\Theta}$. 
 In doing so for $\bs{\mu}$ and $\bs{W}$ the natural objective function would be the marginal posterior  
   $p( \bs{\mu} , \bs{W} | \bs{\hat{y}} )$:
   \be
  p( \bs{\mu} , \bs{W} | \bs{\hat{y}} )=\int p(\bs{\mu}, \bs{W}, \bs{\Theta}, \bs{\tau}| \hat{\bs{y}})~d\bt d\tau.
 \label{eq:mpost}
 \ee
   In such a case  the point estimates for $\bs{\mu} , \bs{W}$ would be the Maximum-a-Posteriori-Estimates (MAP). 
 We note that (up to an additive constant):
 \be
 \begin{array}{ll} 
 \log p( \bs{\mu} , \bs{W} | \bs{\hat{y}} ) & = \log \int p(\bs{\mu}, \bs{W}, \bs{\Theta}, \bs{\tau}| \hat{\bs{y}})~d\bt d\tau \\
 & = \log \int p( \hat{\bs{y}} |\bs{\mu}, \bs{W}, \bs{\Theta}, \bs{\tau})~ p(\bs{\mu})p(\bs{W})p(\bt) p(\tau)~d\bt d\tau \\
 & = \log \int p( \hat{\bs{y}} |\bs{\mu}, \bs{W}, \bs{\Theta}, \bs{\tau})p(\bt) p(\tau)~d\bt d\tau + \log p(\bs{\mu}) +\log p(\bs{W}) \\
 & = \log  \int \left( \frac{\tau}{2\pi}  \right)^{d_y/2} e^{-\frac{\tau}{2} |\hat{\mb{y}}- \mb{y}(\bs{\mu} + \bs{W} \bs{\Theta} )|^2}~p(\bs{\Theta})~p(\tau) ~d\bs{\Theta}~d\tau  \\
 & + \log p(\bs{\mu}) +\log p(\bs{W}).
 \end{array}
 \label{eq:logpost}
\ee
We note that such an integration is analytically impossible primarily due to the nonlinear and implicit nature of $\mb{y}(\bs{\mu} + \bs{W} \bs{\Theta} )$ and secondarily due to the coupling of $\bs{\Theta}$ and $\tau$.  
 To that end we employ a Variational Bayesian approximation \cite{bishop_pattern_2006} to the integral in \refeq{eq:logpost}.  We provide further details in the next section. 
 We note that similar  approximations have been employed in previous works  \cite{jin_hierarchical_2010, jin_variational_2012, chappell_variational_2009} in order to expedite Bayesian inference. The novel element of this work pertains to the dimensionality reduction that can be achieved.

%   We  note here that the (marginal) likelihood or evidence $p(\bs{\hat{y}} | \bs{\mu} , \bs{W})$ is the denominator in  Bayes' formula for the posterior on $(\bt,\tau)$:
%   \be
%   p(\bt, \tau | \bs{\hat{y}} , \bs{\mu} , \bs{W})= \frac{ p(\bs{\hat{y}}| \bs{\Theta}, \tau, \bs{\mu} , \bs{W})~p(\bs{\Theta})~p(\tau) }{ p(\bs{\hat{y}} | \bs{\mu} , \bs{W}) }
%   \label{eq:bayesrule}
%   \ee

 \subsubsection{Variational Bayesian approximation}
\label{sec:vba}
% \tcr{
% In practice it is often analytically intractable to derive Bayesian inference of a probability distribution and to derive the integrals. Sampling methods, as \textit{Markov Chain Monte Carlo} (MCMC) or approximation methods as \textit{Laplace approximation} are often too expensive, compare also Section \ref{sec:Introduction.}
 
% \tcr{On option, which we discusses in the following, is the \textbf{Variational Bayes} method. It approximates the posterior $p(\mb{\Theta}, \tau |\mb{\hat{y}})$ with a simpler probability distribution $q(\mb{\Theta}, \tau)$. $q(\mb{\Theta}, \tau)$, which is selected from a family of distributions, should be as close as possible to the exact posterior. One big  advantage of the variational approach is that convergence can be monitored and that the algorithm is fast.  }
 
\tcreview{ Consider an arbitrary joint density $q(\bt, \tau)$ on the latent variables $\bt, \tau$. Then by employing Jensen's inequality one can construct a lower  bound to the log-marginal-posterior $\log p(\bs{\mu} , \bs{W} | \bs{\hat{y}})$ in \refeq{eq:logpost} as  follows:
\be
 \begin{array}{ll} 
 \log p( \bs{\mu} , \bs{W} | \bs{\hat{y}} ) & = \log \int  p( \bs{\mu}, \bs{W}, \bs{\Theta}, \bs{\tau} | \hat{\bs{y}} )~d\bt d\tau\\
 & = \log \int q(\bt,\tau) \frac{ p( \bs{\mu}, \bs{W}, \bs{\Theta}, \bs{\tau} | \hat{\bs{y}} )}{q(\bt,\tau)} ~d\bt d\tau\\
 &  \ge \int q(\bt,\tau) \log \frac{ p( \bs{\mu}, \bs{W}, \bs{\Theta}, \bs{\tau} | \hat{\bs{y}} )}{q(\bt,\tau)} ~d\bt d\tau\\
 & = \mathcal{F}(q(\bt, \tau), \bs{\mu}, \bs{W}).
%  & =\log \int p( \hat{\bs{y}} |\bs{\mu}, \bs{W}, \bs{\Theta}, \bs{\tau})p(\bt) p(\tau)~d\bt d\tau + \log p(\bs{\mu}) +\log p(\bs{W}) \\
%  & = \log \int q(\bt,\tau) \frac{  p( \hat{\bs{y}} |\bs{\mu}, \bs{W}, \bs{\Theta}, \bs{\tau})p(\bt) p(\tau)}{q(\bt,\tau)}~d\bt d\tau + \log p(\bs{\mu}) +\log p(\bs{W}) \\
%  & \ge \int q(\bt,\tau) \log  \frac{  p( \hat{\bs{y}} |\bs{\mu}, \bs{W}, \bs{\Theta}, \bs{\tau})p(\bt) p(\tau)}{q(\bt,\tau)}~d\bt d\tau + \log p(\bs{\mu}) +\log p(\bs{W}) \\
%  & = E_q\left[ \log \frac{ p(\bs{\hat{y}}| \bs{\Theta}, \tau, \bs{\mu} , \bs{W})~p(\bs{\Theta})~p(\tau)}{q(\bt, \tau)}\right]  + \log p(\bs{\mu}) +\log p(\bs{W}) \\
%  & = \mathcal{F}(q(\bt, \tau), \bs{\mu}, \bs{W})
 \end{array}
   \label{eq:loglike1}
\ee
}

% \be
%   \begin{array}{ll}
%  \log  p(\bs{\hat{y}} | \bs{\mu} , \bs{W}) 
%   & =\log \int p(\bs{\hat{y}}| \bs{\Theta}, \tau, \bs{\mu} , \bs{W})~p(\bs{\Theta})~p(\tau)~d\bs{\Theta}~d\tau \\
%   & =\log \int \frac{ p(\bs{\hat{y}}| \bs{\Theta}, \tau, \bs{\mu} , \bs{W})~p(\bs{\Theta})~p(\tau)}{q(\bt, \tau)} ~q(\bt, \tau)~d\bs{\Theta}~d\tau \\
%     & \ge \int q(\bt, \tau) ~\log \frac{ p(\bs{\hat{y}}| \bs{\Theta}, \tau, \bs{\mu} , \bs{W})~p(\bs{\Theta})~p(\tau)}{q(\bt, \tau)} ~d\bs{\Theta}~d\tau \\
% & = E_q\left[ \log \frac{ p(\bs{\hat{y}}| \bs{\Theta}, \tau, \bs{\mu} , \bs{W})~p(\bs{\Theta})~p(\tau)}{q(\bt, \tau)}\right] \\
% & = \mathcal{F}(q(\bt, \tau), \bs{\mu}, \bs{W}).
%   \end{array}
%   \label{eq:loglike1}
% \ee  
%where $E_q[\,]$ is the expectation with regards to $q$.
\tcreview{
We  note here that the variational lower-bound $\mathcal{F}$ has an intimate connection with the Kullback-Leibler divergence between $q(\bt, \tau)$ and the (conditional) posterior on $(\bt,\tau)$: 
  \be
  p(\bt, \tau | \bs{\hat{y}} , \bs{\mu} , \bs{W})= \frac{p(\bs{\mu} , \bs{W}, \bt, \tau | \bs{\hat{y}}  )}{ p( \bs{\mu} , \bs{W} | \bs{\hat{y}} )}.
%   \frac{ p(\bs{\hat{y}}| \bs{\Theta}, \tau, \bs{\mu} , \bs{W})~p(\bs{\Theta})~p(\tau) }{ p(\bs{\hat{y}} | \bs{\mu} , \bs{W}) }
  \label{eq:bayesrule}
  \ee
%   where $p(\bs{\hat{y}} | \bs{\mu} , \bs{W})$ is the marginal likelihood or evidence.
In particular, if we denote by $E_q[\,]$ is the expectation with regards to $q$: 
% \\The KL-divergence between  $q(\bt, \tau)$ and the posterior $p(\bt, \tau | \bs{\hat{y}} , \bs{\mu} , \bs{W})$ in \refeq{eq:bayesrule}
%  is:
%\be
%  KL(q||p) = -E_q[log\, p] + E_q[log \,q].  \label{eq:KLDiv}
%\ee
%With $q= q(\bt, \tau)$ and $p = p(\bt, \tau | \bs{\hat{y}} , \bs{\mu} , \bs{W})$ it follows:
\be
\begin{array}{ll}
  KL\left( q(\bt, \tau)  || p(\bt, \tau | \bs{\hat{y}} , \bs{\mu} , \bs{W}) \right) & = -E_q \left[ \log \frac{p(\bt, \tau | \bs{\hat{y}} , \bs{\mu} , \bs{W})}{ q(\bt, \tau) } \right] \\
  & = -E_q \left[ \log  \frac{p(\bs{\mu} , \bs{W}, \bt, \tau | \bs{\hat{y}}  ) }{p( \bs{\mu} , \bs{W} | \bs{\hat{y}} )~q(\bt, \tau) } \right] \\
  & = \log p( \bs{\mu} , \bs{W} | \bs{\hat{y}} )-\mathcal{F}(q(\bt, \tau), \bs{\mu}, \bs{W}).
  \end{array}
\ee
}
% Hence the lower bound $\mathcal{F}$ of the log-evidence $\log  p(\bs{\hat{y}} | \bs{\mu} , \bs{W})$ can also be expressed as:
%  \be
%  \begin{array}{ll}
%   \mathcal{F}(q(\bt, \tau), \bs{\mu}, \bs{W}) & = E_q\left[ \log \frac{ p(\bs{\hat{y}}| \bs{\Theta}, \tau, \bs{\mu} , \bs{W})~p(\bs{\Theta})~p(\tau)}{q(\bt, \tau)}\right] \\
%   & = \log  p(\bs{\hat{y}} | \bs{\mu} , \bs{W}) -KL\left( q(\bt, \tau)  || p(\bt, \tau | \bs{\hat{y}} , \bs{\mu} , \bs{W}) \right)
%  \end{array}
%  \label{eq:fb}
% \ee
By definition the KL-divergence is non-negative and it becomes $0$ when $q(\bt, \tau)   \equiv p(\bt, \tau | \bs{\hat{y}} , \bs{\mu} , \bs{W}) $. Hence , for a given $\bs{\mu} , \bs{W}$, constructing a good approximation to the conditional  posterior (in the KL-divergence sense) is equivalent to maximizing the lower bound $\mathcal{F}(q(\bt, \tau), \bs{\mu}, \bs{W})$ with regards to $q(\bt, \tau)$.

\begin{figure}[t]{
        \centering
%         \psfrag{ftm1}{\small $\mathcal{F}(q^{(t-1)},\bs{R}^{(t-1)})$}
%         \psfrag{kltm1}{\small $KL(q^{(t-1)} || p_{aux}(. |\bs{R}^{(t-1)}))$}
%          \psfrag{logptm1}{\small $\log p_{aux}(\bs{R}^{(t-1)})$}
%          \psfrag{estep}{VB-E-step}
% %
%   \psfrag{ft}{\small $\mathcal{F}(q^{(t)},\bs{R}^{(t-1)})$}
%         \psfrag{klt}{\small $KL(q^{(t)} || p_{aux}(. |\bs{R}^{(t-1)}))$}
%          \psfrag{logpt}{\small$\log p_{aux}(\bs{R}^{(t-1)})$}
%          \psfrag{mstep}{VB-M-step}
% %
%   \psfrag{ftp1}{\small $\mathcal{F}(q^{(t)},\bs{R}^{(t)})$}
%         \psfrag{kltp1}{\small $KL(q^{(t)} || p_{aux}(. |\bs{R}^{(t)}))$}
%          \psfrag{logpt1}{\small $\log p_{aux}(\bs{R}^{(t)})$}
      \includegraphics[width=0.90\textwidth]{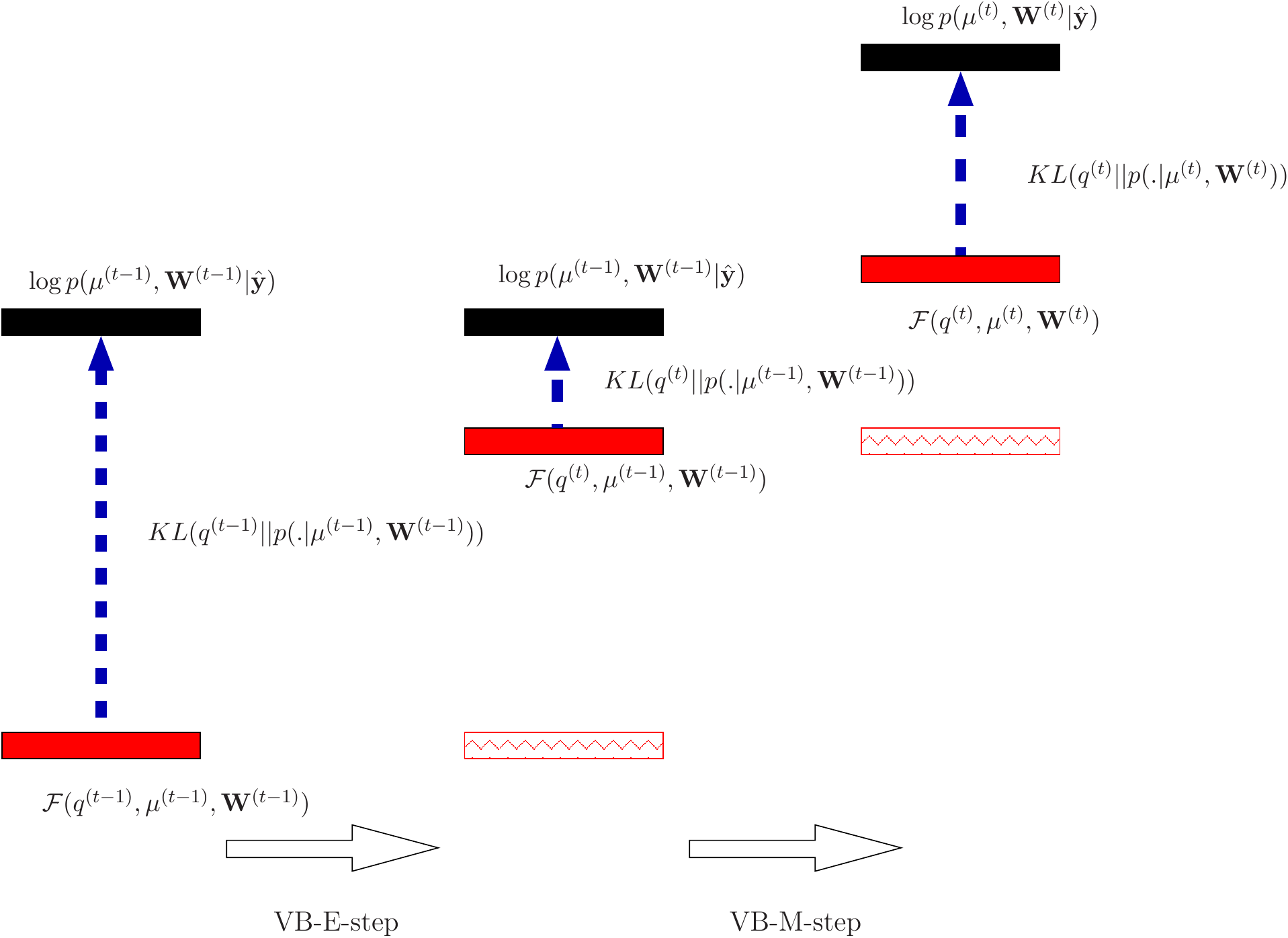}}
      \caption{ During the VB-E step, optimization with respect to the approximating distribution $q$ takes place, whereas during the VB-M step, $\mathcal{F}$ is optimized with respect to the model parameters $\bs{\mu},\bs{W}$ (adapted from \cite{beal_variational_2003})}
      %Fig. 2.5 in his thesis
       \label{fig:vbem}
\end{figure}

\tcreview{The aforementioned discussion suggests an iterative optimization scheme that resembles the  Variational Bayes - Expectation-Maximization (VB-EM) methods that have  appeared  in Machine Learning literature  \cite{beal_variational_2003}. 
At each iteration $t$, one alternates between (Figure \ref{fig:vbem}):
\bi
\item \textbf{VB-Expectation}: Given $(\bs{\mu}^{(t-1)}, \bs{W}^{(t-1)})$, find:
\be
q^{(t)} (\bt, \tau) =\arg \max_q \mathcal{F}(q(\bt, \tau)), \bs{\mu}^{(t-1)}, \bs{W}^{(t-1)})
\ee
\item \textbf{VB-Maximization}: Given $q^{(t)} (\bt, \tau)$, find:
\be
(\bs{\mu}^{(t)}, \bs{W}^{(t)}) =\arg \max_{\bs{\mu},\bs{W}} \mathcal{F}(q^{(t)}(\bt, \tau), \bs{\mu},\bs{W}).
\ee
\ei
}

\begin{figure}[t]{
        \centering
%         \psfrag{q}{$q(\bt, \tau)$}
%         \psfrag{R}{$\{\bs{\mu},\bs{W}\}$}
%                 \psfrag{F}{$\mathcal{F}(q(\bt, \tau),\bs{\mu},\bs{W})$}
\includegraphics[width=0.60\textwidth]{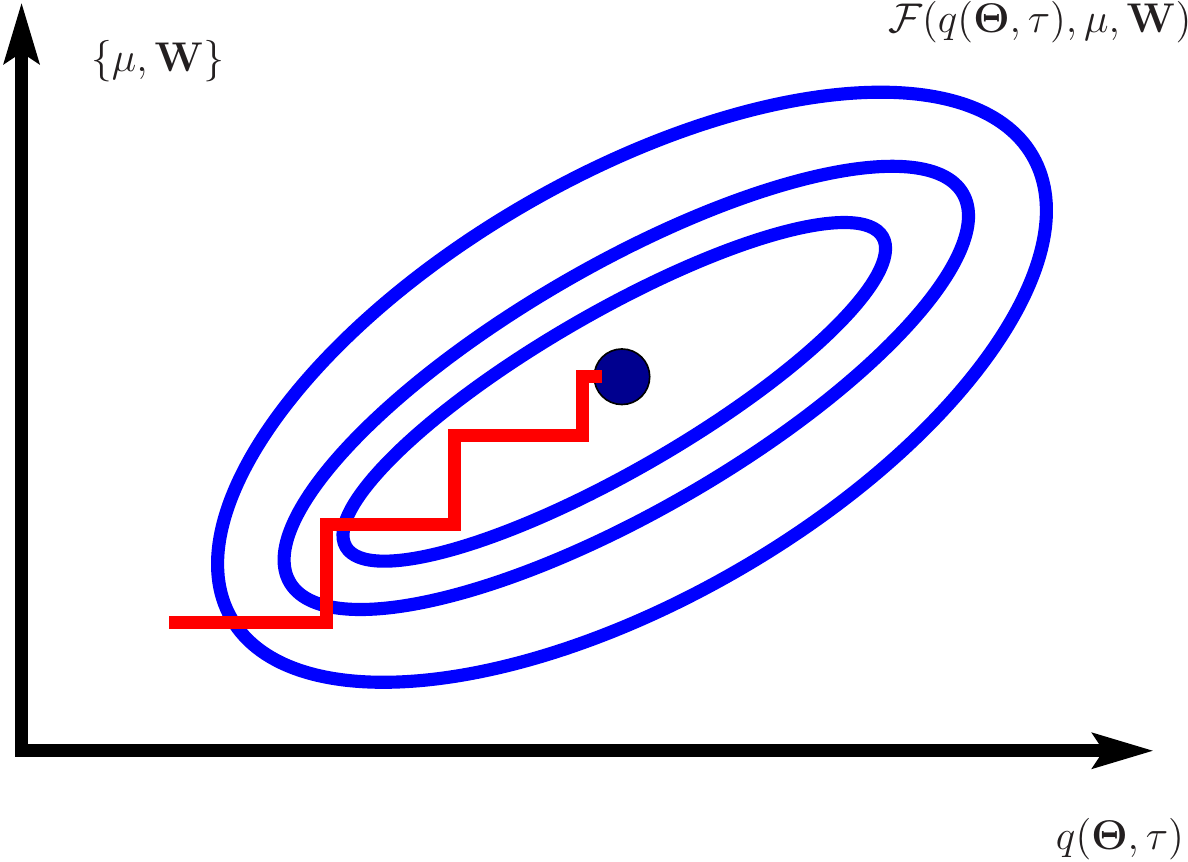}}
        \caption{Variational Bayesian Expectation-Maximization (VB-EM, \cite{beal_variational_2003})}
        \label{fig:maxF}
\end{figure}

\tcreview{In plain terms, the strategy advocated in order to carry out the inference task  can be described as a generalized coordinate ascent with regards to $\mathcal{F}$ (Figure \ref{fig:maxF}).
}

%%%%%%%%%%%%%%%%%%%

From Equations (\ref{eq:jpost}) and (\ref{eq:loglike1}), we have that:
\be
 \begin{array}{ll} 
 \mathcal{F}(q(\bt, \tau), \bs{\mu}, \bs{W}) & =\int q(\bt,\tau) \log \frac{ p( \bs{\mu}, \bs{W}, \bs{\Theta}, \bs{\tau} | \hat{\bs{y}} )}{q(\bt,\tau)} ~d\bt d\tau\\
  & = \int q(\bt,\tau) \log  \frac{  p( \hat{\bs{y}} |\bs{\mu}, \bs{W}, \bs{\Theta}, \bs{\tau})p(\bt) p(\tau)}{q(\bt,\tau)}~d\bt d\tau + \log p(\bs{\mu}) +\log p(\bs{W}) \\
  & = E_q\left[ \log \frac{ p(\bs{\hat{y}}| \bs{\Theta}, \tau, \bs{\mu} , \bs{W})~p(\bs{\Theta})~p(\tau)}{q(\bt, \tau)}\right]  + \log p(\bs{\mu}) +\log p(\bs{W}) \\
  & = \hat{\mathcal{F}}(q(\bt, \tau), \bs{\mu}, \bs{W})  + \log p(\bs{\mu}) +\log p(\bs{W}) 
 \end{array}
   \label{eq:Fvar1}
\ee
where (up to an additive constant):
\be
\begin{array}{ll} 
\hat{\mathcal{F}}(q(\bt, \tau), \bs{\mu}, \bs{W}) & = E_q\left[ \log \frac{ p(\bs{\hat{y}}| \bs{\Theta}, \tau, \bs{\mu} , \bs{W})~p(\bs{\Theta})~p(\tau)}{q(\bt, \tau)}\right] \\
& = E_q\left[ \log \left( \frac{\tau}{2\pi}  \right)^{d_y/2} e^{-\frac{\tau}{2} |\hat{\mb{y}}- \mb{y}(\bs{\mu} + \bs{W} \bs{\Theta} )|^2}\right]+ E_q\left[ \log \frac{p(\bs{\Theta})~p(\tau)}{q(\bt, \tau)}\right]. \\
\end{array}
\label{eq:Fvar2}
\ee

% The latter equation arises due to \refeq{eq:red}. We note that such an integration is analytically impossible primarily due to the nonlinear and implicit nature of $\mb{y}(\bs{\mu} + \bs{W} \bs{\Theta} )$ and secondarily due to the coupling of $\bs{\Theta}$ and $\tau$. 

To alleviate the difficulties with the log-likelihood integral above we employ the following approximations:
 \bi
 \item We linearize the map $\bs{y}(\bs{\mu} + \bs{W} \bs{\Theta} )$ at $\bs{\mu}$. Hence:  
\be
\mb{y} (\bs{\mu} + \mb{W \Theta})  \tcreview{=} \mb{y}(\bs{\mu}) + \mb{G} \mb{W\Theta} \tcreview{+ \mathcal{O}(|\bt|^2)}              \label{eq:FirstTaylorDisp}
\ee
where $ \mb{G}=\frac{\pa \bs{y} }{\pa \bs{\Psi} }|_{\bpsi=\bs{\mu}}$ is the gradient of the map at $\bs{\mu}$.

\tcreview{By keeping the first order terms from \refeq{eq:FirstTaylorDisp} }, the term $|\hat{\mb{y}}- \mb{y}(\bs{\mu} + \bs{W} \bs{\Theta} )|^2$ in the exponent of the likelihood becomes:
\be
\begin{array}{ll}
| \hat{\mb{y}}- \mb{y}(\bs{\mu} + \bs{W} \bs{\Theta} )|^2 &= | \hat{\mb{y}}- \mb{y}(\bs{\mu}) - \mb{G} \mb{W\Theta}   |^2 \\
& = | \hat{\mb{y}}- \mb{y}(\bs{\mu})|^2-2(\hat{\mb{y}}- \mb{y}(\bs{\mu}))^T \mb{G} \mb{W\Theta} \\
 & +
 \bs{W}^T \bs{G}^T \bs{G} \bs{W}: \bt \bt^T.
 \label{eq:l1}
\end{array}
\ee
We note here that a quadratic expression with respect to $\bt$ could also be obtained by considering the $2^{nd}$ order Taylor series of $|\hat{\mb{y}}- \mb{y}(\bs{\mu} + \bs{W} \bs{\Theta} )|^2$ around $\bs{\mu}$ directly. In particular if we denote by $\bs{g}=\frac{ \pa |\hat{\mb{y}}- \mb{y}(\bpsi)|^2 }{ \pa \bpsi  }|_{\bpsi=\bs{\mu}}$ and $\bs{H}=\frac{ \pa |\hat{\mb{y}}- \mb{y}(\bpsi)|^2 }{ \pa \bpsi \pa \bpsi^T }|_{\bpsi=\bs{\mu}}$   \tcreview{and keeping only up to second order terms yields}:
\be
\begin{array}{ll}
| \hat{\mb{y}}- \mb{y}(\bs{\mu} + \bs{W} \bs{\Theta} )|^2 & =| \hat{\mb{y}}- \mb{y}(\bs{\mu})|^2 + \mb{g}^T \mb{W\Theta}   \\
&  +
 \frac{1}{2}\bs{W}^T \bs{H} \bs{W}: \bt \bt^T.
\end{array}
\label{eq:l2}
\ee
The computation of $2^{nd}$ order derivatives $\bs{H}$ can also be addressed within the adjoint framework. We refer the interested reader to \cite{hinze_optimization_2009,bui-thanh_adaptive_2012} as we do not pursue this possibility further in this work. The ensuing expressions are based on \refeq{eq:l1} but can be readily adjusted to include the terms in \refeq{eq:l2} instead \footnote{The only additional requirement is that $\bs{H}$ is semi-positive definite or that a semi-positive approximation $\tilde{\bs{H}} \approx \bs{H}$ is used.}.

We  note that by making use of the linearization of the map $\bs{y}(\bs{\Psi})$ and the Variational Bayesian approximation, 
 one can obtain a tractable approximations  of the  posterior of the latent parameters $\bt$ and $\tau$.  This will enable us to ultimately identify all model parameters and through this process the optimal subspace for approximating the posterior on $\bpsi$. This will be explained in detail when the  final algorithm is presented in section \ref{sec:UpdateEquations}.

\item The aforementioned equations for the VB-Expectation step  imply that probabilistic inference can be expressed in terms of a parametric optimization problem. One can adopt a functional form for  $q(\bt, \tau) $ depending on an appropriate set of parameters and identify their optimal value by minimizing the KL-divergence with the posterior or equivalently maximizing $\mathcal{F}$. We adopt a {\em mean-field} approximation where one looks for factorized densities of the form:
\be 
q(\mb{\Theta}, \tau ) = q(\mb{\Theta})q(\tau). \label{eq:MeanFieldApp}
\ee
Variational mean-field approximations  have their origin in statistical physics \cite{peierls_minimum_1938}. We make these expressions more specific in the next sections where we discuss the prior for $p(\bt)$ as well.

 \ei

 \subsubsection{Prior Specification for $\bt, \bs{\mu}$ and $\bs{W}$ }
\label{sec:priors}
 
 We discuss first the prior specification on $\bs{W}$. Its $\dth$ columns $\bs{w}_i, ~i=1,\ldots , 
 \dth$ span the subspace over which an approximation of $\bpsi$ is sought. 
 We note  that $\bpsi$ depends on the product $\bs{W} \bt$ which would remain invariant by appropriate rescaling of each pair of $\bs{w}'_i=\alpha_i ~\bs{w}_i$ and $\Theta'_i=\frac{1}{\alpha_i} \Theta_i$ for any $\alpha_i$. Hence, to resolve identifiability issues we require that $\bs{W}$ is {\em orthogonal} i.e. $\bs{W}^T \bs{W}=\bs{I}_{\dth}$ where $\bs{I}_{\dth}$ is the $\dth-$dimensional identity matrix. This is equivalent to employing a uniform prior on $\bs{W}$ on the Stiefel manifold $V_{\dth}(\RR^{\dpsi})$ \cite{muirhead_aspects_1982}.

 The latent, reduced coordinates $\bt \in \RR^{\dth}$ capture the variation of $\bpsi$ around its mean $\bs{\mu}$ along the directions of $\bs{W}$ as implied by \refeq{eq:red}. It is therefore reasonable to assume that, a priori, these should have zero mean and should be uncorrelated \cite{tipping_probabilistic_1999}. For that purpose we adopt a multivariate Gaussian prior (denoted by $p(\bt)$ in the Equations of the previous section) with a diagonal covariance denoted by \tcreview{$\bs{\Lambda}_0^{-1}=diag(\lambda_{0,i}^{-1}), i=1, \ldots \dth$}. We select prior variances $\lambda_{0,r}^{-1}$ such that $\lambda_{0,1}^{-1} >   \lambda_{0,2}^{-1}> \ldots >\lambda_{0,\dth}^{-1}$. This induces a natural (stochastic) ordering to the reduced coordinates $\bt$  since $\bpsi$ is invariant to permutations of the entries of the $\bt$ and the columns of $\bs{W}$ (\refeq{eq:red}). As a result of this ordering, $\Theta_1$ is associated with the direction along which the largest variance in $\bs{\Psi}$ is attained, $\
Theta_2$ 
with the 
direction with the second largest 
variance and so on. We discuss the particular   values given to prior hyperparameters  $\lambda_{0,i}$ in the sequel (Section \ref{sec:NumericalExamples}) and in Section \ref{sec:card} the possibility of an adaptive  decomposition is also presented. This enables the sequential addition of reduced coordinates until a sufficiently good  approximation to the posterior is attained.
% Due to the aforementioned reasons, we note that the result on $\bpsi$  is invariant under rotations of $\bt$. In particular if $\bs{V}$ is an orthogonal $R\times R$ matrix, $\bt'=\bs{V} \bt$ and $\bs{W}'=\bs{W} \bs{V}

 The final aspect of the prior model pertains to $\bs{\mu}$. We use a hierarchical prior    that induces the requisite smoothness given that $\bpsi$ represents the spatial variability of the material parameters. In particular the prior model employed penalizes the jumps in the values of $\Psi_k$ and $\Psi_l$ which correspond to neighboring sites/locations $k,l$. The definition of a neighborhood can be adjusted depending on the problem, but in this work we assume that sites/locations belong to the neighborhood if the they correspond to adjacent pixels/voxels. Suppose $d_L$ is the total number of jumps or neighboring  pairs. Then for  $j=1,\ldots, d_L$ if $k_j$ and $l_j$ denote the corresponding neighboring pair:
 \be
 p( \mu_{k_j}-\mu_{l_j}| \phi_j)= \sqrt{ \frac{\phi_j}{2\pi}} e^{ -\frac{\phi_{j}}{2} (\mu_{k_j}-\mu_{l_j})^2}.
 \label{eq:priormuj}
 \ee
 The strength of the penalty is proportional to the hyperparameter $\phi_j$, i.e. smaller values of $\phi_{j}$ induce a weaker penalty and vice versa. Let $\bs{L}$ the $d_L \times d_{\Psi}$ denote the Boolean matrix that can be used to produce the vector of all $d_L$ jumps (as the one above) between all neighboring sites from the vector $\bs{\mu}$ as $\bs{L} \bs{\mu}$, and $\bs{\Phi}=diag(\phi_j)$ the {\em diagonal matrix} containing  all the hyperparameters $\phi_j$ associated with each of these jumps.  We can represent the combined prior on $\bs{\mu}$ as:
 \be
 p(\bs{\mu} | \bs{\Phi}) \propto |\bs{\Phi}|^{1/2} e^{-\frac{1}{2} \bs{\mu}^T \bs{L}^T \bs{\Phi} \bs{L} \bs{\mu} }.
 \ee
A \tcreview{conjugate  prior} of the hyperparameters $\bs{\Phi}$ is a product of Gamma distributions:
\be
p(\bs{\Phi})=\prod_{j=1}^{d_L} Gamma(a_{\phi},b_{\phi}).
\label{eq:priorphi}
\ee
As in \cite{bardsley_hierarchical_2010} the independence is motivated by the absence of correlation (a priori) with regards to the locations of the jumps. In this work we use $a_{\phi} = b_{\phi} = 0$   which corresponds to a limiting case of a Jeffreys prior that is scale invariant. We note that in contrast to previous works where such priors have been employed for the vector of unknowns  $\bs{\Psi}$ and MAP estimates have been obtained \cite{kaipio_statistical_2006}, we employ this here for $\bs{\mu}$ which is only part of the overall decomposition in \refeq{eq:red}.
We discuss in the following section the update equations for $\bs{\mu}$ and the associated hyper-parameters $\bs{\Phi}$ as well as for the remaining model variables.

 \subsubsection{Update equations for $q(\bt),q(\tau), \bs{\mu,W}$}\label{sec:UpdateEquations}
 
% This section discusses the optimization with respect to $q(\bt$) and $q(\tau)$ for the latent variables as well as  the model parameters $\bs{\mu,W}$.  In the absence of a prior for $\bs{\mu,W}$ the objective function is the lower bound $\mathcal{F}$ to the log-evidence in \refeq{eq:loglike1}. Maximizing $\mathcal{F}$ with regards to $q(\bt)$ and $q(\tau)$ is equivalent to minimizing their KL-divergence from the actual posterior on $\bt$ and $\tau$ as discussed in \refeq{eq:fb}. Maximizing $\mathcal{F}$ with regards to $\bs{\mu,W}$ leads to  approximate  MLEs for these parameters. 
% 
% To perform the optimization we propose a coordinate ascent algorithm where we alternate between optimizing $\mathcal{F}$ with regards to one of these variables while keeping the rest fixed as illustrated in Figure \ref{fig:qandWupdate}.     
%   We note that in the presence of the priors $p(\bs{\mu})=\int p(\bs{\mu}|\bs{\Phi}) p(\bs{\Phi})~d\bs{\Phi}$ and $p(\bs{W})$, the optimization should be carried with the objective $\mathcal{F}+\log p(\bs{\mu}) + \log p(\bs{W})$. This is a lower bound on true log-posterior $\log p( \bs{\mu} , \bs{W} | \bs{\hat{y}} )=  \log  p(\bs{\hat{y}} | \bs{\mu} , \bs{W})+ \log p(\bs{\mu})+ \log p(\bs{W})$.
%    \begin{figure}[H]{
%         \centering
%         {\includegraphics[width=0.60\textwidth]{FiguresIF/qandWupdate2.eps}}
%         }
%         \caption{Coordinate ascent: Alternating between optimizations with regards to $\bs{\mu}$, $\mb{W}$ and $q(\bs{\Theta},\tau)$ .}
%         \label{fig:qandWupdate}
% \end{figure}
 We postulate first that the reduced coordinates should a posteriori have zero mean as they capture variability around $\bs{\mu}$. For that purpose we confine our search for $q(\bt)$ to distributions with zero mean.
  Given the aforementioned prior $p(\bt)$ and  the linearization discussed in the previous section, we can readily deduce from \refeq{eq:Fvar1} that the optimal approximate posteriors $q^{opt}(\bt)$ and $q^{opt}(\tau)$ under the mean-field Variational Bayesian scheme adopted will be:
 \be
 q^{opt}(\bt) \equiv \mathcal{N}(\bs{0}, \bs{\Lambda}^{-1}), \quad q^{opt}(\tau)\equiv Gamma(a,b).
 \label{eq:qopt1}
 \ee
 The associated parameters are given by the following {\em iterative} Equations:
 \be
 \begin{array}{c}
  a = a_0 + d_y/2  \\
  b = b_0 + \frac{1}{2} |\mb{\hat{y}} - \mb{y}(\bs{\mu})|^2 + \frac{1}{2} tr( \mb{W}^T \mb{G}^T \mb{G W}  \mb{\Lambda}^{-1}) 
  %b = b_0 + \frac{1}{2} |\mb{\hat{y}} - \mb{y}(\bs{\mu})|^2 -  (\mb{\hat{y}} -\mb{y}(\bs{\mu}))^T \mb{G W m}  + \frac{1}{2} tr( \mb{W}^T \mb{G}^T \mb{G W} (\mb{mm}^T + \mb{\Lambda}^{-1})) \\
  \label{eq:qupdtau}
  \end{array}
\ee
\be
 \begin{array}{c}
  \bs{\Lambda}=\bs{\Lambda}_0+<\tau> \bs{W}^T \bs{G}^T \bs{G} \bs{W} \\
 %\bs{\Lambda} \bs{m}=  <\tau> \mb{W}^T\mb{G}^T  [\mb{\hat{y}} - \mb{y}(\bs{\mu})] 
  \end{array}
  \label{eq:qupdt}
\ee
where $<\tau>=E_{q^{opt}(\tau)}[\tau]=\frac{a}{b}$.

As a result of the aforementioned Equations and \refeq{eq:red}, one can establish that the {\em posterior} of $\bpsi$ is approximated by a  Gaussian with mean and covariance given by:
\be
 \begin{array}{l}
  E[\bpsi]=E[\bs{\mu}+\bs{W}\bt]=\bs{\mu}+\bs{W}\bt \\
  Cov[\bpsi]=\bs{W} \bs{\Lambda}^{-1} \bs{W}^T.
 \end{array}
\ee
 We note that if we diagonalize $\bs{\Lambda}^{-1}$ i.e. $\bs{\Lambda}^{-1}=\bs{V} \bs{D} \bs{V}^T$ where $\bs{D}$ is diagonal and $\bs{V}$ is orthogonal with columns equal to the eigenvectors of $\bs{\Lambda}^{-1}$, then:
 \be
 \begin{array}{ll}
 Cov[\bpsi] & =\bs{W} \bs{V} \bs{D} \bs{V}^T \bs{W}^T \\
 & = \bs{\tilde{W}} \bs{D}\bs{\tilde{W}}^T
 \end{array}
 \ee
 where $\bs{\tilde{W}} $ is also orthogonal (i.e. $\bs{\tilde{W}}^T\bs{\tilde{W}}=\bs{I}_{\dth}$)  
 and contains the $\dth$ principal directions of the posterior covariance of $\bpsi$. Hence it  suffices to consider approximate posteriors $q(\bt)$ with covariance $\bs{\Lambda}^{-1}$ that is {\em diagonal} i.e. $\bs{\Lambda}=diag(\lambda_i), ~i=1,\ldots, \dth$. In this case the update equations for  $\lambda_i$ in \refeq{eq:qupdt} reduce to:
 \be
 \begin{array}{c}
  \lambda_i=\lambda_{0,i}+<\tau> \bs{w}_i^T \bs{G}^T \bs{G} \bs{w}_i. %\\
  %m_i=  \frac{<\tau>}{\lambda_i} \mb{w}_i^T\mb{G}^T  [\mb{\hat{y}} - \mb{y}(\bs{\mu})] 
  \end{array}
  \label{eq:qupd1}
  \ee
 %the aforementioned density of $\bpsi$ is invariant under rotations of $\bt$ i.e. for any orthogonal matrix $\bs{V} \in \RR^{R\times R}$, the vector $\bt'=\bs{m}+\bs{V}(\bt-\ms{m})$ will have the same mean $\bs{m}$ and covariance $\bs{\Lambda}^{-1}
 We note that despite the  prior assumption on uncorrelated  $\bt$, the posterior on $\bpsi$ exhibits correlation and captures  the principal directions along which the variance is largest. Furthermore, implicit to the aforementioned derivations is the assumption of a unimodal posterior on $\bt$ and subsequently on $\bpsi$. This assumption can be relaxed by employing a mixture of Gaussians (e.g. \cite{choudrey_variational_2003}) that will enable the approximation of highly non-Gaussian and potentially multi-modal posteriors. Such approximations could also be combined with the employment of different basis sets $\bs{W}$ for  each of the mixture component which would provide a wide range of possibilities. We defer further discussions along these lines to future work. In the examined elastography applications, the unimodal assumption seems to be a reasonable one due to the generally large of amounts of data/observations obtained from various imaging modalities.

 Given the aforementioned results one can obtain an expression for the variational lower bound $\mathcal{F}$  in \refeq{eq:Fvar1}. For economy of notation we use $<.>$ to denote expectations with respect to $q^{opt}(\bt)$ and/or $q^{opt}(\tau)$ as implied by the arguments:
\be
 \begin{array}{ll}
  \mathcal{F}(q(\bt)q(\tau), \bs{\mu}, \bs{W}) & = E_q\left[ \log \frac{ p(\bs{\hat{y}}| \bs{\Theta}, \tau, \bs{\mu} , \bs{W})~p(\bs{\Theta})~p(\tau)}{q(\bt, \tau)}\right] + \log p(\bs{\mu}) +\log p(\bs{W})  \\
  & =-\frac{d_y}{2} \log2\pi+ \frac{d_y}{2} <\log \tau>-\frac{<\tau>}{2} <|\bs{\hat{y}} - \bs{y}(\bs{\mu})-\bs{G~W} \bt|^2>  %\quad (E_q \left[ \log  p(\bs{\hat{y}}| \bs{\Theta}, \tau, \bs{\mu} , \bs{W})\right]) 
  \\
  & +\frac{1}{2} \log |\bs{\Lambda}_0|-\frac{1}{2} \bs{\Lambda}_0 : <\bt \bt^T>\\ 
  & +(a_0-1)<\log \tau>-b_0<\tau>-\log Z(a_0,b_0) \\
  & -\frac{1}{2} \log |\bs{\Lambda}|+\frac{\dth}{2} \\
  & -(a-1)<\log\tau>+b<\tau>+\log Z(a,b)  \\
  & + \log p(\bs{\mu}) +\log p(\bs{W}) 
 \end{array}
\ee
where $Z(\gamma,\delta)=\frac{\Gamma(\gamma)}{\delta^{\gamma}}$ is the normalization constant of a $Gamma$ distribution with parameters $x,y$. 
The aforementioned equation can be further simplified by making use of the following expectations: $<\bt>= \bs{0} $, $<\bt \bt^T>=\bs{\Lambda}^{-1}$:
\be
 \begin{array}{ll}
  \mathcal{F}(q^{opt}(\bt)q^{opt}(\tau), \bs{\mu}, \bs{W}) & =-\frac{d_y}{2} \log2\pi+ \frac{d_y}{2} < \log \tau> -\frac{<\tau>}{2} |\bs{\hat{y}} - \bs{y}(\bs{\mu})|^2 \\
  %& +<\tau>(\bs{\hat{y}} - \bs{y}(\bs{\mu}))^T \bs{GW} \bs{m} \\
  %&-\frac{<\tau>}{2}\bs{W}^T\bs{G}^T\bs{G~W}:(\bs{m}\bs{m}^T+\bs{\Lambda}^{-1})    \\
  &-\frac{<\tau>}{2}\bs{W}^T\bs{G}^T\bs{G~W}:\bs{\Lambda}^{-1}    \\
  & +\frac{1}{2} \log |\bs{\Lambda}_0|-\frac{1}{2} \bs{\Lambda}_0 :\bs{\Lambda}^{-1}\\ 
  & +(a_0-1)<\log \tau>-b_0<\tau>-\log Z(a_0,b_0) \\
  & -\frac{1}{2} \log |\bs{\Lambda}|+\frac{\dth}{2} \\
  & -(a-1)<\log\tau>+b<\tau>+\log Z(a,b) \\
  & + \log p(\bs{\mu}) +\log p(\bs{W}). 
 \end{array}
 \label{eq:fbmw}
\ee

 %W-updates
 In order to update $\bs{W}$ in the VB-Maximization step, it suffices to consider only the terms of $\mathcal{F}$ that depend on it which we denote by $\mathcal{F}_W(\bs{W})$  i.e.:
 \be
 \begin{array}{ll}
 \mathcal{F}_W(\bs{W})& = -\frac{<\tau>}{2}\bs{W}^T\bs{G}^T\bs{G~W}:\bs{\Lambda}^{-1}  +\log p(\bs{W}) 
  % \mathcal{F}_W(\bs{W})& =<\tau>(\bs{\hat{y}} - \bs{y}(\bs{\mu}))^T \bs{GW} \bs{m} \\
  %&-\frac{<\tau>}{2}\bs{W}^T\bs{G}^T\bs{G~W}:(\bs{m}\bs{m}^T+\bs{\Lambda}^{-1})  
  \end{array}
  \label{eq:fw}
 \ee
 As discussed earlier the prior  $\log p(\bs{W})$  enforces the orthogonality constraint on $\bs{W}$. To address this constrained optimization problem,   we employ the iterative  algorithm proposed in \cite{wen_feasible_2013} which has proven highly efficient in terms of the number of iterations and the cost per iterations in several settings.  It employs   the Cayley transform to preserve the constraint during the optimization and  makes use only of first order derivatives:
   \be
  \frac{\pa \mathcal{F}_W}{\pa \bs{W}}= -<\tau>\bs{G}^T\bs{G} \bs{W} \bs{\Lambda}^{-1} +\log p(\bs{W}) 
   %\frac{\pa \mathcal{F}_W}{\pa \bs{W}}=<\tau> \bs{G}^T \left( (\bs{\hat{y}} - \bs{y}(\bs{\mu})) \bs{m}^T-\bs{G}^T\bs{G} \bs{W} ~(\bs{m}\bs{m}^T+\bs{\Lambda}^{-1}) \right)
  \label{eq:dFdW}
  \ee
  In brief, if  $\bs{B}$ is the skew-symmetric matrix:
 \be
  \bs{B} = \frac{\pa \mathcal{F}_W}{\pa \bs{W}} \bs{W}^T - \bs{W}\frac{\pa \mathcal{F}_W}{\pa \bs{W}} ^T
 \ee
 the update equations are based on a Crank-Nicholson-like scheme:
 \be
  \bs{W}_{new} = (\bs{I}+ \frac{\alpha_W}{2}\bs{B})^{-1}(\bs{I}+ \frac{\alpha_W}{2}\bs{B})\bs{W}_{old} 
  \label{eq:Wupd}
 \ee
 where $\alpha_W$ is the step size. One notes that the aforementioned update preserves the orthogonality of $\bs{W}_{new}$ (\cite{wen_feasible_2013}). In order to derive a good step size we use the Barzilai-Borwein scheme \cite{barzilai_2-point_1988} which results in a non-monotone line search algorithm:
\be
  \alpha_W = \frac{|tr(\Delta \bs{W} \Delta \frac{\pa \mathcal{F}_W}{\pa \bs{W}} )|}   {tr(\Delta\frac{\pa \mathcal{F}_W}{\pa \bs{W}}^T \Delta \frac{\pa \mathcal{F}_W}{\pa \bs{W}}  )} \label{eq:alphaW}
\ee
where $\Delta$ represents the difference between  the current parameter values  as compared to the previous step. As discussed in detail in  \cite{wen_feasible_2013} the  inversion of the $d_{\Psi} \times d_{\Psi}$ matrix $(\bs{I}+ \frac{\alpha_W}{2}\bs{B})$ in \refeq{eq:Wupd} can be efficiently performed by inverting a matrix of dimension $2 d_{\Theta}$ which is much smaller than $d_{\Psi}$. We note that the updates of $\bs{W}$ require no forward calls for the computation of $\bs{y}(\bs{\mu})$ or its derivatives $\bs{G}$. The updates/iterations are terminated when no further improvement to the objective is possible.

The final component involves the optimization of $\bs{\mu}$. As with $\bs{W}$ we consider only the terms of $\mathcal{F}$ (\refeq{eq:fbmw}) that depend on $\bs{\mu}$ which we denote by $\mathcal{F}_{\mu}(\bs{\mu})$  i.e.: 
 \be
 \begin{array}{ll}
 \mathcal{F}_{\mu}(\bs{\mu})& =-\frac{<\tau>}{2} |\bs{\hat{y}} - \bs{y}(\bs{\mu})|^2 + \log p(\bs{\mu}). 
  %\mathcal{F}_{\mu}(\bs{\mu})& =-\frac{<\tau>}{2} \left( |\bs{\hat{y}} - \bs{y}(\bs{\mu})|^2  -2 (\bs{\hat{y}} - \bs{y}(\bs{\mu}))^T \bs{GW} \bs{m} \right)   
  \end{array}
  \label{eq:fmu}
 \ee

 Due to the analytical unavailability of $\log p(\bs{\mu})$ and its derivatives $\frac{ \pa \log p(\bs{\mu})}{\pa \bs{\mu}}$ we employ here an Expectation-Maximization scheme \cite{dempster_maximum_1977,neal_view_1998} which we describe in \ref{app:mu} for completeness. The output of this algorithm is also the posterior on the hyperparameters $\phi_j$  in \refeq{eq:priormuj} which capture the locations of jumps in $\bs{\mu}$ as well as the probabilities associated with them.
The cost of the numerical operations  is minimal and scales linearly with the number of neighboring pairs $d_L$. In the following we simply make use of Equations (\ref{eq:pmu}) without further explanation.

Formally the determination of the optimal $\bs{\mu}$ would require the derivatives $\frac{\pa \mathcal{F}_{\mu}(\bs{\mu})}{\pa \bs{\mu}}$ in \refeq{eq:fmu}.  We note   that $\bs{G}=\frac{\pa \bs{y} }{\pa \bs{\Psi} }|_{\bpsi=\bs{\mu}}$ depends on $\bs{\mu}$. Hence finding $\frac{\pa \mathcal{F}_{\mu}(\bs{\mu})}{\pa \bs{\mu}}$ would require the computation of second-order derivatives of $\bs{y}(\bpsi)$ which poses significant computational  difficulties in the high-dimensional setting considered. To avoid this and {\em only} for the purpose of the $\bs{\mu}$ updates, we linearize \refeq{eq:fmu} around the current guess by ignoring the dependence of $\bs{G}$ on $\bs{\mu}$ or equivalently by assuming that $\bs{G}$ remains constant in the vicinity of the current guess. In particular, let $\bs{\mu}^{(t)}$ denote the value of $\bs{\mu}$ at iteration $t$, then in order to find the increment $\Delta \bs{\mu}^{(t)}$, we define the new objective ${F}_{\mu}^{(t)}(\Delta \bs{\mu}^{(t)})$ as follows:
\be
\begin{array}{ll}
 {F}_{\mu}^{(t)}(\Delta \bs{\mu}^{(t)}) = & {F}_{\mu}(\bs{\mu}^{(t)}+\Delta \bs{\mu}^{(t)})+\log p(\bs{\mu}^{(t)}+\Delta \bs{\mu}^{(t)}) \\
 & = -\frac{<\tau>}{2} |\bs{\hat{y}} - \bs{y}(\bs{\mu}^{(t)}+\Delta \bs{\mu}^{(t)})|^2   \\
 & -\frac{1}{2} (\bs{\mu}^{(t)}+\Delta \bs{\mu}^{(t)})^T \bs{L}^T <\bs{\Phi}> \bs{L} (\bs{\mu}^{(t)}+\Delta \bs{\mu}^{(t)})\\
 & \approx -\frac{<\tau>}{2} |\bs{\hat{y}} - \bs{y}(\bs{\mu}^{(t)})-\bs{G}^{(t)} \Delta \bs{\mu}^{(t)}|^2 \\
 & -\frac{1}{2} (\bs{\mu}^{(t)}+\Delta \bs{\mu}^{(t)})^T \bs{L}^T <\bs{\Phi}> \bs{L} (\bs{\mu}^{(t)}+\Delta \bs{\mu}^{(t)}).\\
\end{array}
\ee
We note that there is no approximation with regards to the $p(\bs{\mu})$ prior term.
By keeping only the terms depending on $\Delta \bs{\mu}^{(t)}$ in the Equation above we obtain:
\be
\begin{array}{ll}
 {F}_{\mu}^{(t)}(\Delta \bs{\mu}^{(t)}) = & -\frac{<\tau>}{2} (\Delta \bs{\mu}^{(t)} )^T (\bs{G}^{(t)} )^T \bs{G}^{(t)}~\Delta \bs{\mu}^{(t)}  +<\tau> (\bs{\hat{y}} - \bs{y}(\bs{\mu}^{(t)}))^T \bs{G}^{(t)}~\Delta \bs{\mu}^{(t)} \\
 & - \frac{1}{2}(\Delta \bs{\mu}^{(t)})^T \bs{L}^T <\bs{\Phi}> \bs{L} ~\Delta \bs{\mu}^{(t)} \\
 & - (\bs{\mu}^{(t)})^T\bs{L}^T <\bs{\Phi}> \bs{L} ~\Delta \bs{\mu}^{(t)}.
\end{array}
\ee
This is a concave and quadratic with respect to the unknown $\Delta \bs{\mu}^{(t)}$. The maximum can be found by setting $ \frac{\pa {F}_{\mu}^{(t)}(\Delta \bs{\mu}^{(t)}) }{\pa \Delta \bs{\mu}^{(t)}}=\bs{0}$ which yields:
\be
(<\tau>  (\bs{G}^{(t)} )^T \bs{G}^{(t)}+\bs{L}^T <\bs{\Phi}> \bs{L}) \Delta \bs{\mu}^{(t)} = <\tau> (\bs{G}^{(t)} )^T (\bs{\hat{y}} - \bs{y}(\bs{\mu}^{(t)})) -\bs{L}^T <\bs{\Phi}> \bs{L} \bs{\mu}^{(t)}.
\label{eq:muupd}
\ee
We  note that the {\em exact} objective ${F}_{\mu}(\bs{\mu})+\log p(\bs{\mu})$ is evaluated at $ \bs{\mu}^{(t+1)}= \bs{\mu}^{(t)}+\Delta \bs{\mu}^{(t)}$  and $\bs{\mu}^{(t+1)}$ is accepted only if the value of the  objective is larger than that at $\bs{\mu}^{(t)}$. Iterations at terminated when no further improvement is possible.
Finally it was found that activating the regularization term ($\log p(\bs{\mu})$)
after five updates/iterations during which the optimization is performed solely on the basis of 
${F}_{\mu}(\bs{\mu})$, enabled better exploration of the feasible solutions.

%Apart from the obvious possibility of continuing the iterations until no further improvement in the exact objective can be achieved, one can  perform a fixed number of iterations (e.g. one).   We note that total number of iterations $t$ can be alternating maximization scheme advocated (Figure \ref{fig:qandWupdate}) can accommodate  a fixed 

% When comparing the lowerbound for the old and the new $\bs{\mu}$ we take the corresponding, new evaluated parameters of $q(\bs{D})$ corresponding to the optimal $\bs{D}$ in each setting.
% \\Updating $\bs{\mu}$ is the most expensive part of the algorithm. The cost for the forward run, which is necessary to derive the $\bs{y}[\bs{\mu}_{n+1}]$ and  the derivatives $\bs{G}$ is of order $O(N^3 + N^2 T )$. In addition also the matrix product $\bs{G}^T \bs{G}$ needs to
% be derived and the system, equation \ref{eq:newMy}, to be solved, which results in additional costs of
% $O(N^T 2 + T^3)$. 

 We summarize below the basic steps of the iterative Variational Bayesian scheme proposed in Algorithm \ref{alg:GenerealVB}. 
\begin{algorithm}[H]
\caption{Variational Bayesian Approach Including Dictionary Learning for fixed $\dth$} \label{alg:GenerealVB}
\begin{algorithmic}[1]
\State $\text{Initialize $\bs{\mu}$, $\mb{W}$,  $\mb{\Lambda}_0$ and the hyperparameters  $a_0$, $b_0$, $a_{\phi}$, $b_{\phi}$}$
\State Update $\bs{\mu}$ using Equation (\ref{eq:muupd})
\While{$\mathcal{F}$ (\refeq{eq:fbmw}) has not  converged}
\State Update $\bs{W}$ using Equations (\ref{eq:fw}-\ref{eq:alphaW})
\State Update $q(\bt)\equiv \mathcal{N}(\bs{0}, \bs{\Lambda}^{-1})$ using \refeq{eq:qupd1} and $q(\tau) \equiv Gamma(a,b)$ using \refeq{eq:qupdtau}
\EndWhile
%\EndProcedure
\end{algorithmic}
\end{algorithm}

With regards to the overall computational cost we note that the updates of $\bs{\mu}$ are the most demanding as they require  calls to the forward model to evaluate $\bs{y}(\bs{\mu}^{(t)})$  and the derivatives $\bs{G}^{(t)} =\frac{\pa \bs{y}}{\pa \bs{\Psi} }|_{\bpsi=\bs{\mu}^{(t)}}$. 
The updates were terminated when no further increase in $\mathcal{F}$ (\refeq{eq:fbmw}) can be attained.

\subsection{Adaptive learning - Cardinality of reduced coordinates }
\label{sec:card}

The presentation  thus far was based on a fixed number $\dth$ of reduced coordinates $\bt$. A natural question that arises is how many should one consider. In order to address this issue we propose an adaptive learning scheme.  According to this  the analysis is first performed  with a few (even one) reduced coordinates and upon convergence additional reduced coordinates are introduced, either in small batches or even one-by-one.
Critical to the implementation of such a scheme is a metric for the progress achieved by the addition of reduced coordinates and basis vectors which can also be used as a termination criterion.

In this work we advocate the use of an information-theoretic criterion which measures the information gain between the prior beliefs on $\bt$ and the corresponding posterior. To measure such gains we employ again the KL-divergence between the aforementioned distributions \cite{itti_bayesian_2009}. In particular if $p_{\dth}(\bt)$ (section \ref{sec:priors}) and $q_{\dth}(\bt)$ (\refeq{eq:qupd1}) denote the $\dth-$dimensional prior and posterior respectively, we define    the quantity  $I(\dth)$ as follows: 

\be
 I(\dth) = \frac{KL(p_{\dth}(\bs{\Theta})|| q_{\dth}(\bs{\Theta})) - KL(p_{{\dth}-1}(\bs{\Theta})|| q_{{\dth}-1}(\bs{\Theta}))}   {KL(p_{\dth}(\bs{\Theta})|| q_{\dth}(\bs{\Theta}))} \label{eq:IGain}
\ee
which measures the (relative) information gain from $\dth-1$ to $\dth$ reduced coordinates. 
The KL divergence between $p_{\dth}(\bt)$ and $q_{\dth}(\bt)$ with  $p_{\dth}(\bt)\equiv \mathcal{N}(\bs{0}, \bs{\Lambda}_0^{-1})$  and  $q_{\dth}(\bt)\equiv \mathcal{N}(\bs{0}, \bs{\Lambda}^{-1})$ where $\bs{\Lambda_0, \Lambda}$ are diagonal as explained previously follows with:
\be
  KL(p_{\dth}(\bs{\Theta})|| q_{\dth}(\bs{\Theta})) = \frac{1}{2} \sum_{i=1}^{\dth} (-log(\frac{\lambda_i}{\lambda_{0,i}})+\frac{\lambda_i}{\lambda_{0,i}} -1)
\ee
and equation (\ref{eq:IGain}) becomes:
\be
I(\dth) = \frac{\sum_{i=1}^{\dth} (-log(\frac{\lambda_i}{\lambda_{0,i}})+\frac{\lambda_i}{\lambda_{0,i}}-1)   -  \sum_{i=1}^{\dth-1} (-log(\frac{\lambda_i}{\lambda_{0,i}})+\frac{\lambda_i}{\lambda_{0,i}} -1) }   {\sum_{i=1}^{\dth} (-log(\frac{\lambda_i}{\lambda_{0,i}})+\frac{\lambda_i}{\lambda_{0,i}}-1)}.
\ee

% 
% 
% When the variance of $q(\bs{\Theta})$ of a system with $R-1$ bases is the same as the variance of the first $R-1$-entries of the variance a system with $R$ basis then the information gain simplifies to:
% \be
% I(R) = \frac{-log(\frac{\lambda_R}{\lambda{0,R}})+\frac{\lambda_R}{\lambda_{0,R}} -1 }   {\sum_{i=1}^{\dth} (-log(\frac{\lambda_i}{\lambda{0,i}})+\frac{\lambda_i}{\lambda_{0,i}}-1)}
% \ee

In the simulations performed in section \ref{sec:NumericalExamples}, we demonstrate the evolution of this metric as  reduced-coordinates/basis vectors  are added one-by-one.
The addition of reduced coordinates was terminated when $I(\dth)$ was below $1\%$ for at least five consecutive $\dth$.
\tcreview{In Figure \ref{fig:Flowchart} an overview flowchart of the proposed  algorithm is shown which incorporates the VB algorithm including dictionary learning from Algorithm \ref{alg:GenerealVB} and the information gain assessment to identify the necessary number of basis vectors from this subsection.}
\vspace{-0.0cm}

%For information about writing a flow chart look at: $http://www.texample.net/tikz/examples/simple-flow-chart/$

\tikzstyle{decision} = [diamond, aspect=2.5, draw, fill=blue!20, %fill=blue!20 is the color
    text width=5.5em, text badly centered, node distance=2.5cm, inner sep=0pt] % node distance means to the node before
\tikzstyle{block} = [rectangle, draw, fill=blue!20,
    text width=15em, text centered, rounded corners, minimum height=3em]
\tikzstyle{blocksmall} = [rectangle, draw, fill=blue!20, 
    text width=8em, text centered, rounded corners, minimum height=2em]  
 \tikzstyle{blockverysmall} = [rectangle, draw, fill=blue!20, 
    text width=5em, text centered, rounded corners, minimum height=0.6em]     
\tikzstyle{line} = [draw, very thick, color=black!50, -latex']
\tikzstyle{cloud} = [draw, ellipse,fill=red!20, node distance=2.5cm,
    minimum height=2em]
    
\begin{figure}[H]
    
    \makebox[\textwidth][c]{ %used to center figures within the lage
\begin{minipage}{0.8\textwidth}    
      \begin{tikzpicture}[scale=2, node distance = 2.0cm, auto] %node distance dercribes the vertical distance between nodes 
	  % Place nodes
	  \node [blocksmall] (init) {\small initialize values};
	%  \node [cloud, left of=init] (expert) {expert};
	%  \node [cloud, right of=init] (system) {system};
	  \node [blocksmall, below of=init] (myupdate) {\small \textbf{fix W, q, \tcr{update $\bs{\mu}$}} \\with smoothing prior $p(\bs{\mu})$};
	
	  \node [blocksmall, below of=myupdate] (identify) {\small \textbf{fix $\bs{\mu}$, q, \tcb{update W}} \\with $\bs{W}^T\bs{W}=\bs{I}$ };
	  
	  \node [blocksmall, below of=identify] (evaluate) {\small \textbf{fix W, $\bs{\mu}$, \tcc{update q}} \\update iteratively latent variables %\footnote{\tiny approximation used, no forward calls are necessary} 
	  };
	  
	  \node [blockverysmall, right of=identify, node distance=4.0cm] (update) {\small do another iteration};
	  \node [blockverysmall, right of=evaluate, node distance=6.0cm] (update2) {\small add a new basis $\bs{w}$, do another iteration};
	  \node [decision, below of=evaluate] (decide) {\small Has $\mathcal{F}$ converged?};
	  \node [decision, below of=decide] (decide2) {\small Has $I(d_{\Theta})$ converged?};
	  \node [blockverysmall, below of=decide2, node distance=1.8cm] (stop) {stop};
	  % Draw edges
	  \path [line] (init) -- (myupdate);
	  \path [line] (myupdate) -- (identify);
	  \path [line] (myupdate) -- (identify) coordinate[midway] (midpoint1);	  
	  \path [line] (identify) -- (evaluate);
	  \path [line] (evaluate) -- (decide);
	  \path [line] (decide) -| node [near start, color=black] {no} (update);
	  \path [line] (update) |- (midpoint1); % point to the midline
	  \path [line] (decide) -- node [, color=black] {yes}(decide2);
	  \path [line] (decide2) -| node [near start, color=black] {no} (update2);
	  \path [line] (update2) |- (midpoint1); % point to the midline
	  \path [line] (decide2) -- node [, color=black] {yes}(stop);
	%  \path [line,dashed] (expert) -- (init);
	%  \path [line,dashed] (system) -- (init);
	%  \path [line,dashed] (system) |- (evaluate);
      \end{tikzpicture}
\end{minipage}%
\begin{minipage}{0.5\textwidth}

 \footnotesize
	\tcr{\textbf{$\bs{\mu}$-update:}}
	\be
	    arg\,\underset{\bs{\mu}}{max} \,  \mathcal{F}_{\mu} =  -\frac{<\tau>}{2} |\bs{\hat{y}} - \bs{y}(\bs{\mu})|^2 + log\, p(\bs{\mu}) \nonumber
	\ee
    \vspace{+0.5cm}
	\\ \tcb{\textbf{W-update:}}
	\be
	  arg\,\underset{\bs{W}}{max} \,  \mathcal{F}_W  = - \frac{<\tau>}{2}\bs{W}^T \bs{G}^T\bs{G} \bs{W} :\bs{\Lambda}^{-1} \nonumber
	\ee
  \vspace{+0.5cm}
	\\ \tcc{\textbf{q-update:}}
	\be
	  \bs{\Lambda} = \bs{\Lambda}_{0}+<\tau> \bs{W}^T \bs{G}^T \bs{G} \bs{W} \nonumber
	\ee
	\be
	  a = a_0 + d_y/2  \nonumber
	  \ee
	\be 
	  b = b_0 + \frac{1}{2} |\bs{\hat{y}} - \bs{y}(\bs{\mu})|^2 + \frac{1}{2} tr( \bs{W}^T \bs{G}^T \bs{G W}  \bs{\Lambda}^{-1}) \nonumber
	\ee
\end{minipage}%  
}
\caption{Flowchart for the new algorithm. As the $\bs{\mu}$-update does not depend on the $\bs{W}$ just one $\bs{\mu}$-update (which is the expensive part of the full algorithm) is necessary during the calculations.}  
\label{fig:Flowchart}
\end{figure}    
 
%%%%%%%%%%%%%%%%%%%%%%%%%%%%%%%%%%%%%%%%%%%%%%%%%%%

\subsection{Validation - Combining VB approximations with Importance Sampling}
\label{sec:is}

\tcreview{Thus far  we have employed the  variational lower bound in order to identify the optimal dimensionality reduction and to infer the latent variables that approximate the posterior.
The goal of this section is twofold. Firstly to show how the biased VB approximation can be used in order to obtain efficiently, {\em (asymptotically) unbiased} estimates with regards to the true posterior and secondly to assess (quantitatively) the accuracy of the VB approximation.
To that end we employ Importance Sampling  (IS) with the variational posterior as the importance sampling
distribution.
 We can thusly obtain consistent estimators of several exact posterior quantities as well as to measure the efficiency of IS.
}
 
\tcreview{Consider the {\em exact} posterior $p(\bt | \hat{\bs{y}}, \bs{\mu,W})=\frac{ p(\bs{\hat{y}}| \bs{\Theta},  \bs{\mu} , \bs{W})~p(\bs{\Theta}) }{ p(\bs{\hat{y}} | \bs{\mu} , \bs{W}) }$. We note that when $\tau$ is unknown as in the cases considered herein, the (marginal) likelihood  $p(\bs{\hat{y}}| \bs{\Theta},  \bs{\mu} , \bs{W})$ can be determined by integrating with respect to $\tau$. With the conjugate Gamma prior adopted (\refeq{eq:tauPrior}) this can be done analytically and would yield:
\be
\begin{array}{ll}
p(\bs{\hat{y}}| \bs{\Theta},  \bs{\mu} , \bs{W}) & = \int p(\bs{\hat{y}}, \tau | \bs{\Theta},  \bs{\mu} , \bs{W}) ~d\tau  \\
	& = \int p(\bs{\hat{y}} | \tau, \bs{\Theta},  \bs{\mu} , \bs{W})~p(\tau)  ~d\tau \\
	& \propto  \frac{\Gamma(a_0+ d_y/2)}{(b_0+\frac{ |\hat{\mb{y}}- \mb{y}(\bs{\mu} + \bs{W} \bs{\Theta} )|^2}{2} )^{a_0+d_y/2} }.
\end{array}
\ee
In cases where non-conjugate priors for $\tau$ are employed, the IS procedure detailed here has to be performed in the joint space ($\bt,\tau)$.
}

\tcreview{The evidence:
\be
p(\bs{\hat{y}} | \bs{\mu} , \bs{W}) =\int p(\bs{\hat{y}}| \bs{\Theta},  \bs{\mu} , \bs{W})~p(\bs{\Theta})  ~d\bt
\ee
as well as the expectation of any function $g(\bs{\Psi})=g(\bs{\mu}+\bs{W}\bt)$ with regards to the {\em exact posterior} $p(\bt | \hat{\bs{y}}, \bs{\mu,W})$:
\be
\begin{array}{ll}
E[g(\bs{\Psi})] & =\int g(\bs{\mu}+\bs{W}\bt) ~p(\bt | \hat{\bs{y}}, \bs{\mu,W})~d\bt \\
& = \int  g(\bs{\mu}+\bs{W}\bt) ~\frac{ p(\bs{\hat{y}}| \bs{\Theta},  \bs{\mu} , \bs{W})~p(\bs{\Theta}) }{p(\bs{\hat{y}} | \bs{\mu} , \bs{W}) } ~d\bt \\
\end{array}
\ee
can be estimated using IS with respect to the IS density $q(\bt)$ as follows:
\be
\begin{array}{l}
 \frac{1}{M} \sum_{m=1}^M w(\bt^{(m)}) \to p(\bs{\hat{y}} | \bs{\mu} , \bs{W}) \\
 \frac{1}{{\sum_{m=1}^M w(\bt^{(m)})}} \sum_{m=1}^M g(\bs{\mu} + \bs{W}\bt^{(m)})~ w(\bt^{(m)})
 \to E[g(\bs{\Psi})]
 \end{array}
\ee
where the samples $\{ \bt^{(m)}\}_{m=1}^M $ are independent draws from $q(\bt)$ and the IS weights are given by:
\be
w(\bt)=\frac{ p(\bs{\hat{y}}| \bs{\Theta},  \bs{\mu} , \bs{W})~p(\bs{\Theta}) }{q(\bt)}.
\ee
}
% 
% Similarly  the KL-divergence between the approximate $q(\bt)$ and the exact posterior $p(\bt | \hat{\bs{y}}, \bs{\mu,W})$:
% \be
% \begin{array}{ll}
%  KL(q(\bt) || p(\bt | \hat{\bs{y}}, \bs{\mu,W})) & = - \int q(\bt) \log \frac{  p(\bt | \hat{\bs{y}}, \bs{\mu,W}) }{ q(\bt)}~d \bt \\
%  & = - \int q(\bt) \log \frac{ p(\bs{\hat{y}}| \bs{\Theta},  \bs{\mu} , \bs{W})~p(\bs{\Theta}) }{ q(\bt)}~d \bt +\log p(\bs{\hat{y}} | \bs{\mu} , \bs{W}) 
% \end{array}
% \ee
% can be estimated as follows \cite{beal_variational_2003}:
% \be
% -\frac{1}{M} \sum_{m=1}^M \log w(\bt^{(m)}) +\log \left(  \frac{1}{M} \sum_{m=1}^M w(\bt^{(m)}) \right) \to KL(q(\bt) || p(\bt | \hat{\bs{y}}, \bs{\mu,W}))
% \ee
% The latter, normalized by the entropy of $q(\bt)$ \footnote{$Entropy(q(\bt))=-\int q(\bt) \log q(\bt)~d\bt$}
% could be used to assess the quality of the approximation, particularly for $\bt$ of varying dimensionality.

\tcreview{An indicator of the efficacy of the IS density is the (normalized) Effective Sample Size (ESS) 
 which provides a measure of the degeneracy in the population of particles/samples as quantified by their variance \cite{kong_sequential_1994}:
 \be
 ESS=\frac{ (\sum_{m=1}^M w(\bt^{(m)}) )^2 }{M \sum_{m=1}^M w^2(\bt^{(m)}) }.
 \label{eq:ess}
 \ee
 }
 
 \tcreview{The latter attains values between the following extremes. If $q(\bt)$ coincides with the exact posterior then all the importance weights $w(\bt^{(m)})$ would be equal and $ESS=1$. On the other hand if $q(\bt)$ provides a poor approximation then the expression for the $ESS$ is dominated by the largest weight $w(\bt^{(m)})$ and  would yield $ESS=1/M \to 0$ (as $M \to \infty$).
 The normalized ESS can be compared with that of  MCMC \cite{robert_monte_2004}:
 \be
 ESS_{MCMC}= \frac{1}{1+2\sum_{k=1}^M (1-\frac{k}{M})\rho(k)} \to \frac{1}{1+2\sum_{k=1}^{\infty} \rho(k)}
 \label{eq:mcmcess}
\ee
where $\rho(k)$ is the autocovariance between MCMC states that are $k$ steps apart.
}

\tcrereview{In summary, the VB framework advocated introduces approximations due to the linearization of the response (\refeq{eq:FirstTaylorDisp}) and the mean field approximation (\refeq{eq:MeanFieldApp}). To assess the bias of these approximations  in the posterior inferred, we employ IS as explained above. This can lead to accuracy metrics (e.g. ESS) but more  importantly can produce (asymptotically) unbiased statistics with regards to the exact posterior i.e. the one  obtained without the approximations mentioned earlier. These metrics can be readily compared with those of alternative strategies (e.g. MCMC as in \refeq{eq:mcmcess}).  
Unequivocally, another important source of error is due to  model discrepancies. That is, if the difference between observables and model predictions in \refeq{eq:objective}  is not valid due  missing physics, discretization errors etc, then the inference results will deviate from reality, irrespectively of the numerical tools one employs \cite{arridge_approximation_2006, kaipio_statistical_2007,koutsourelakis_novel_2012}. We emphasize that the methodology proposed, as most strategies for the solution of inverse problems, is based on the  assumption that model errors are zero or in any case much smaller than the observation errors.
}

\section{Numerical Illustrations} \label{sec:NumericalExamples}
The  examples presented are concerned with the probabilistic identification of material parameters from displacement data. We demonstrate the efficacy of the proposed methodology in two, two-dimensional cases where synthetic displacement data are utilized. The data are contaminated with noise as discussed in the sequence.   The first example is based on
a linear elastic material model.  
The second example is based on the Mooney-Rivlin material model which is used to model nonlinear and  incompressible response.

In the computations we use $a_0=b_0=a_{\phi} = b_{\phi} = 0$.
We  employ the adaptive learning scheme discussed in section \ref{sec:card} whereby  reduced-coordinates/basis vectors are added one-by-one.
The first reduced coordinate is assigned the broadest prior i.e.   $\lambda_{0,1}$ is the smallest of all other $\lambda_{0,i}$ and captures  the largest expected (a priori) variance.   %The precision parameter $\lambda_{0,1}$ in the prior for for $q(\Theta_1)$ is set to  $\lambda_{0,1}=10^{-10}$ which corresponds to a vague prior. 
For subsequent bases $i=2,3,\ldots$ we assign values to the precision parameters $\lambda_{0,i}$ as follows:
\be
\lambda_{0,i}=max(\lambda_{0,1},\lambda_{i-1}-\lambda_{0,i-1}), \quad i=2,3,\ldots
\ee
We note that $\lambda_{i-1}$ corresponds to the {\em posterior} precision for the {\em previous} reduced coordinate $\Theta_{i-1}$ as found in \refeq{eq:qupd1} according to which $\lambda_{0,i}=<\tau> \bs{w}_{i-1}^T \bs{G}^T \bs{G} \bs{w}_{i-1}$. This essentially implies that, a priori, the next reduced coordinate $\Theta_i$ will have the precision of the previous one as long as it is larger than the threshold $\lambda_{0,1}$. Since by construction $\bs{w}_{i}^T \bs{G}^T \bs{G} \bs{w}_{i}>\bs{w}_{i-1}^T \bs{G}^T \bs{G} \bs{w}_{i-1}$, we have that $ \lambda_{0,i+1} \ge \lambda_{0,i}$.

\tcreview{The most important quantities and dimensions of the ensuing  two examples are summarized in table \ref{tab:NumberOverview}.}

\begin{table}[h]
\begin{center}
  \begin{tabular}{ | c | c | c |}
    \hline
      & Example 1 & Example 2 \\ \hline\hline
    Dimension of observables: $d_{y}$ & $198$ & $5100$ \\ \hline
    Dimension of latent variables: $\dpsi$ & $90$ & $2500$ \\ \hline
    Dimension of reduced latent variables: $\dth$ & $5-10$ & $15-25$ \\ \hline
    Nr. of forward calls & $<25$ & $<35$ \\
    \hline
 
  \end{tabular}   
  \caption{Summary of the number of observables, forward calls and the dimensionality reduction in the following two examples.}
  \label{tab:NumberOverview}  
\end{center}  
\end{table}

% 
% 
% 
% 
% 
% Then when adding a next basis $i$ the precision of the prior is set equal to the posterior value of the previous updated bases minus its prior precision: $\lambda_{0,i} = \lambda_{new,i-1} - \lambda_{0,i-1}$. Our prior knowledge is that every new basis will have a larger precision as the previous basis. Hence this information is directly included in the prior of the new basis. Each additional basis will capture less additional variance compared to the previous bases. This is due to the updated/optimized $\bs{W}$, $\bs{W}$ will first find the optimal basis, then the second optimal, then the third optimal basis, etc... Therefore each additional basis will have a larger $\lambda$ then the bases beforehand.
% 
% 

\subsection{Example 1: Linear elastic material}
\begin{figure}
  \centering
  \includegraphics[width=0.5\textwidth]{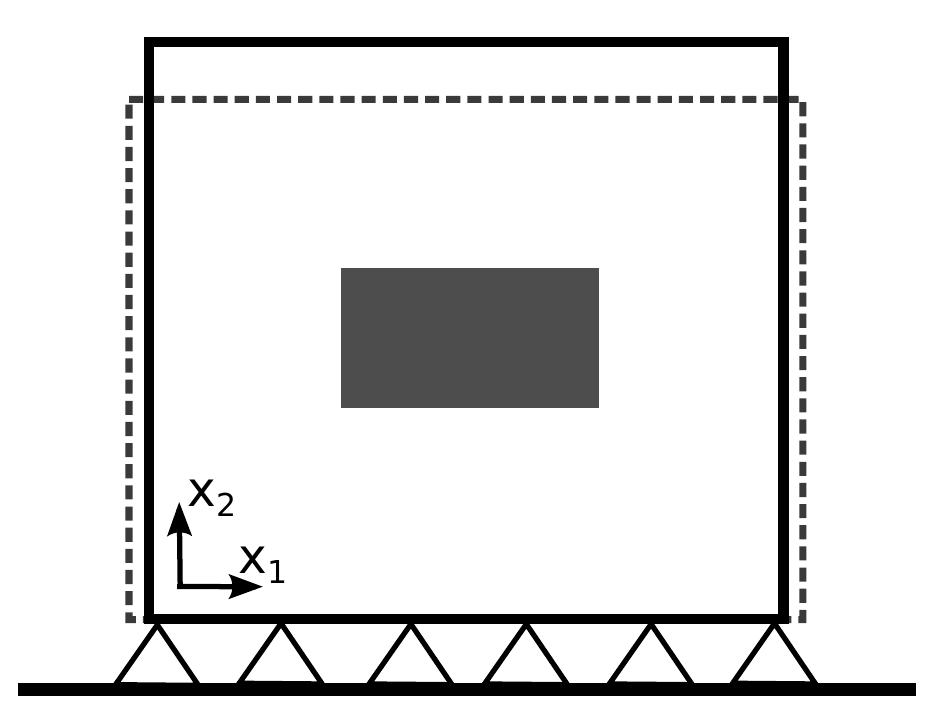} 
  \caption{Problem configuration for example 1.}
  \label{fig:SmallExampleSetting}
\end{figure}

\begin{figure}
  \centering
  \includegraphics[width=0.4\textwidth]{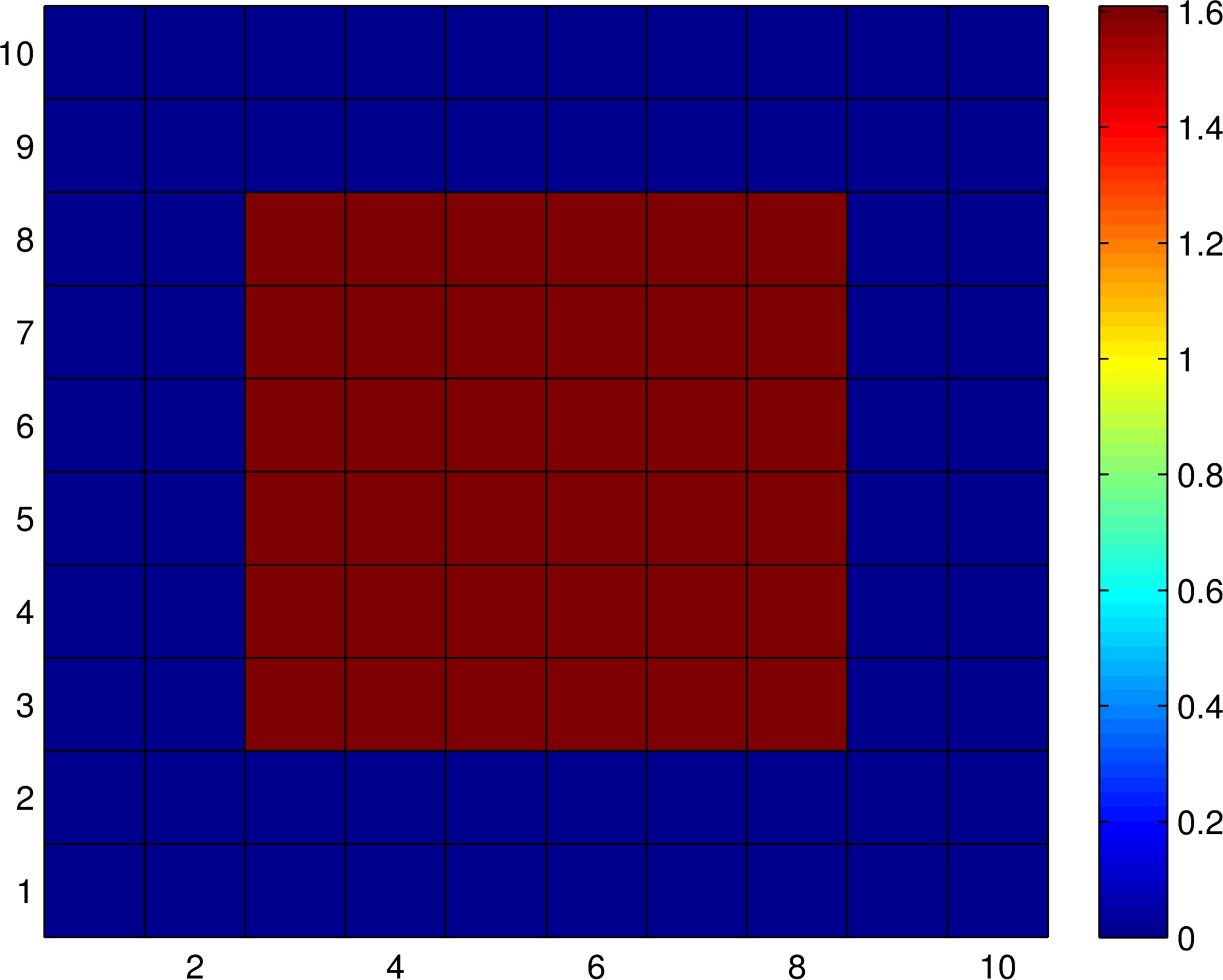} 
  \caption{Reference configuration of the material parameters $E$ in log-scale.}
  \label{fig:SmallExampleReference}
\end{figure}

The primary objective of the first example is to assess the performance of the proposed framework in terms of accuracy and dimensionality reduction in a simple problem with the absence of model errors. For that purpose  we consider a linear, isotropic  elastic material model where the  stress-strain relation is given by: 
\be
  \bs{S} = \mathbb{C} : \bs{E}
\ee
where $\mathbb{C}$ is the elasticity tensor \cite{mase_continuum_2009}. It is given by:
\be
  \mathbb{C} = \frac{E}{(1+\nu)} ( \bs{I} + \frac{\nu}{(1-2\nu)} \bs{1} \otimes \bs{1})
\ee
where $E$ is  the elastic modulus. The second material parameter is the Poisson's ratio $\nu$ which in this example is assumed known ($\nu = 0$).  The vector of  unknown parameters $\bpsi$ consists of the values of the elastic moduli at each finite element.
We assume that the elastic modulus can take two values $E_{inclusion}$ and $E_{matrix}$  such that $\frac{E_{inclusion} }{E_{matrix}}=5$.  The ratio is representative of ductal carcinoma in situ in glandular tissue in the breast under a strain of $15\%$ \cite{wellman_breast_1999}. The spatial distribution of the material is shown in \reffig{fig:SmallExampleReference}.  The  problem  is $\Omega = (0,L) \times (0,L)$ with $L=10$ units. We employ a $10 \times 10$ FE mesh.  Displacement boundary conditions are employed which resemble those encountered when static pressure is applied on a tissue with the ultrasound  transducer invoking a  $1\%$ strain as depicted in \reffig{fig:SmallExampleSetting}.
 In particular the boundary displacements at the bottom ($x_2=0$) are set to zero and at the top ($x_2=10$) the vertical displacements are set to $-0.1$ and the horizontal displacements equal to zero. The vertical edges ($x_1=0, 10$) are traction-free.

 The parameter values are at the top row of elements are assumed known and equal to the exact values ($E_{matrix}$)  otherwise any  solutions for which  $\frac{E_{inclusion} }{E_{matrix}}=5$ would yield the same likelihood \cite{gokhale_solution_2008}.
The  displacements generated using the reference configuration were subsequently contaminated with Gaussian noise such that the resulting  Signal-to-Noise Ratio (SNR) was $SNR = 10^5$. 
We adopt a very vague prior i.e. $\lambda_{0,1}=10^{-10}$. %The starting of $\bs{\mu}_0$ are: $\bs{\mu}_0=1.65$.

%In the following figures the results are shown in the log-scale.

\begin{figure}[H]{
	\centering
	%\captionsetup[subfigure]{labelformat=empty}
		%\vspace{-0.5cm}
		\subfloat[][{ Mean, $d_{\bs{\Theta}}=90$}] 
		{\includegraphics[width=0.32\textwidth]{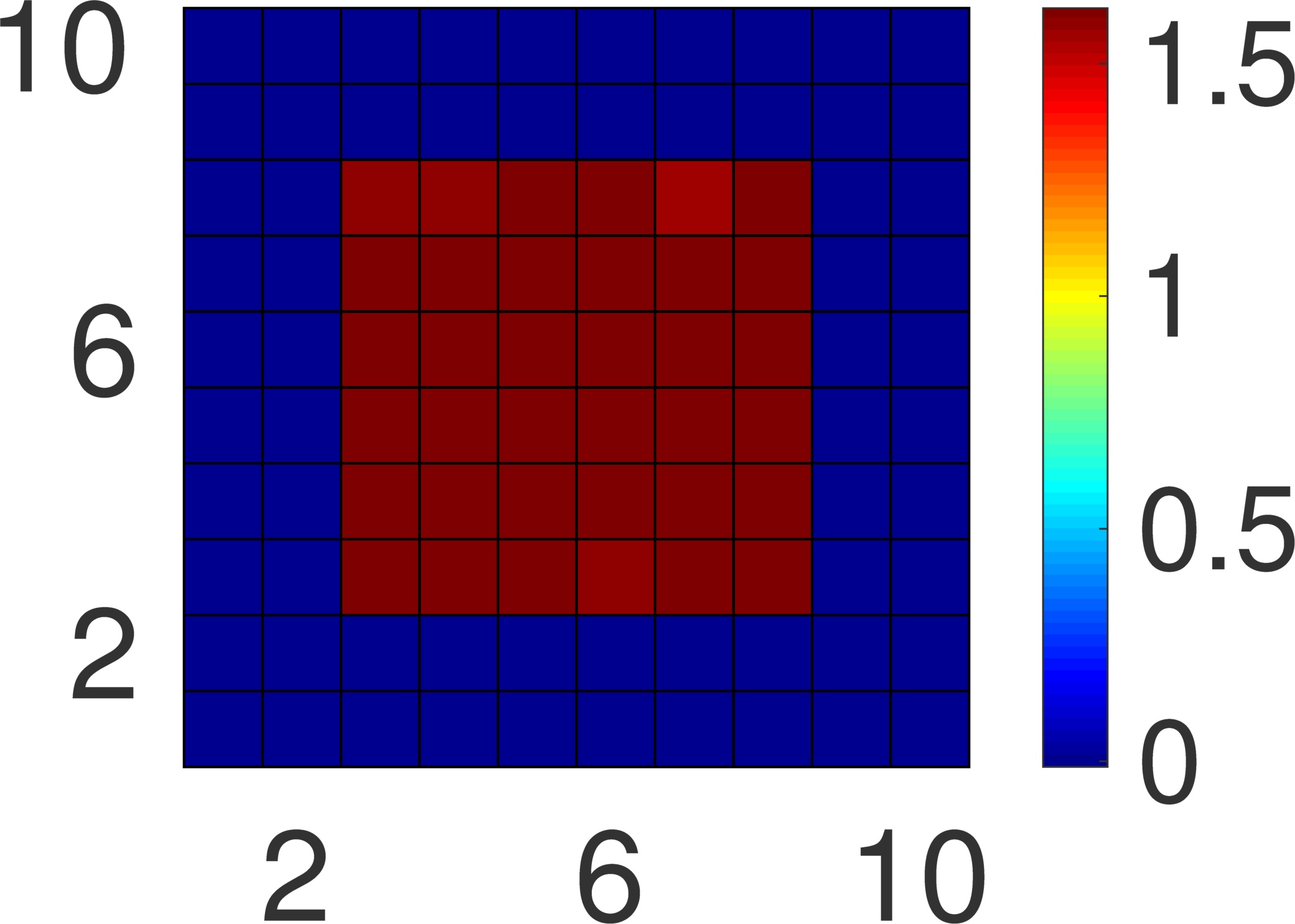}} 
		\hspace{0.1cm}
		\subfloat[][{ Diagonal cut, $d_{\bs{\Theta}}=90$ }]
		{\includegraphics[width=0.30\textwidth]{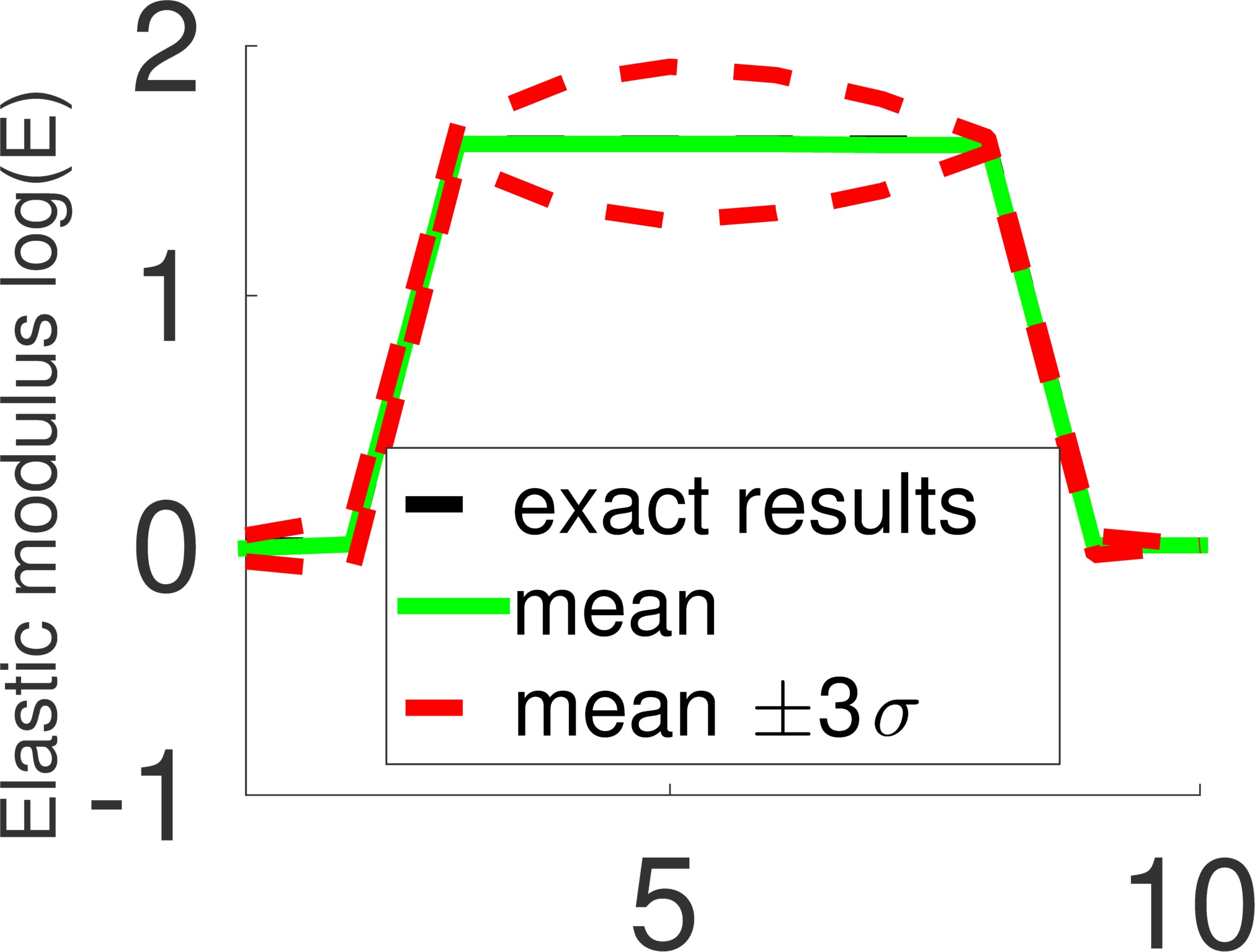} } 
		\hspace{0.1cm}
		\subfloat[][{ St. dev., $d_{\bs{\Theta}}=90$}]
		  {\includegraphics[width=0.32\textwidth]{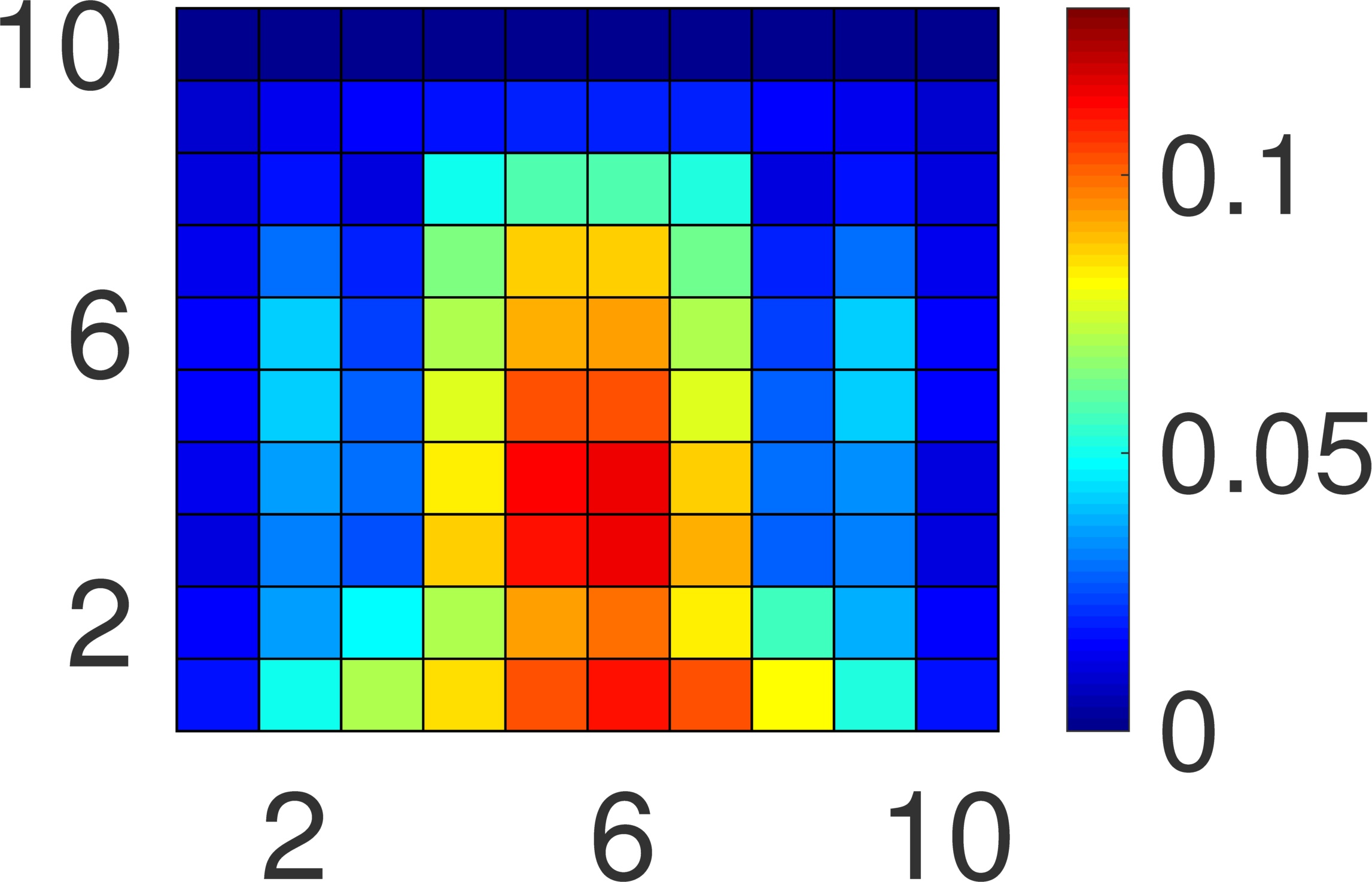} } 
		\vspace{0.3cm}\\
		\subfloat[][{Mean, $d_{\bs{\Theta}}=9$}] 
		{\includegraphics[width=0.32\textwidth]{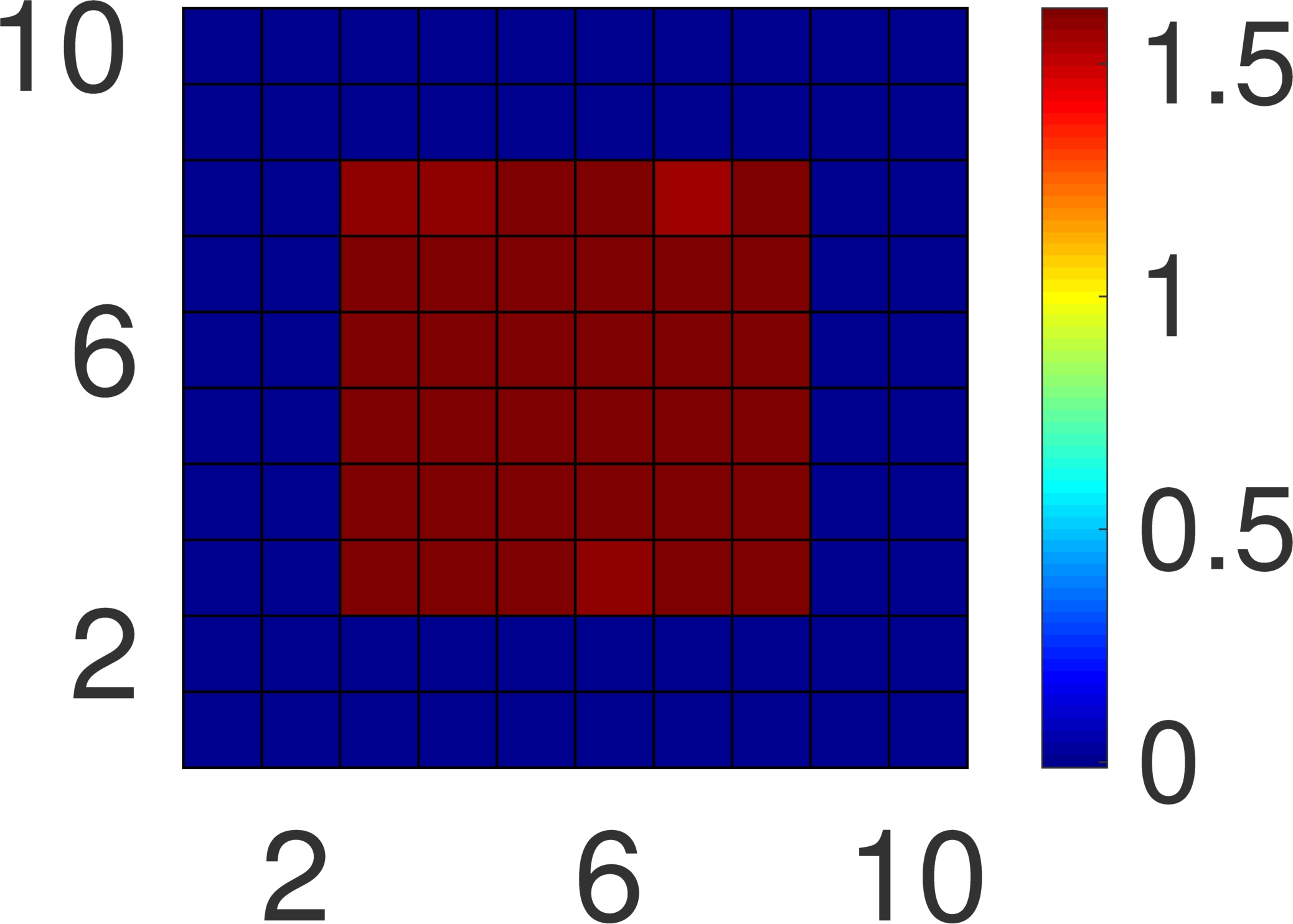}} 
		\hspace{0.1cm}
		\subfloat[][{ Diagonal cut, $d_{\bs{\Theta}}=9$}]
		{\includegraphics[width=0.30\textwidth]{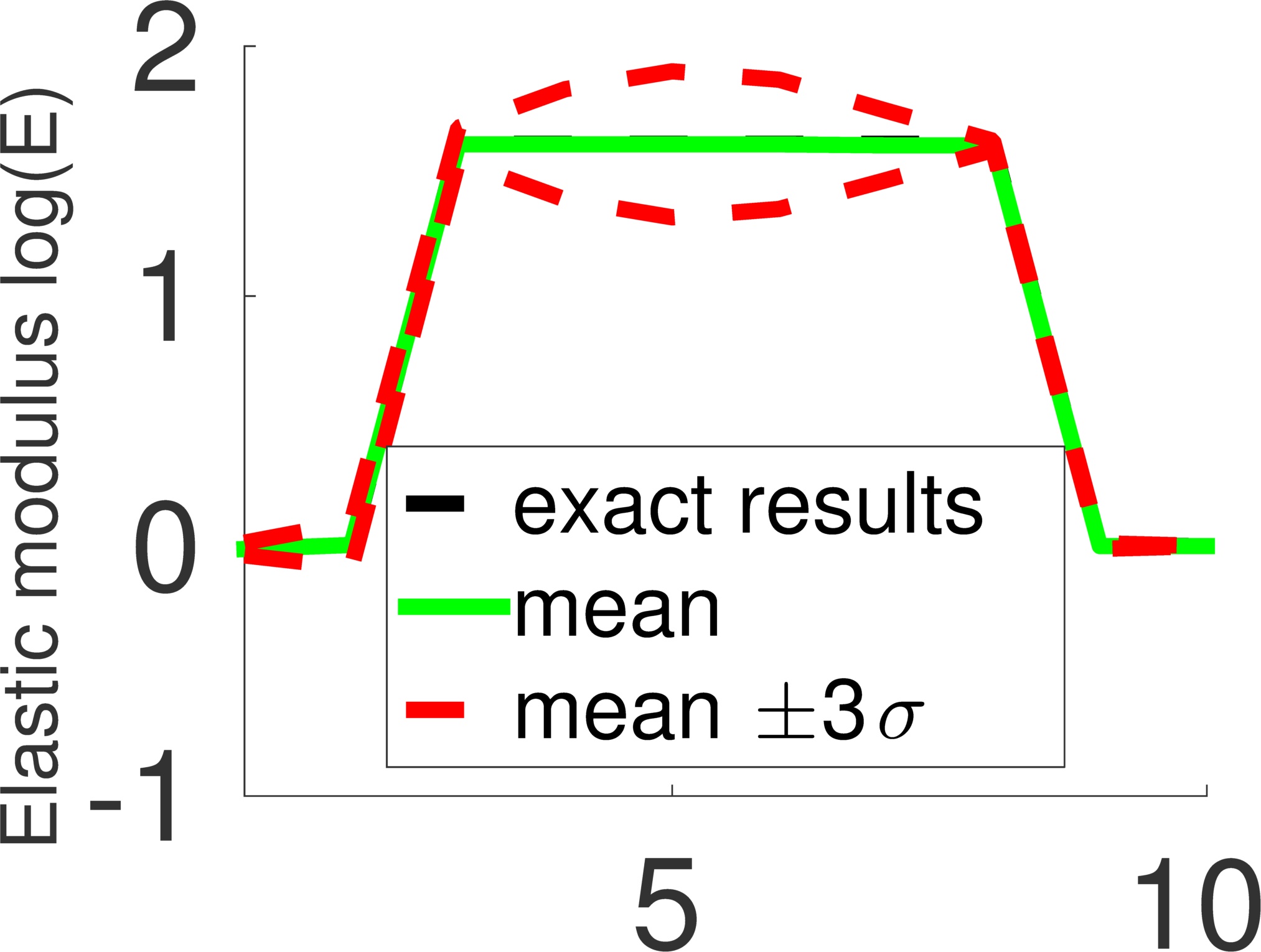} } 
		\hspace{0.1cm}
		\subfloat[][{ St. dev., $d_{\bs{\Theta}}=9$}]
		  {\includegraphics[width=0.32\textwidth]{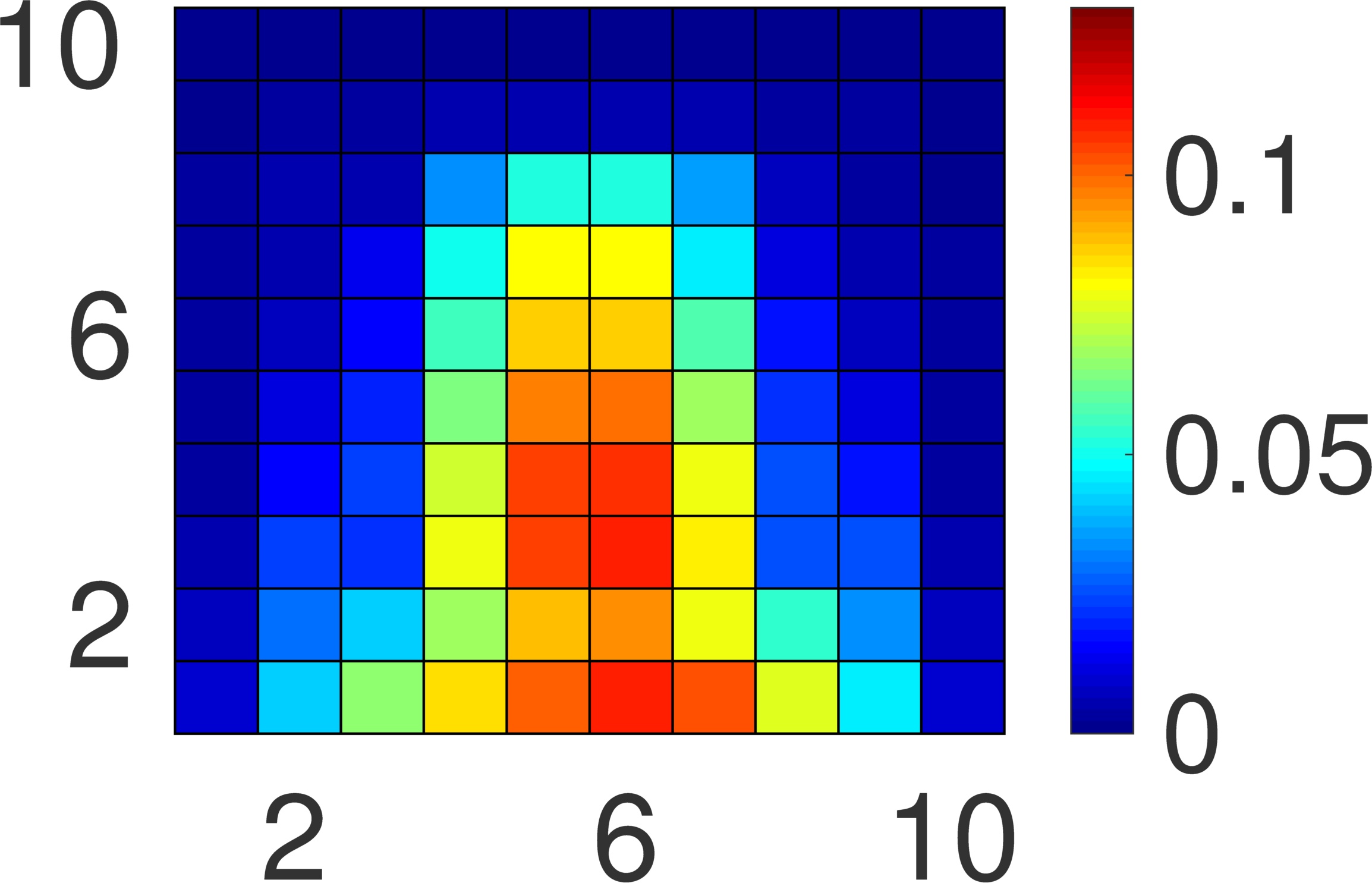} } 
		 
	  }
	  \caption{The first row corresponds to results derived with $d_{\bs{\Theta}}=90$ and the second row to $d_{\bs{\Theta}}=9$. Figures (a), (d) depict the posterior mean $\bs{\mu}$ of the elastic moduli $E$ in log-scale which is shown to be independent of  of the  number of reduced coordinates $d_{\bs{\Theta}}$. Figures (b), (e) show the posterior mean and posterior quantiles ($\pm 3$ standard deviations) along the diagonal from $(0,0)$ to $(10,10)$. Figures  (c), (f) depict the posterior  standard deviation. The differences are indistinguishable which implies that the full posterior ($d_{\bs{\Theta}}=90$) can be very well approximated  with only $d_{\bs{\Theta}}=9$ reduced coordinates/basis vectors.}
	 \label{fig:PosteriorSmall}
\end{figure}

In the top row of \reffig{fig:PosteriorSmall} various aspects of the posterior of the elastic moduli using 90 basis vectors, $d_{\bs{\Theta}}=90$ (equal to the total number of unknowns), are depicted and are compared with the corresponding results $d_{\bs{\Theta}}=9$ (second row).  One can see that the inferred posterior means are practically identical and coincide with the ground truth. The same can be said for the posterior variances which can be capture to a large extent by employing only   $d_{\bs{\Theta}}=9$ reduced coordinates. 

\begin{figure}[H]
	\centering
	\includegraphics[width=0.60\textwidth]{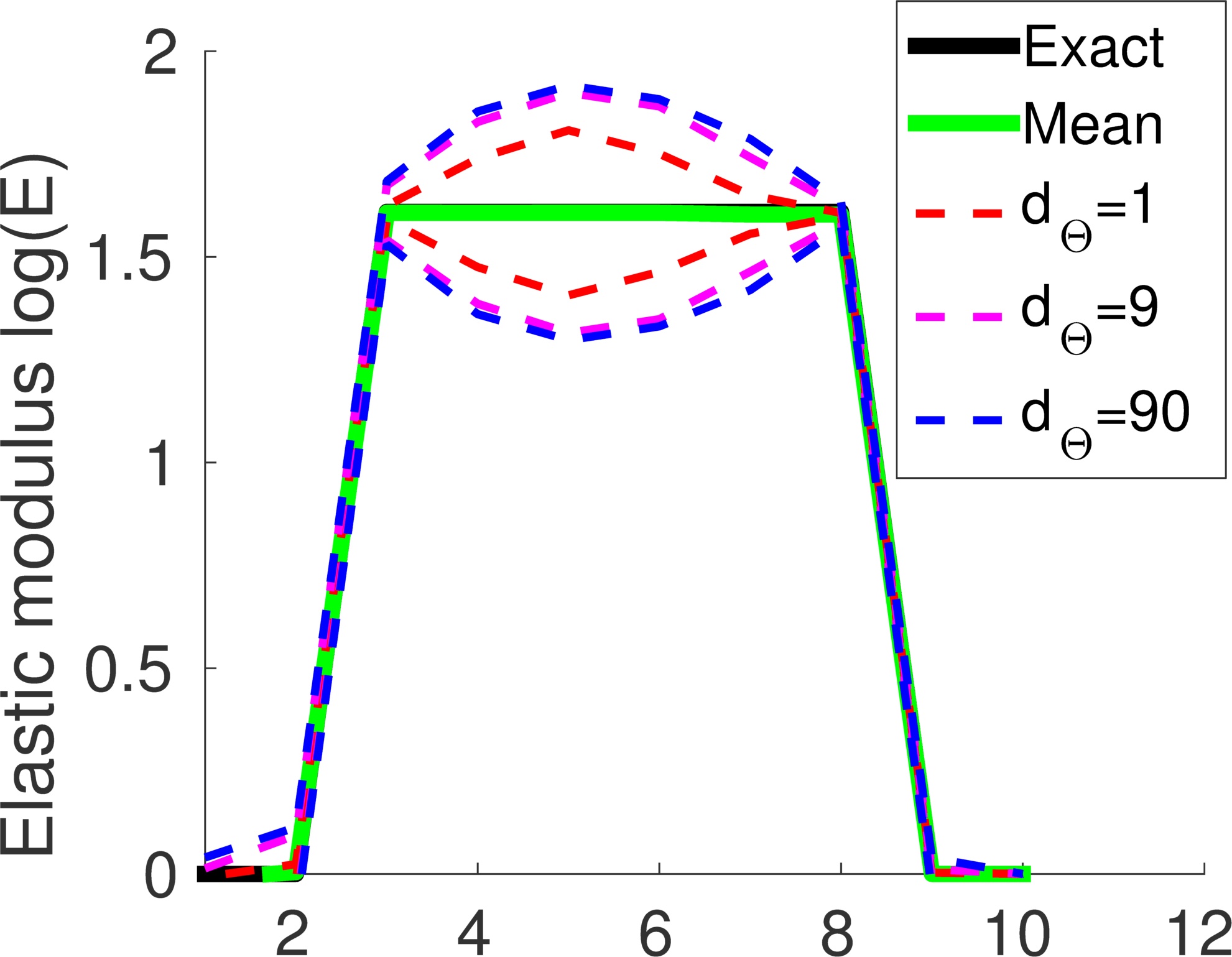} 
	  \caption{Posterior mean and credible intervals at $\pm 3$ standard deviations (dashed lines) along the diagonal from $(0,0)$ to $(10,10)$ for various values of $d_{\Theta}$.}\label{fig:VarianceSmall}
\end{figure}

A more detailed comparison of the inferred posterior for various $d_{\bs{\Theta}}$ is depicted in \reffig{fig:VarianceSmall}. 

\begin{figure}[H]
	\centering
	  \includegraphics[width=0.60\textwidth]{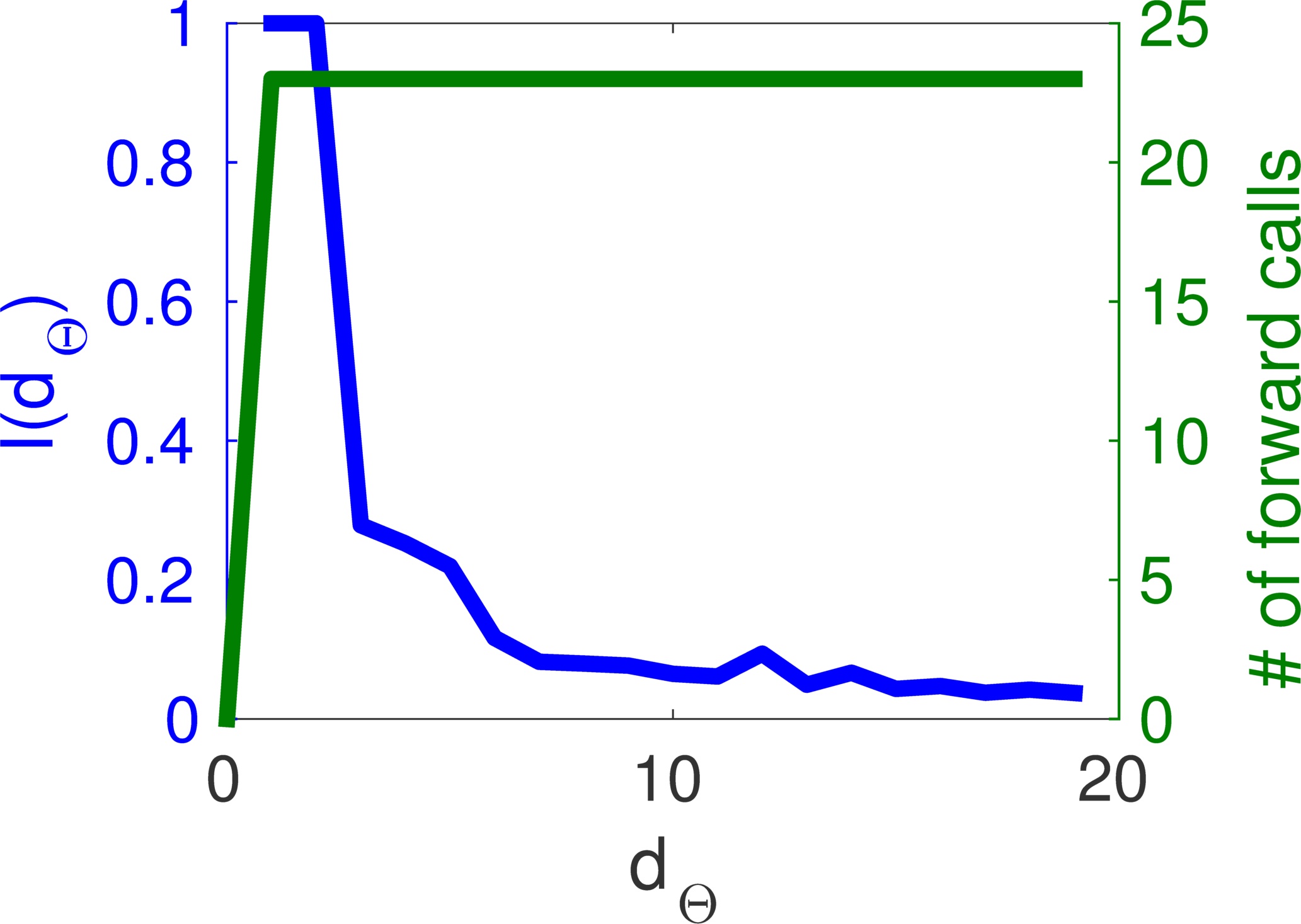} 
	  \caption{Information gain $I(\dth)$  {\color{blue} ---}  and computational cost {\color[rgb]{0,0.5,0}{---}} as measured by the number of forward calls over the number of dimensions, $\dth$. 
	  }\label{fig:IGSmall}
\end{figure}

\begin{figure}[H]{
	% \centering
	%\captionsetup[subfigure]{labelformat=empty}
		%\vspace{-0.5cm}
		\subfloat[][{$F_{{\mu}}$ over all bases}] 
		{\includegraphics[width=0.43\textwidth]{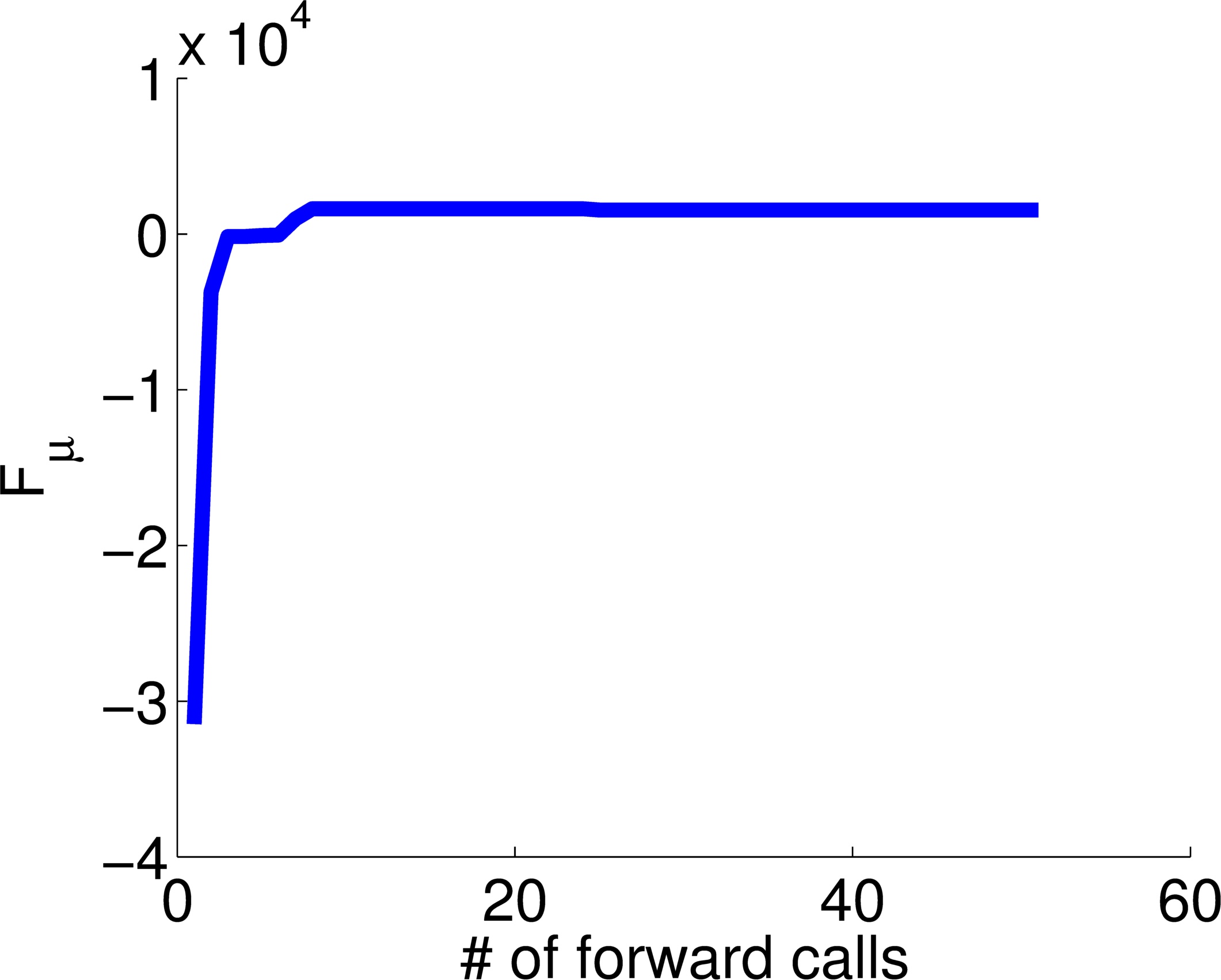}}  
		\vspace{0.1cm}
		\subfloat[][{$F_{{W}}$ with 9 bases}]
		{\includegraphics[width=0.43\textwidth]{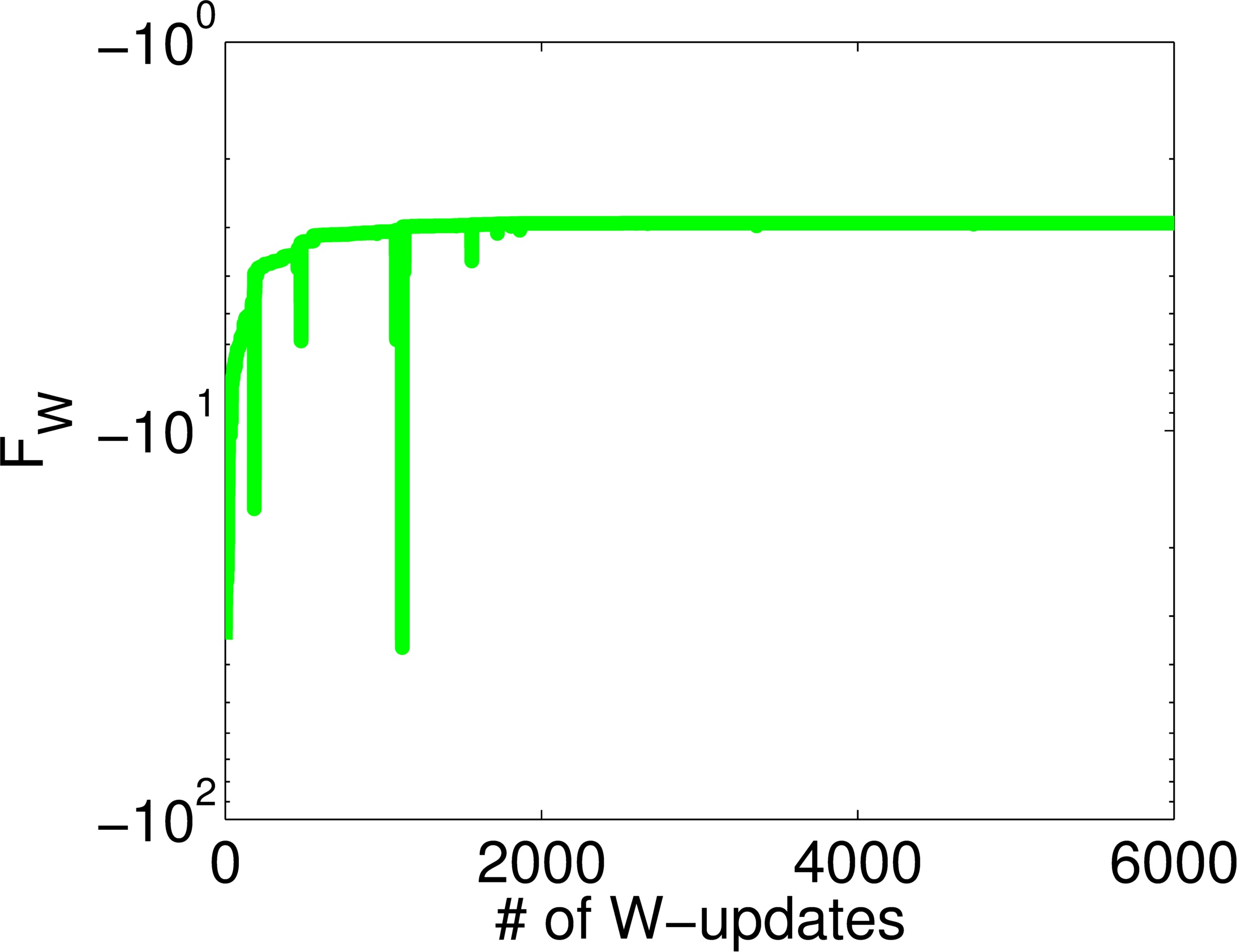} } 
		%\hspace{0.3cm}		

	  }
	  \caption{(a): $F_{{\mu}}$ over the total number of $\bs{\mu}$-updates. (b): $F_{{W}}$ during the $\bs{W}$-update, after adding the ninth basis.}\label{fig:Lowerbound}
\end{figure}

  \reffig{fig:IGSmall} depicts  the relative information gain (as defined in Section \ref{sec:card}) and  the number of forward calls (which determines the computational cost) as a function of the number of reduced coordinates/basis vectors. One can notice that the information gain drops to relatively small values only after a small number of reduced coordinates (after the $d_{\bs{\Theta}}=6$, it drops below  $10\%$). For the posterior approximation obtained  with $d_{\bs{\Theta}}=9$ (which as shown earlier is practically indistinguishable from the full-order result with $d_{\bs{\Theta}}=90$) only $23$ forward calls are needed. These forward calls, are performed at $d_{\bs{\Theta}}=1$ and for additional reduced coordinate no further forward calls are needed. 
  A more detailed account of the optimization with regards to the model parameters $\bs{\mu}$ and $\bs{W}$ can be seen in \reffig{fig:Lowerbound} where the evolution of  the corresponding variational objectives $F_{\mu}$ and $F_W$ (Section \ref{sec:UpdateEquations}) is plotted.  We note again that the $\bs{\mu}$ updates are the only ones that require forward calls. The  optimization results with regards to $F_W$ are shown for $d_{\bs{\Theta}}=9$. These are performed using the Barzilai-Borwein step size selection discussed previously which results in a non-monotone but robust optimization.

  %We note that within our implementation, we check at least three times if the lowerbound would improve for an updated $\bs{\mu}$. However, $\bs{\mu}$ did not changed/improved when adding a new basis so the three forward calls were done to check if any further optimization is possible/needed.

\begin{figure}[H]{
	% \centering
	%\captionsetup[subfigure]{labelformat=empty}
		\vspace{-1.5cm}
		\subfloat[][{ $\lambda_1^{-1} = 9.8 \times  10^{-2}$}] 
		{\includegraphics[width=0.30\textwidth]{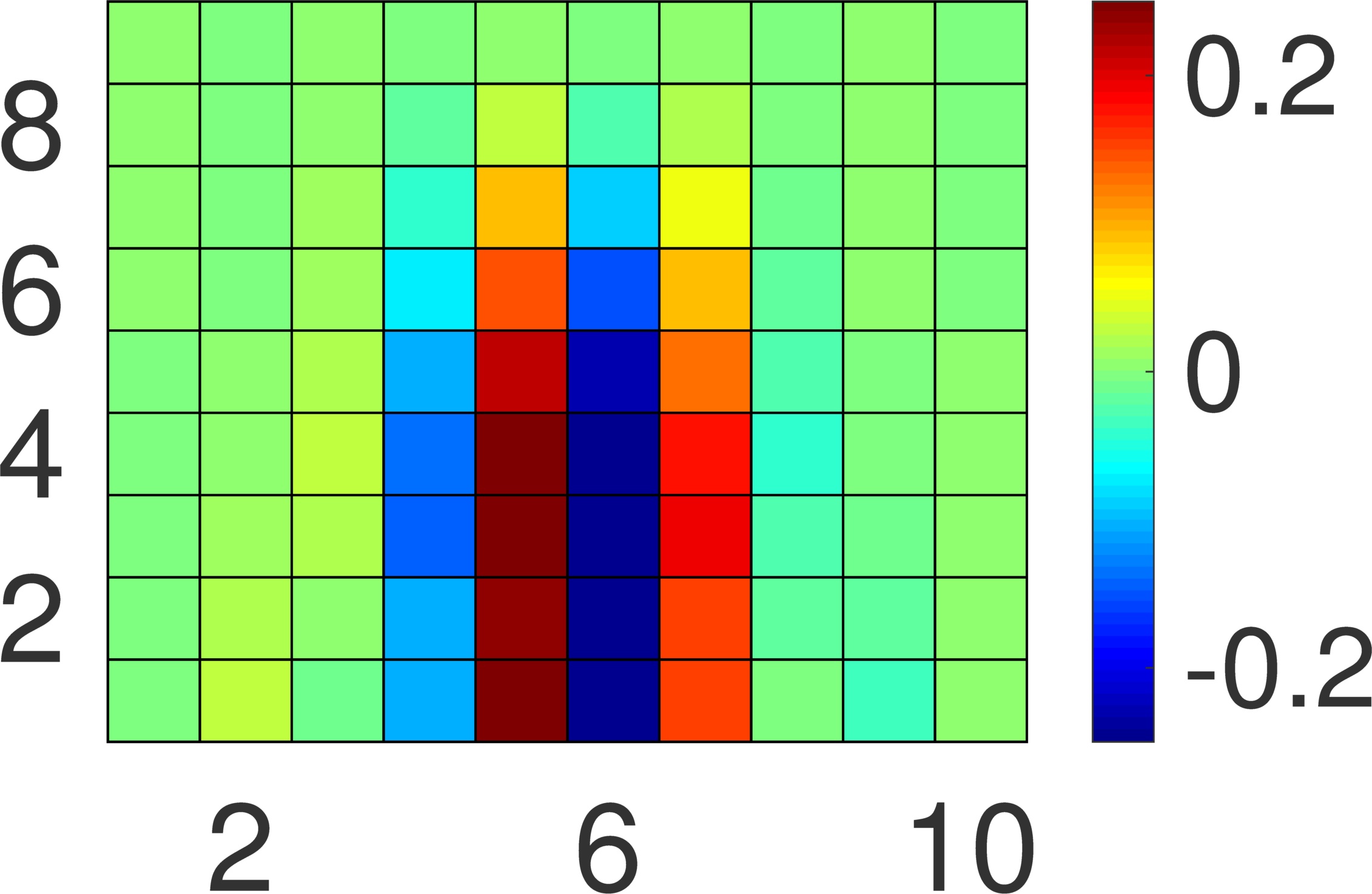}} 
		\hspace{0.1cm}
		\subfloat[][{$\lambda_2^{-1} = 4.0\times  10^{-2}$ }]
		{\includegraphics[width=0.30\textwidth]{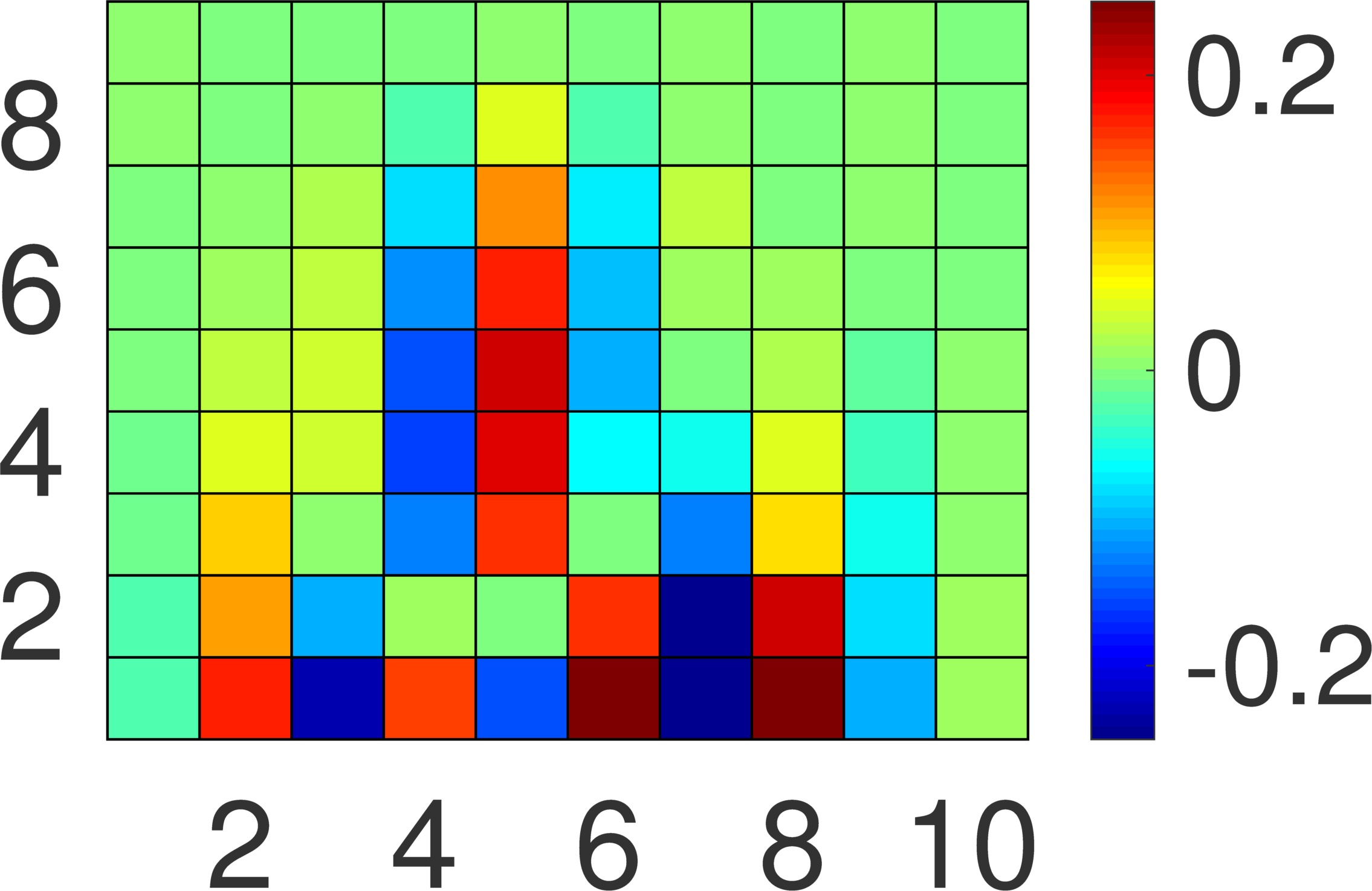} } 
		\hspace{0.1cm}
		\subfloat[][{$\lambda_3^{-1} = 3.0\times  10^{-2}$ }]
		{\includegraphics[width=0.30\textwidth]{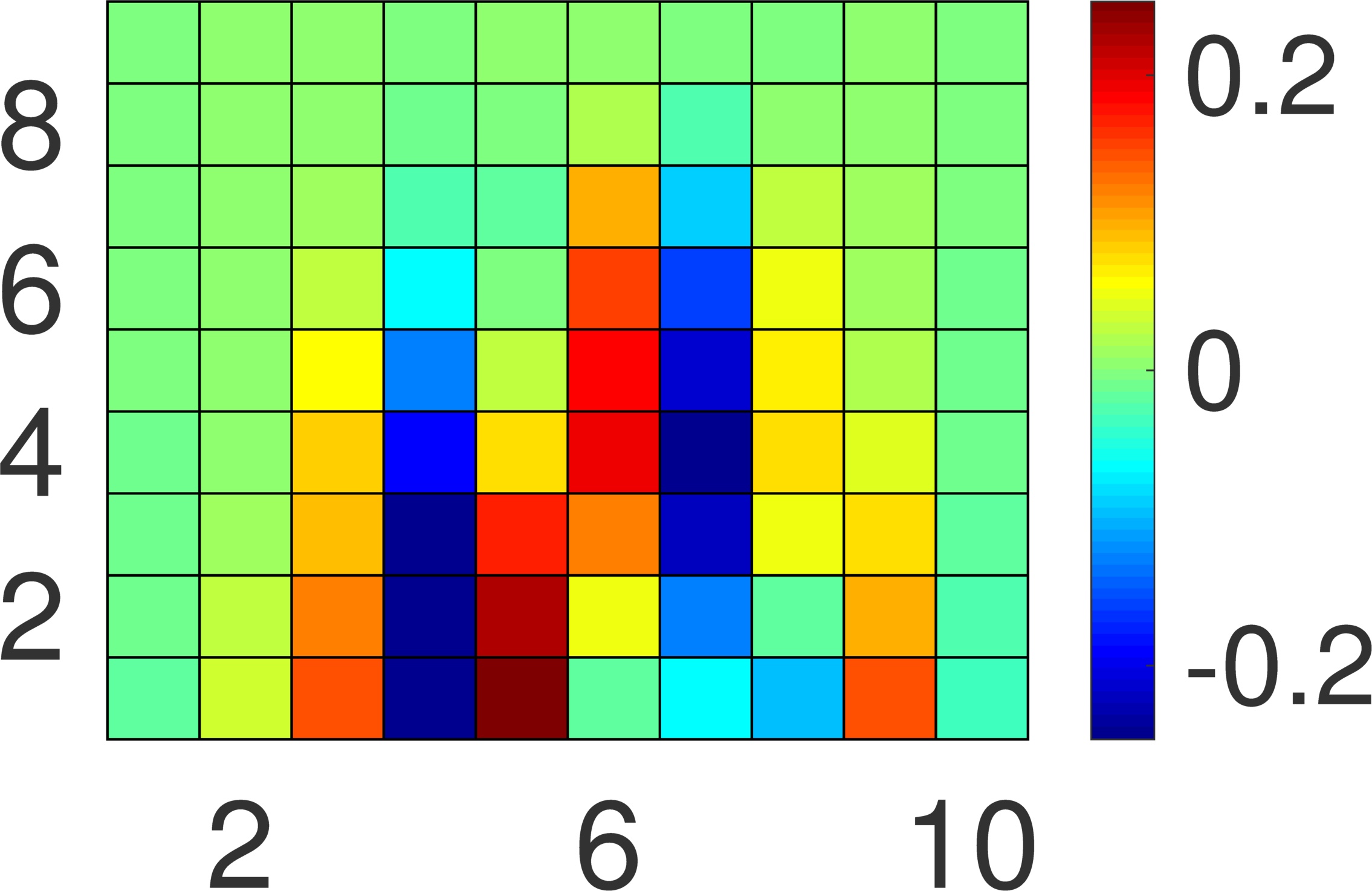} } 
		\hspace{0.3cm}
		\\
		\subfloat[][{$\lambda_4^{-1} = 2.3 \times  10^{-2}$ }] 
		{\includegraphics[width=0.30\textwidth]{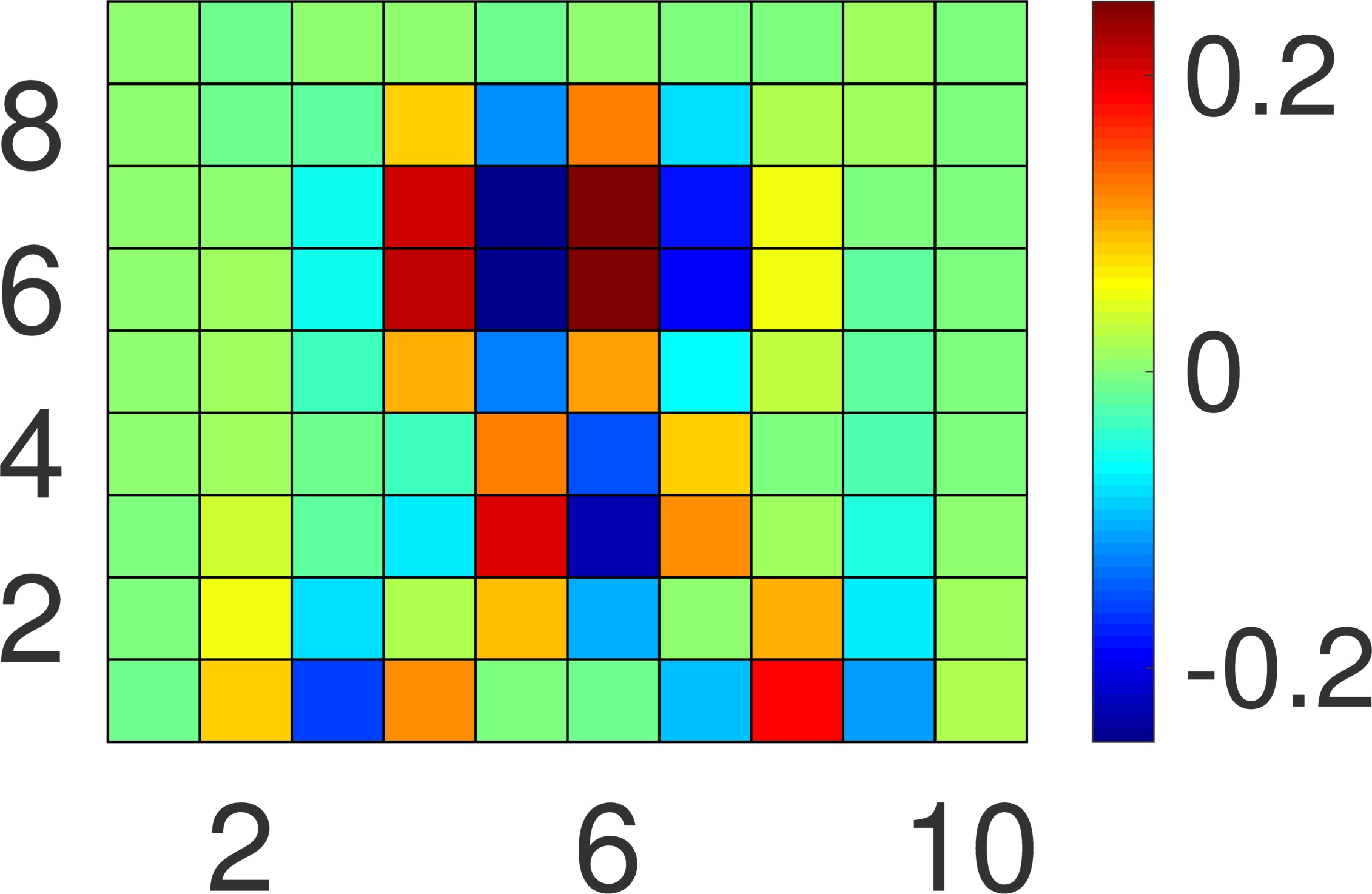}} 
		\hspace{0.1cm}
		\subfloat[][{$\lambda_5^{-1} = 1.4 \times  10^{-2}$ }]
		{\includegraphics[width=0.30\textwidth]{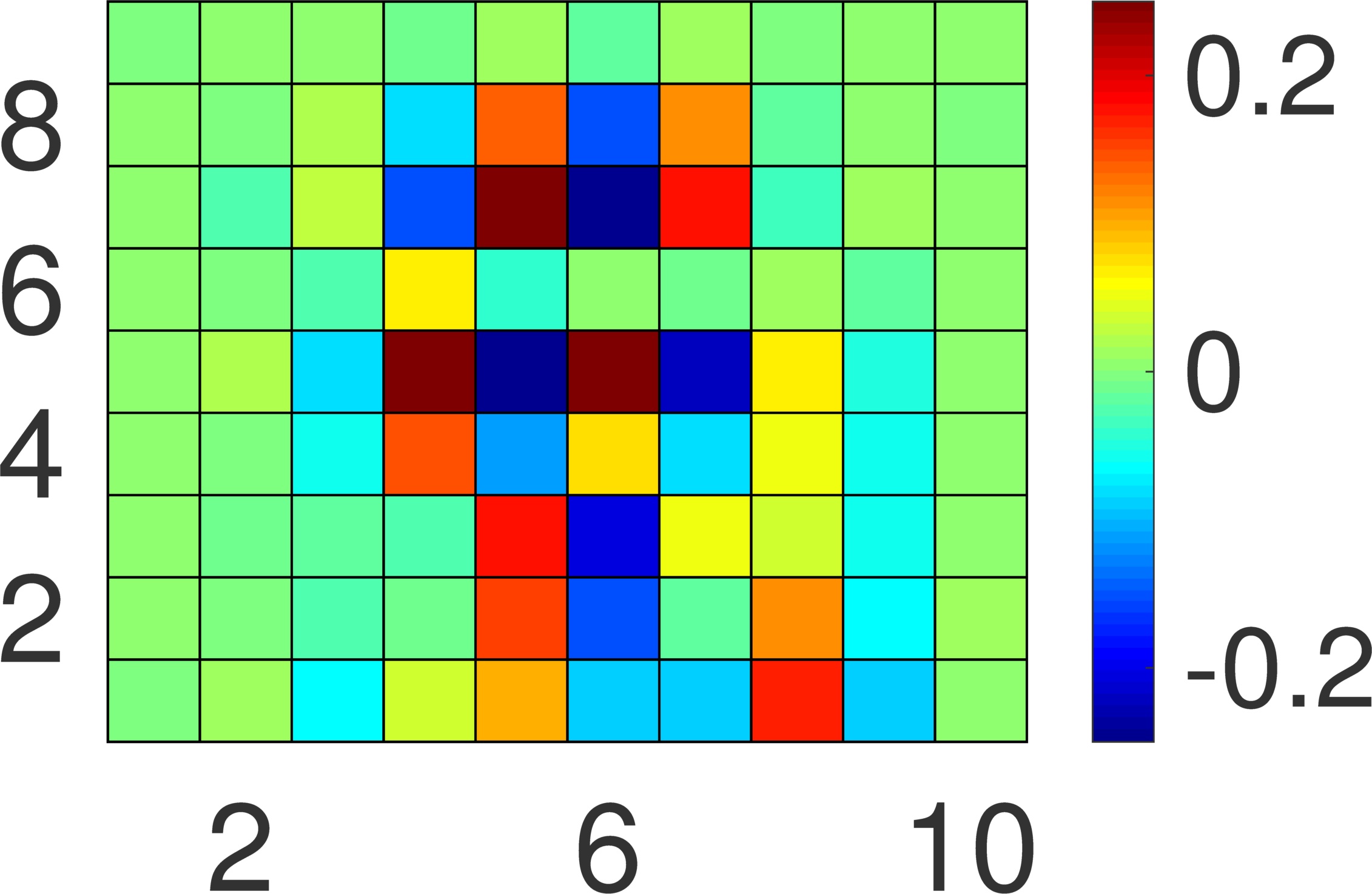} } 
		\hspace{0.1cm}
		\subfloat[][{$\lambda_6^{-1}= 9.5 \times  10^{-3}$ }]
		{\includegraphics[width=0.30\textwidth]{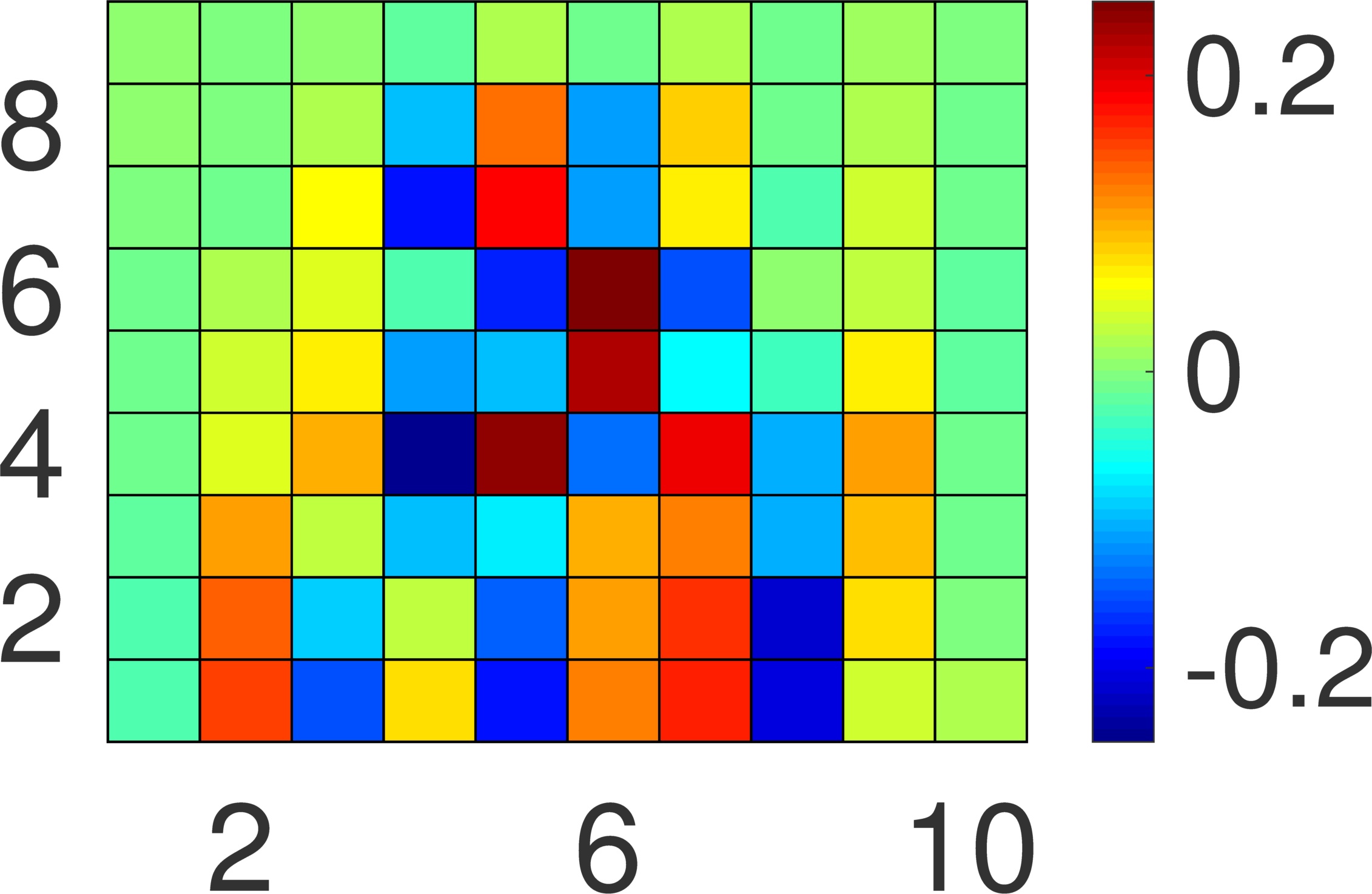} } 
		\hspace{0.3cm}
		\\
		\subfloat[][{$\lambda_7^{-1}= 8.0\times   10^{-3}$}] 
		{\includegraphics[width=0.30\textwidth]{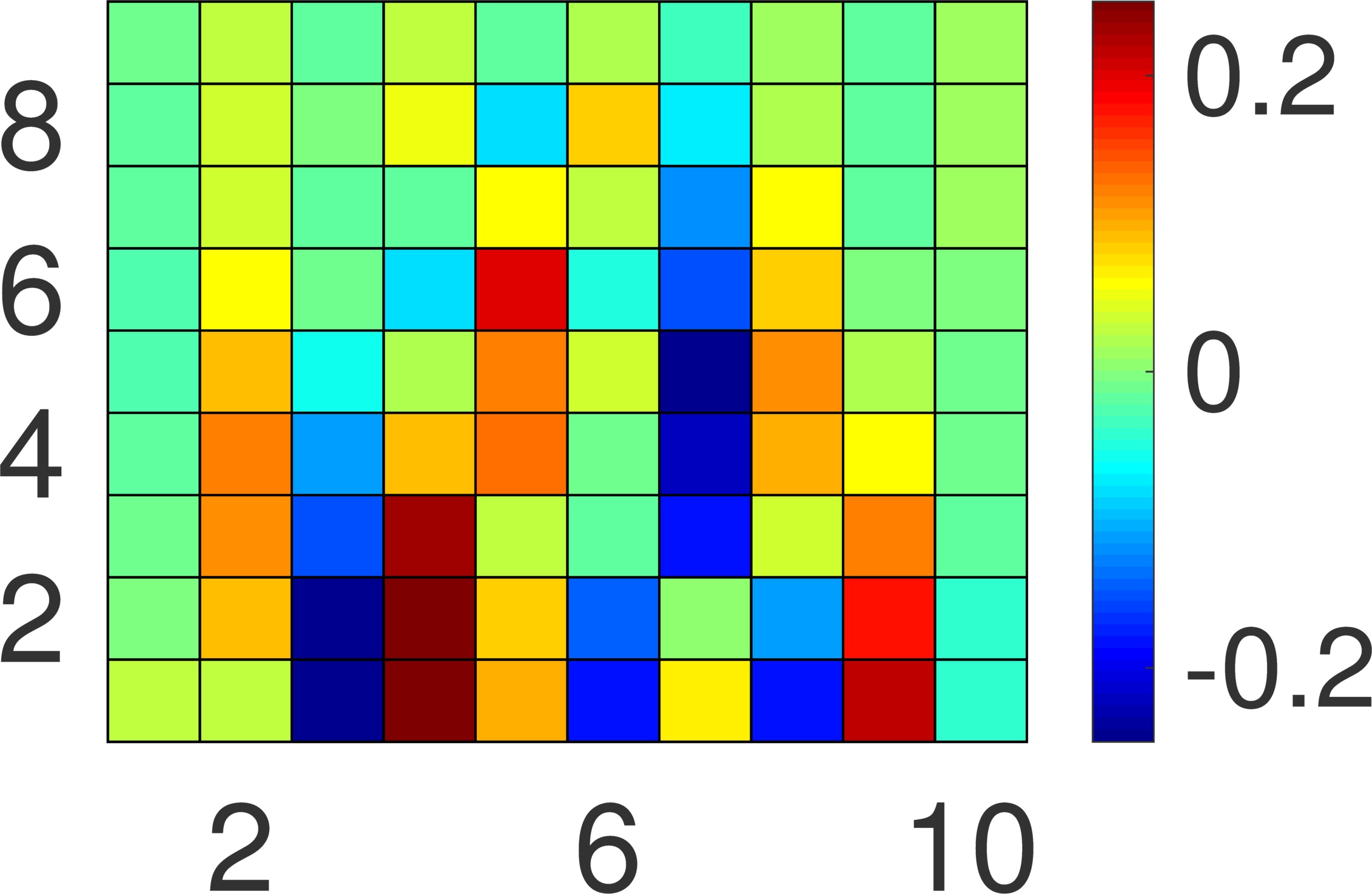}} 
		\hspace{0.1cm}
		\subfloat[][{$\lambda_8^{-1} = 7.5\times  10^{-3}$ }]
		{\includegraphics[width=0.30\textwidth]{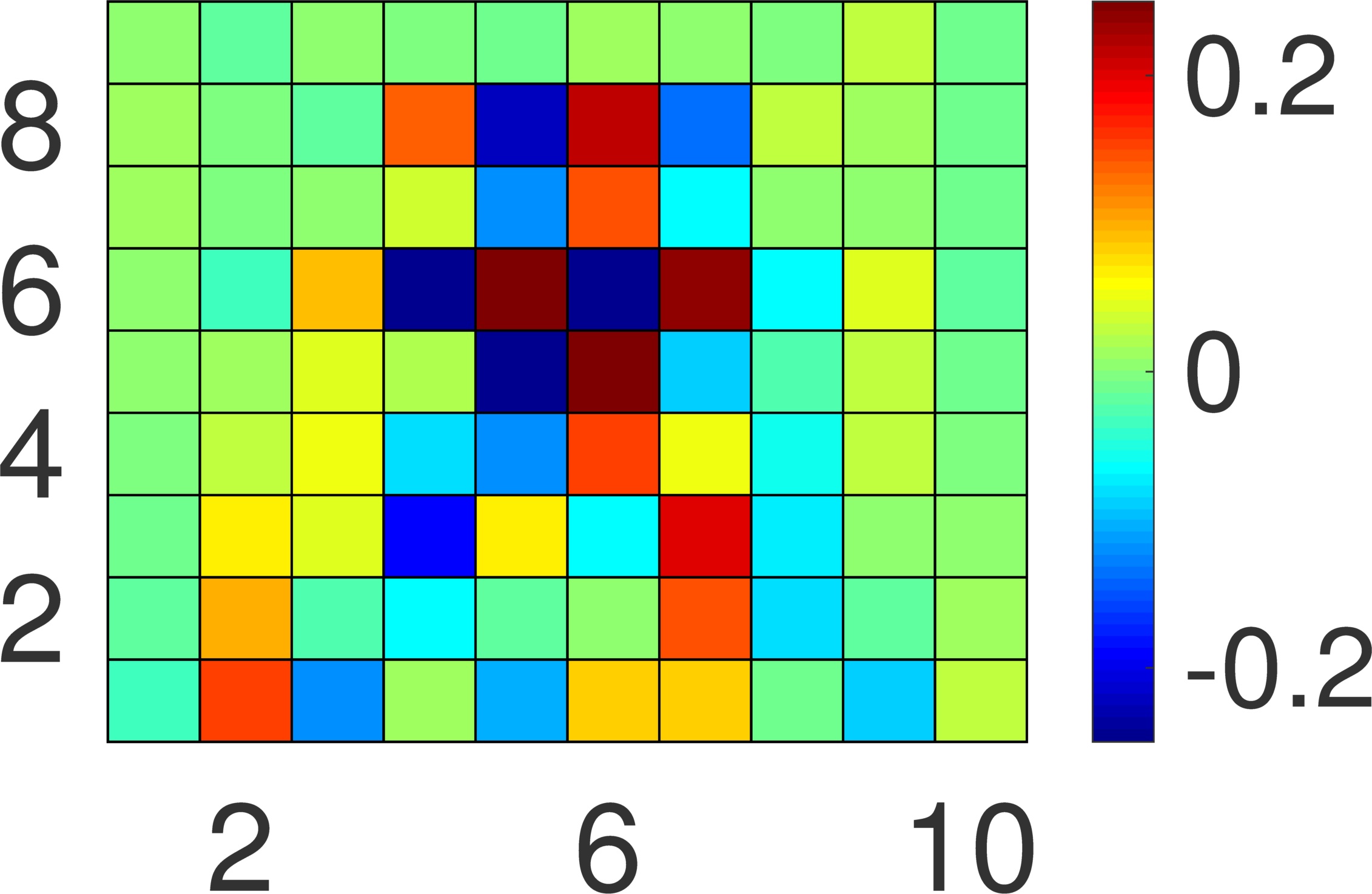} } 
		\hspace{0.1cm}
		\subfloat[][{$\lambda_9^{-1}  = 6.6\times  10^{-3}$ }]
		{\includegraphics[width=0.30\textwidth]{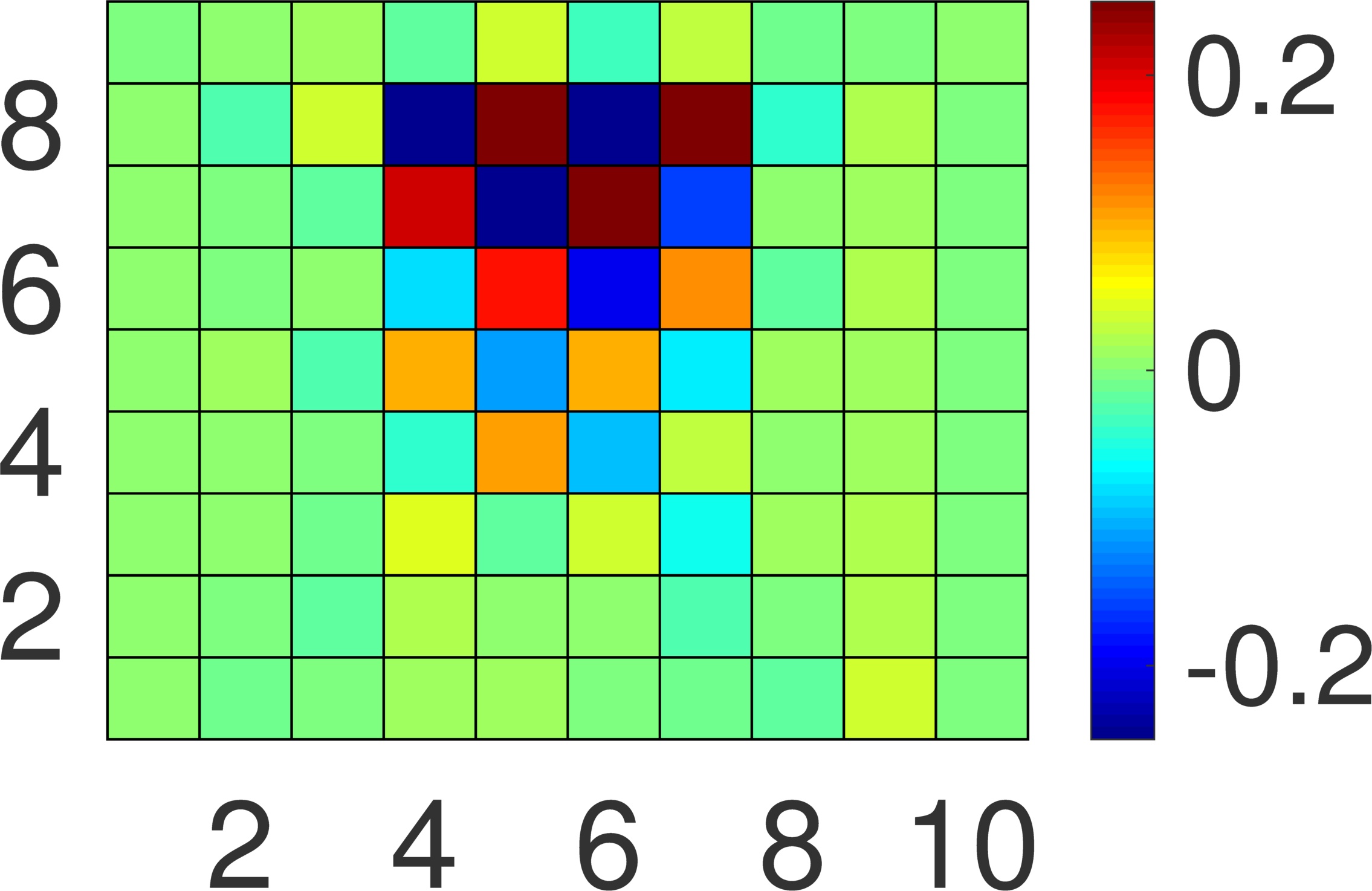} } 
		\hspace{0.3cm}
		
	  }
	  \caption{The first $9$  basis vectors  $\bs{w}_i$ in decreasing order, based on the corresponding variance $\lambda_i^{-1}$. One notes that the variance captured by the $9^{th}$ reduced coordinate is more than one magnitude smaller than that of the $1^{st}$  reduced coordinate.}\label{fig:BASES}
\end{figure}

\begin{figure}[H]
	\centering
	\includegraphics[width=0.60\textwidth]{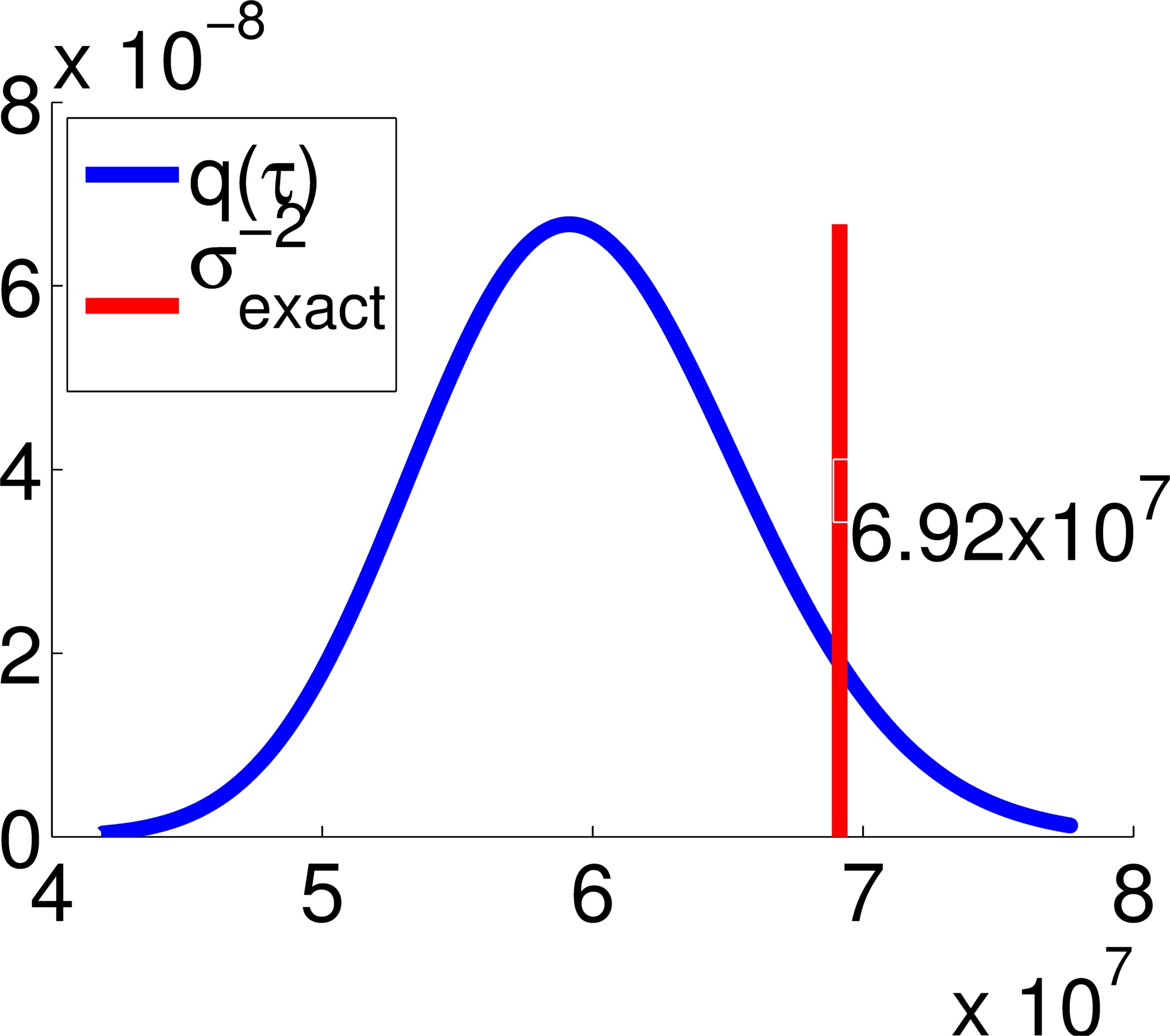}
	  \caption{Posterior distribution  $q(\tau)$ for $9$ bases and the exact value.}\label{fig:Tau9}
\end{figure}

The $9$ most important basis vectors $\bs{w}_i$  can be seen in  \reffig{fig:BASES}, in decreasing order,  based on the corresponding variance $\lambda_i^{-1}$. %These are the same nine bases used to derive the posterior in \reffig{fig:PosteriorSmall}d, \reffig{fig:PosteriorSmall}e, \reffig{fig:PosteriorSmall}f. Below each sub-figure the variance of each basis can be seen.
Finally,  the posterior of $\tau$ is depicted in \reffig{fig:Tau9}. One can observe that the magnitude is captured correctly, compared to the exact value, i.e. the corresponding variance of the Gaussian noise with which the data was contaminated. %One needs to mention that the exact value $\sigma^{-2}_{exact}$ does not necessarily go coincide with the standard deviation one detects comparing the noisy displacements and the ones without noise. This can especially occur for a small number of data.

\begin{figure}[H]{
	\centering
	%\captionsetup[subfigure]{labelformat=empty}
		%\vspace{-0.5cm}
		\subfloat[][{Mean, $d_{\bs{\Theta}}=9$}] 
		{\includegraphics[width=0.32\textwidth]{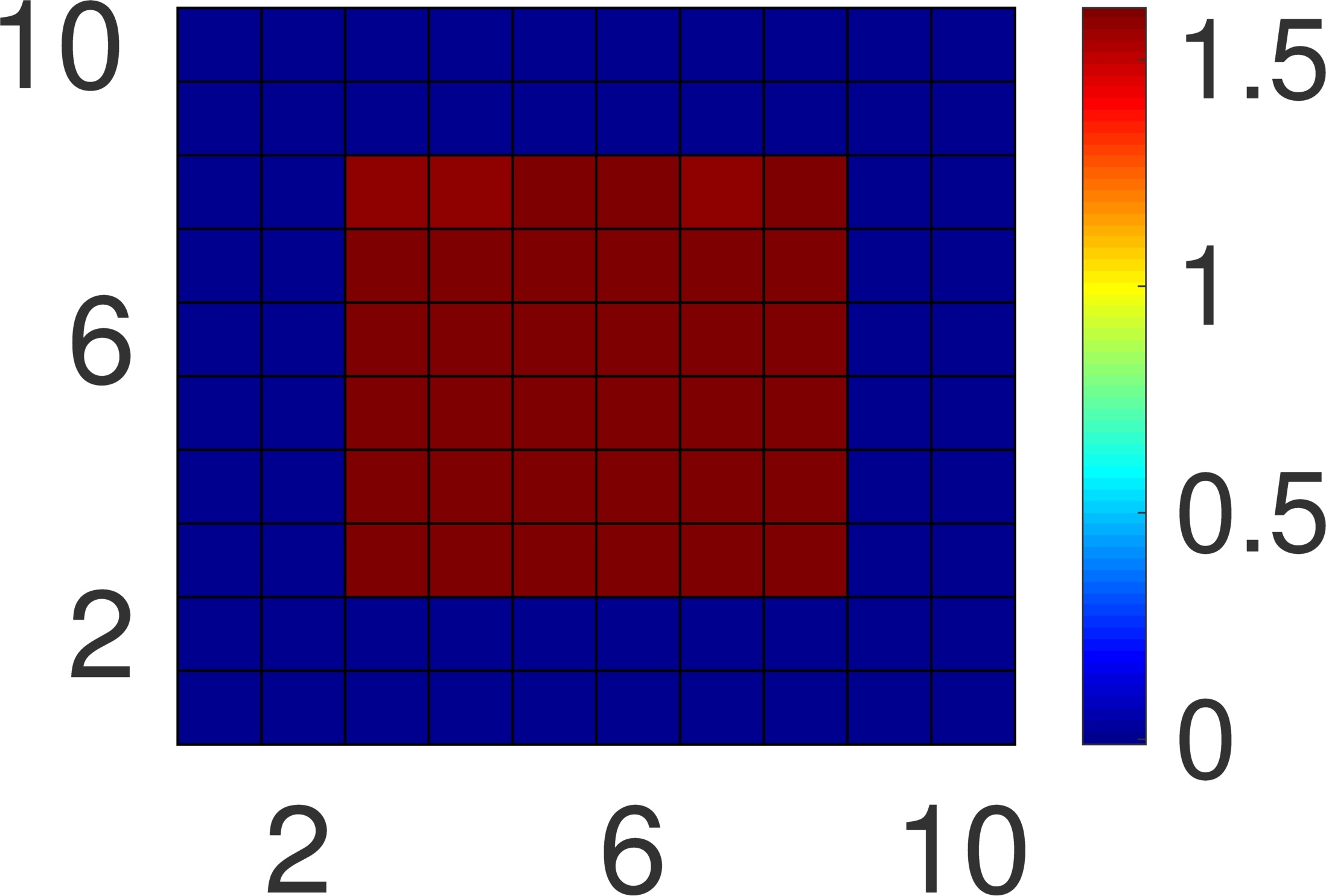}} 
		\hspace{0.1cm}
		\subfloat[][{ Diagonal cut, $d_{\bs{\Theta}}=9$}]
		{\includegraphics[width=0.30\textwidth]{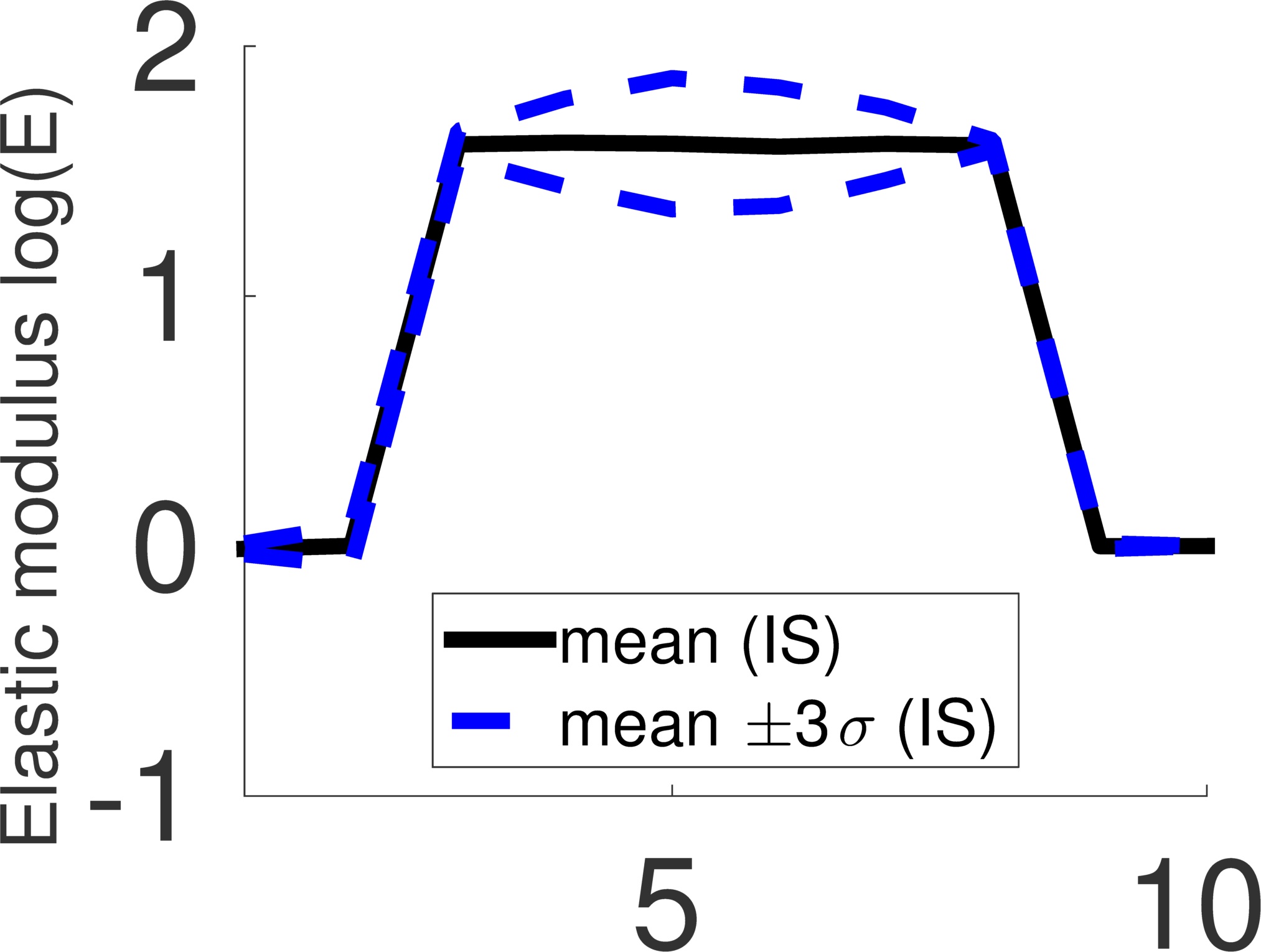} } 
		\hspace{0.1cm}
		\subfloat[][{ St. dev.}]
		  {\includegraphics[width=0.32\textwidth]{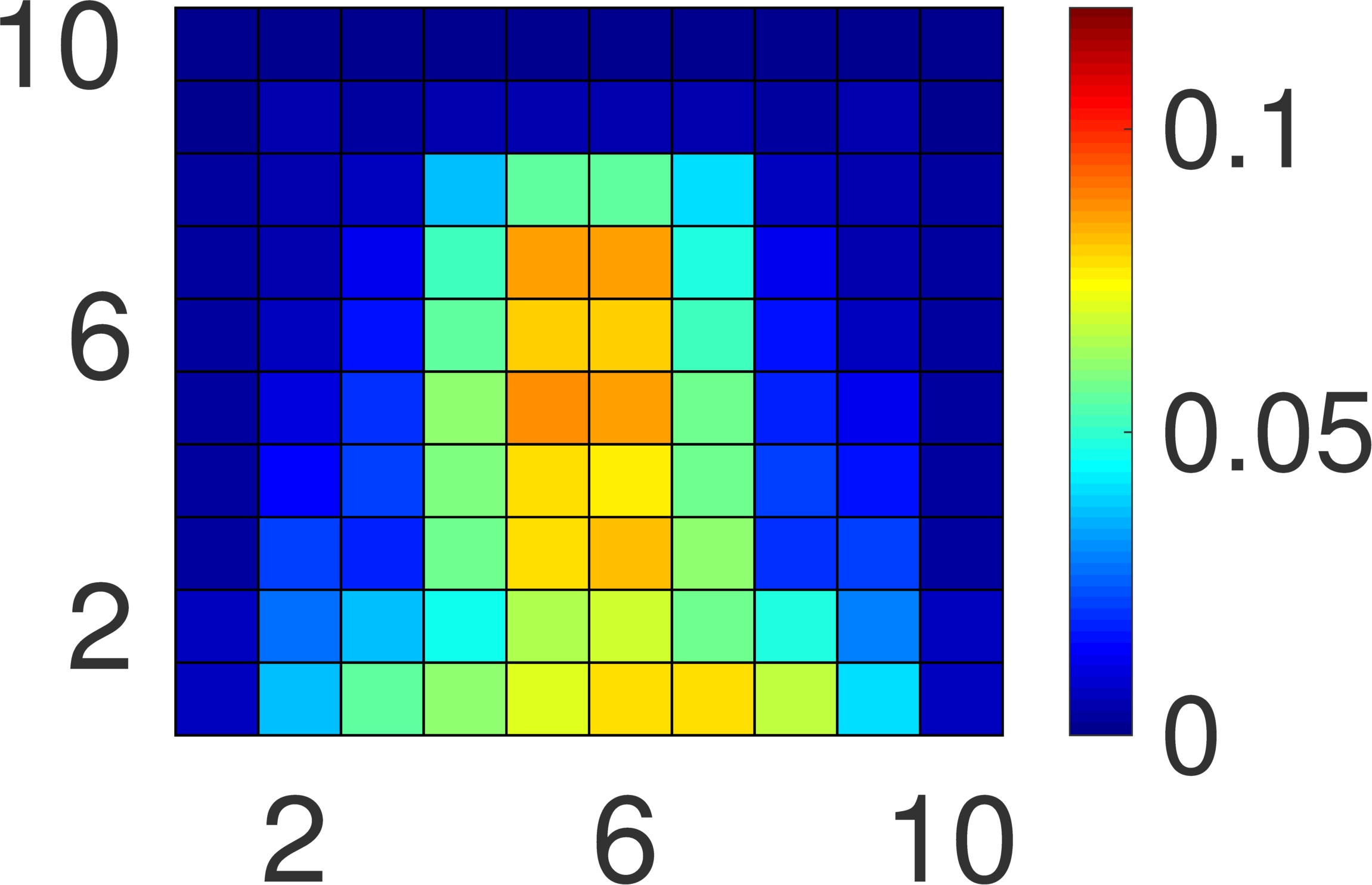} } 	 
	  }
	  \caption{First and second order statistics of the exact posterior as estimated with Importance Sampling. Figure (a) depicts the posterior mean $\bs{\mu}$ of the elastic moduli $E$ in log-scale. Figure (b) shows the posterior mean and posterior quantiles ($\pm 3$ standard deviations) along the diagonal from $(0,0)$ to $(10,10)$ and Figure (c) depicts the posterior standard deviation. These  should be compared with the VB approximations in \reffig{fig:PosteriorSmall}.}
	 \label{fig:PosteriorSmallIS}
\end{figure}

\begin{figure}[H]
	\centering
	\includegraphics[width=0.60\textwidth]{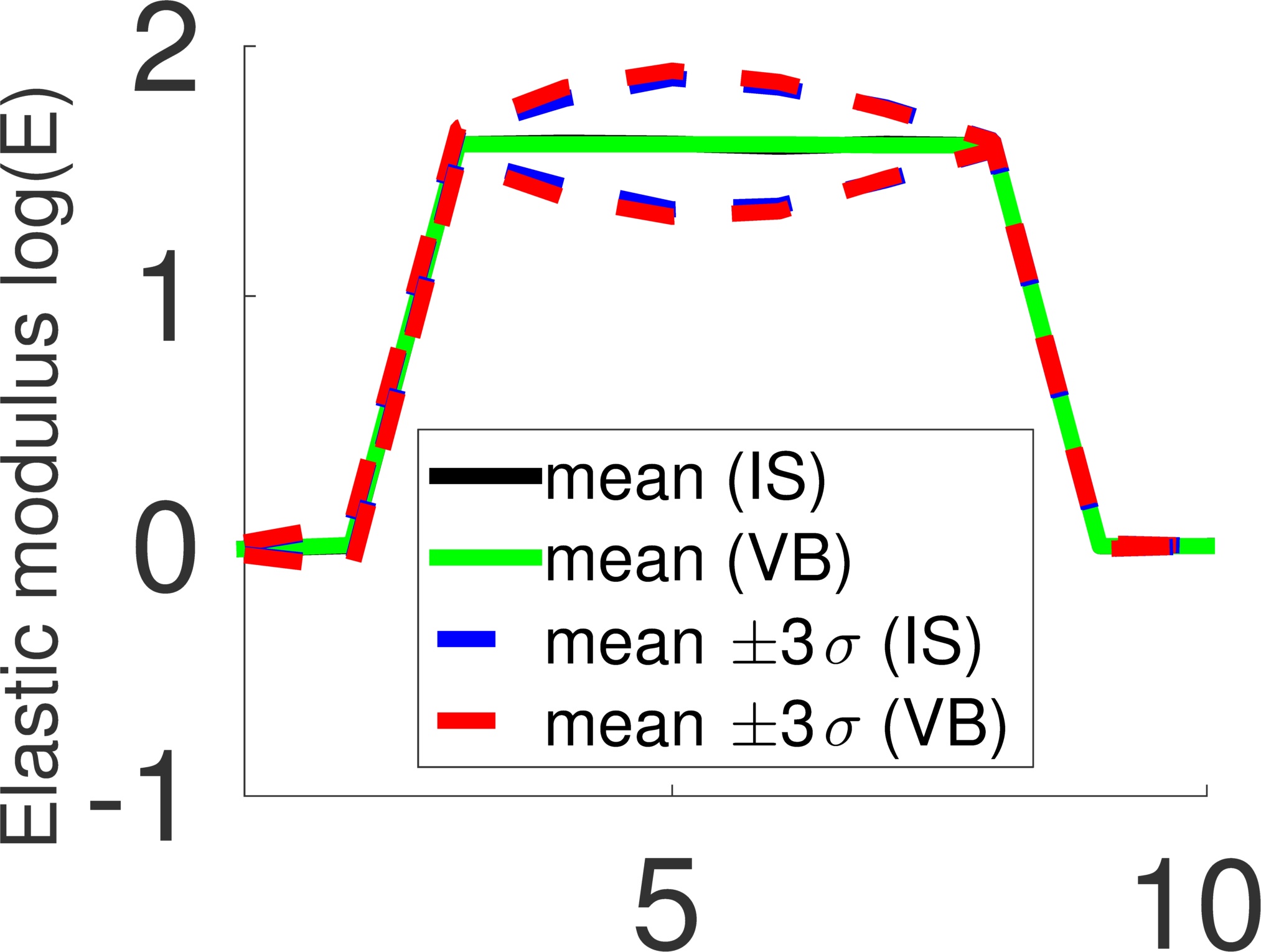}
	  \caption{Posterior mean and posterior quantiles ($\pm3$ standard deviations) along the diagonal from
(0, 0) to (10, 10) for VB and Importance Sampling (IS).}\label{fig:CutThetaVBIS}
\end{figure}

\tcreview{The aforementioned results were validated by employing Importance Sampling as discussed in Section \ref{sec:is}. The Effective Sample Size (ESS , \refeq{eq:ess})  was $0.25$ (for $d_{\Theta}=9$) which suggests a good approximation to the actual posterior is provided by then VB result \cite{liu_monte_2001}. More importantly, as it is shown in Figures \ref{fig:PosteriorSmallIS} and \ref{fig:CutThetaVBIS}, the first and second-order statistics of the exact posterior (estimated with Importance Sampling) and the VB approximation.}
% there is very good agreement BBased on our results we derived Importance Sampling to validate the variational results and their accuracy. The results are built on the same configuration as above. The resulting normalized effective sample size (divided by the number of samples) is around $20-25 \%$ which shows the good approximation of the posterior. In \reffig{fig:PosteriorSmallIS} the posterior mean and the standard deviation, derived by the Importance Sampling (IS) algorithm, is shown. In \reffig{fig:CutThetaVBIS} VB and IS are directly confronted and it is visible that the results are in accordance with each other.

\subsection{Incompressible Mooney-Rivlin material}
Nonlinear, hyperelastic  models have been successfully used in the past to describe the  behavior of several biomaterials \cite{samani_method_2004, ohagan_measurement_2009, oberai_linear_2009}. In this example we employ the  Mooney-Rivlin model  \cite{mooney_theory_1940, rivlin_large_1948} that is characterized by the following strain energy density function $w$ (\refeq{eq:const}): 
\be
 w =  c_1(\hat{I}_1-3)+c_2(\hat{I}_2-3)+\frac{1}{2} \kappa (log J)^2
 \label{eq:mr}
\ee
where $\kappa$ is the bulk modulus, $J = det(\bs{F})$ and $\hat{I}_1=\frac{I_1}{J^{2/3}}, \hat{I}_2=\frac{I_2}{J^{4/3}}$  where $I_1,I_2$ are the first and second invariants of the of the left Cauchy-Green deformation tensor $\bs{b}=\bs{FF}^T$. The last term in \refeq{eq:mr} is related to volumetric deformations whereas the first two terms to distortional. We consider here an {\em incompressible} material, i.e.  $J=1$, in which case the bulk modulus $\kappa$ plays the role of  a penalty parameter  that enforces this constraint. 
We employ the three-field Hu-Washizu  principle in order to enforce incompressibility and  suppress volumetric locking  \cite{simo_quasi-incompressible_1991,zienkiewicz_finite_1977}. The three-field formulation requires a 
separate integration rule for the dilatational stiffness contribution. The bulk modulus is chosen as a function of $c_1$ with $\kappa = \kappa_o c_1$. We use $\kappa_0 = 1000$ \cite{simo_quasi-incompressible_1991, schoeberl_comparison_2013}. The higher $\kappa_0$ is, the stronger is the incompressibility constraint.

\begin{figure}[H]
	\centering
	\includegraphics[width=0.45\textwidth]{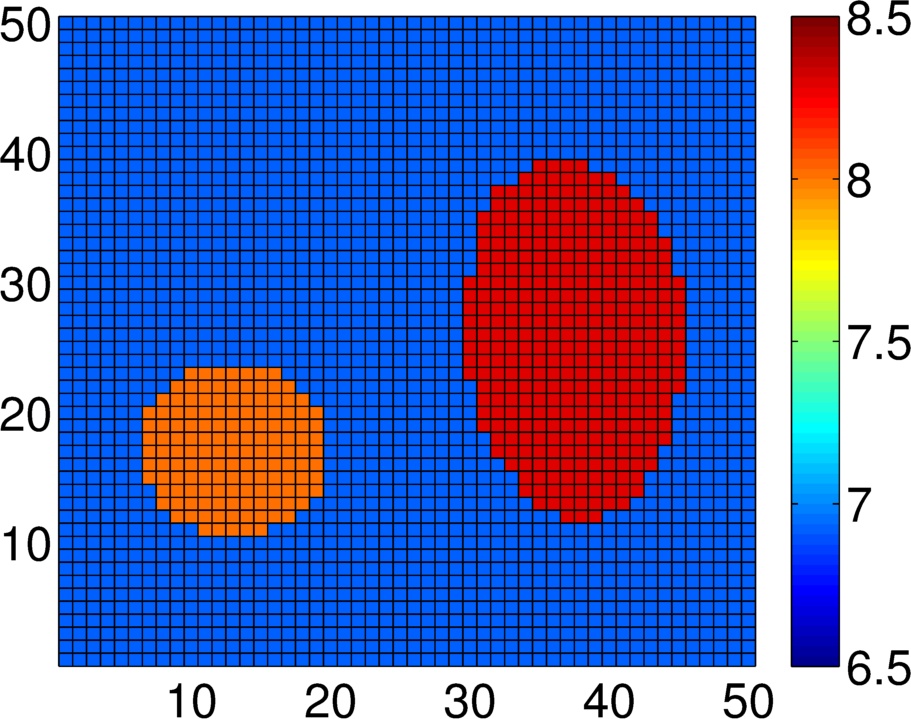} 
	\caption{Reference $c_1$ distribution in the log-scale.}
	\label{fig:ForwardLarge}
	%With AdditionalVariance
\end{figure}

In this example we assume $c_2=0$ which reduces the model to an uncoupled version of the incompressible neo-Hookean model \cite{mase_continuum_2009}.  The remaining parameter  $c_1$ is assumed to vary in the problem domain which can be seen in \reffig{fig:ForwardLarge}. In this example we have two inclusions, an elliptic and a circular inclusion, with different material properties. In the larger, elliptic inclusion $c_1 = 4000$ (red),  in the smaller, circular inclusion $c_1 = 3000$ (orange) and in the remaining material $c_1 =1000$ (blue).
 The problem domain is $\Omega =(0,L)\times (0,L)$ with $L=50$. It is discretized with $200 \times 200$ finite elements of equal size and the governing equations are solved under plane strain conditions. The following boundary conditions are employed: both displacements are set to zero at the bottom ($x_2=0$) and a vertical distributed load $f=-100$ in the vertical direction (pointing downwards) is applied along the  top i.e. $x_2=50$. The vertical edges  ($x_1 = 0, 50$) are traction-free.

The forward model for the Bayesian identification employed a regular $50 \times 50$ mesh and only the corresponding (noisy) displacements at the nodes were used as data ($\hat{\bs{y}}$). We note that due to the different meshes employed the data will also contain model (discretization) errors. The SNRs reported in the sequence include also these errors. We further assumed that $c_1$ was constant within each of the elements which resulted in {\em $d_{\Psi}=2500$ unknowns}. Using the displacements obtained from the fine $200 \times 200$ mesh we consider three settings: 
\bi
\item Case A (high SNR/low noise): without additional noise resulting in a SNR $1.93\times10^3$.
\item Case B (medium SNR/medium noise): the data are contaminated with relatively smaller Gaussian noise resulting in a total SNR $1.89\times10^3$.  
%\item Case C (small SNR/high noise): the data  are contaminated with relatively larger Gaussian noise resulting in a total SNR $1.60 \times 10^3$.
\item Case C (small SNR/high noise): the data  are contaminated with relatively larger Gaussian noise resulting in a total SNR $6.9 \times 10^2$.
\ei

The results presented in the sequence were obtained for $\lambda_{0,1}=5\times 10^{-1}$ and the  material parameters are plotted in the log-scale. 

\begin{figure}[H]{
	\centering
	\captionsetup[subfigure]{labelformat=empty}
		%\vspace{-0.5cm}
% 		\subfloat[][{ With SNR 692}] 
% 		{\includegraphics[width=0.30\textwidth]{FiguresIF/PlotYoungModulusMeanVBSNR3.jpg}} 
% 		\hspace{0.1cm}
		\subfloat[][{ SNR $1.93\times10^3$}] 
		{\includegraphics[width=0.30\textwidth]{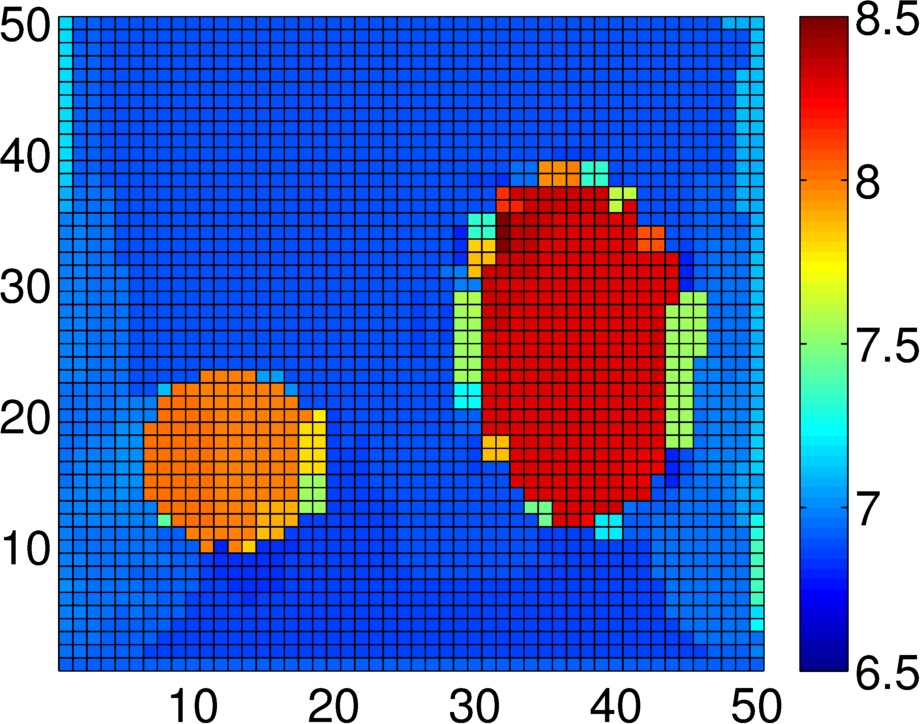}} 
		\hspace{0.1cm}
		\subfloat[][{ SNR $1.89\times10^3$}] 
		{\includegraphics[width=0.30\textwidth]{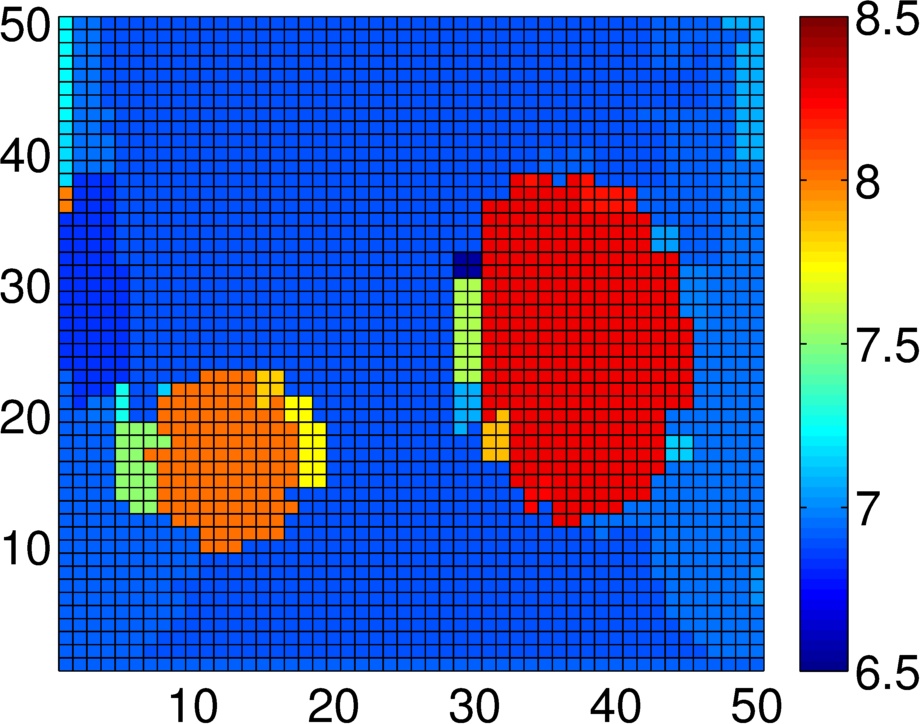}} 
% 		\hspace{0.1cm}
% 		\subfloat[][{ SNR $1.60\times10^3$}] 
% 		{\includegraphics[width=0.30\textwidth]{FiguresIF/PlotYoungModulusMeanVBSNR4.jpg}} 
		\hspace{0.1cm}
		\subfloat[][{ SNR $6.9\times10^2$}] 
		{\includegraphics[width=0.30\textwidth]{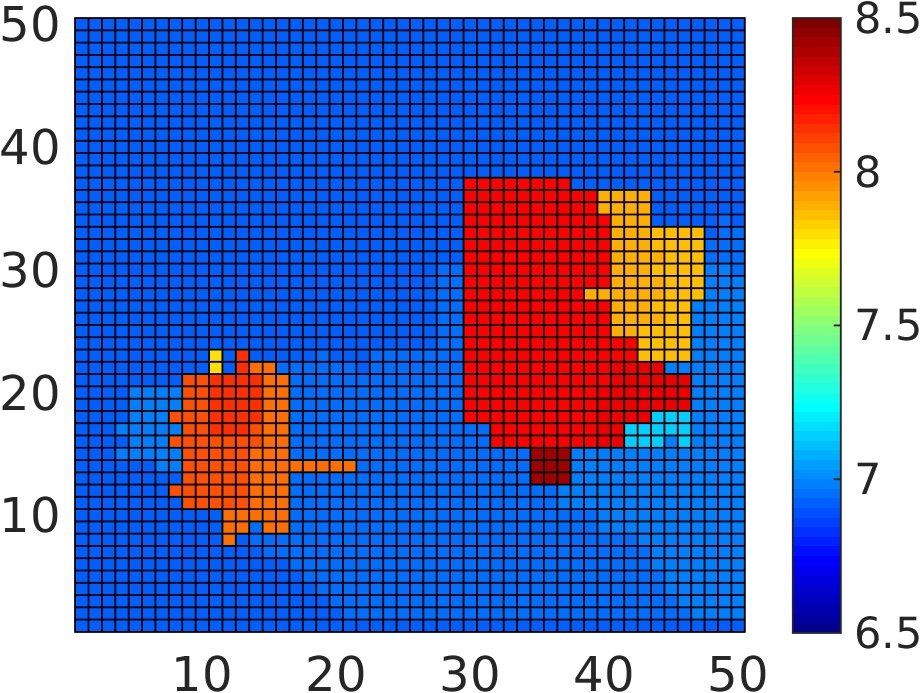}} 
		  \hspace{0.3cm}
	  }
	  \caption{Posterior mean of $c_1$ in log-scale for Cases A (large SNR), B (medium SNR) and C (small SNR).}
	 \label{fig:PosteriorMean}
\end{figure}

\begin{figure}[H]{
	\centering
	\captionsetup[subfigure]{labelformat=empty}
% 		\vspace{-0.5cm}
% 		\subfloat[][{ With SNR 692}] 
% 		{\includegraphics[width=0.30\textwidth]{FiguresIF/StdSNR3.jpg}} 
% 		\hspace{0.1cm}
		\subfloat[][{ SNR $1.93\times10^3$}] 
		%{\includegraphics[width=0.30\textwidth]{FiguresIF/StdModel.jpg}} 
		{\includegraphics[width=0.30\textwidth]{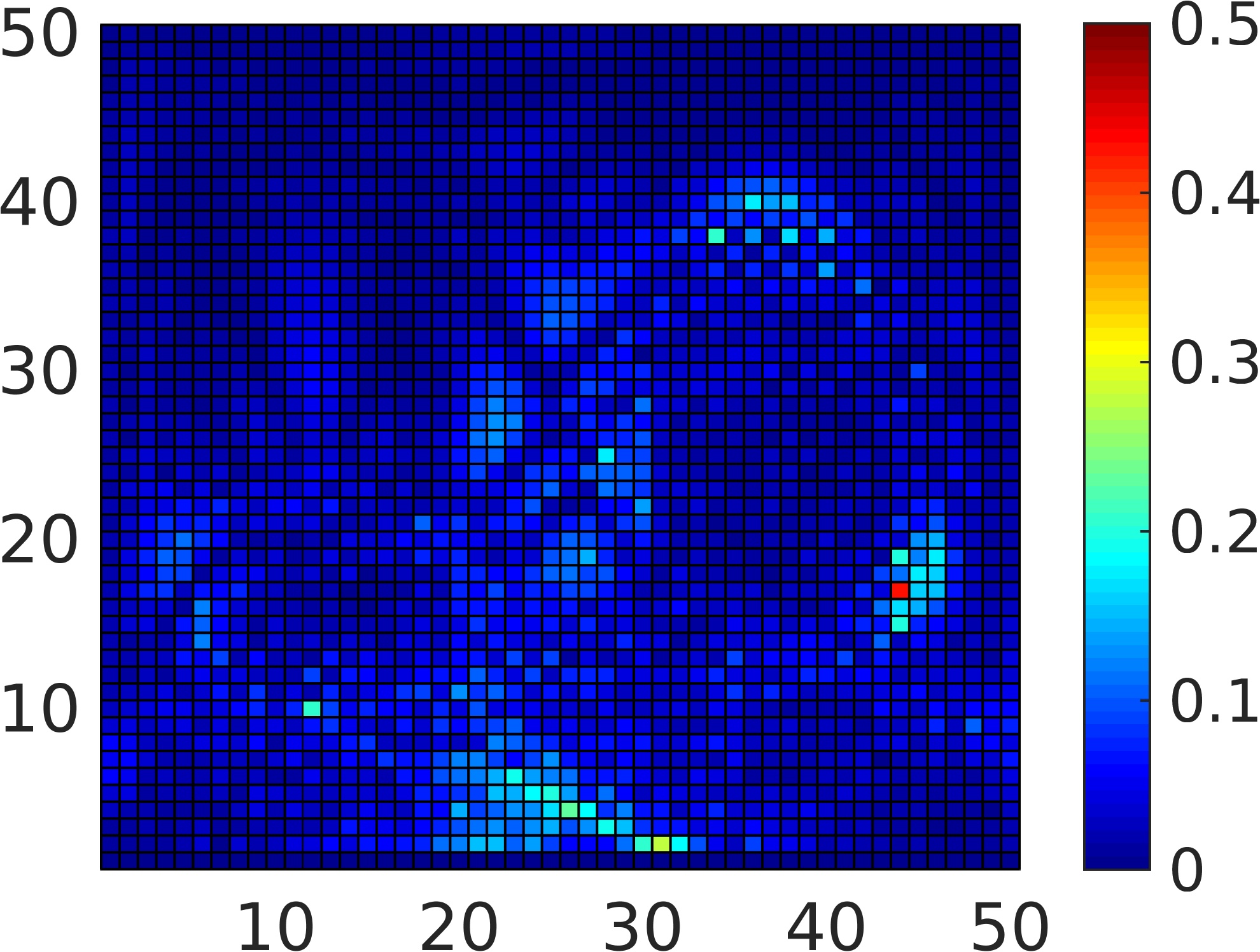}} 
		\hspace{0.1cm}
		\subfloat[][{ SNR $1.89\times10^3$}] 
		%{\includegraphics[width=0.30\textwidth]{FiguresIF/StdSNR5.jpg}} 
		{\includegraphics[width=0.30\textwidth]{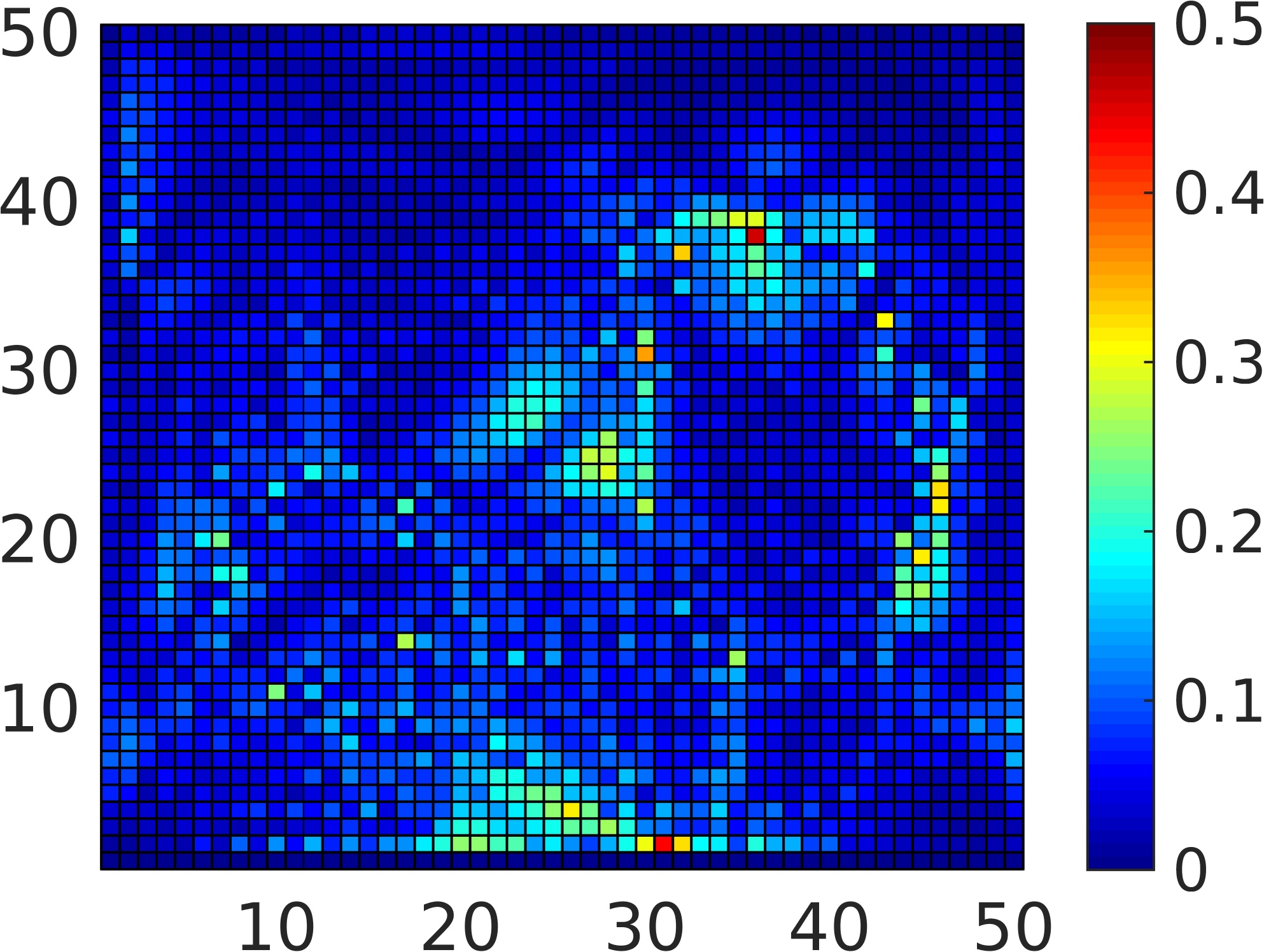}} 
		\hspace{0.1cm}
% 		\subfloat[][{ SNR $1.60\times10^3$}] 
% 		{\includegraphics[width=0.30\textwidth]{FiguresIF/StdSNR4.jpg}} 
% 		\hspace{0.1cm}
		\subfloat[][{ SNR $6.9\times10^2$}] 
		%{\includegraphics[width=0.30\textwidth]{FiguresIF/StdSNR3.jpg}} 
		{\includegraphics[width=0.30\textwidth]{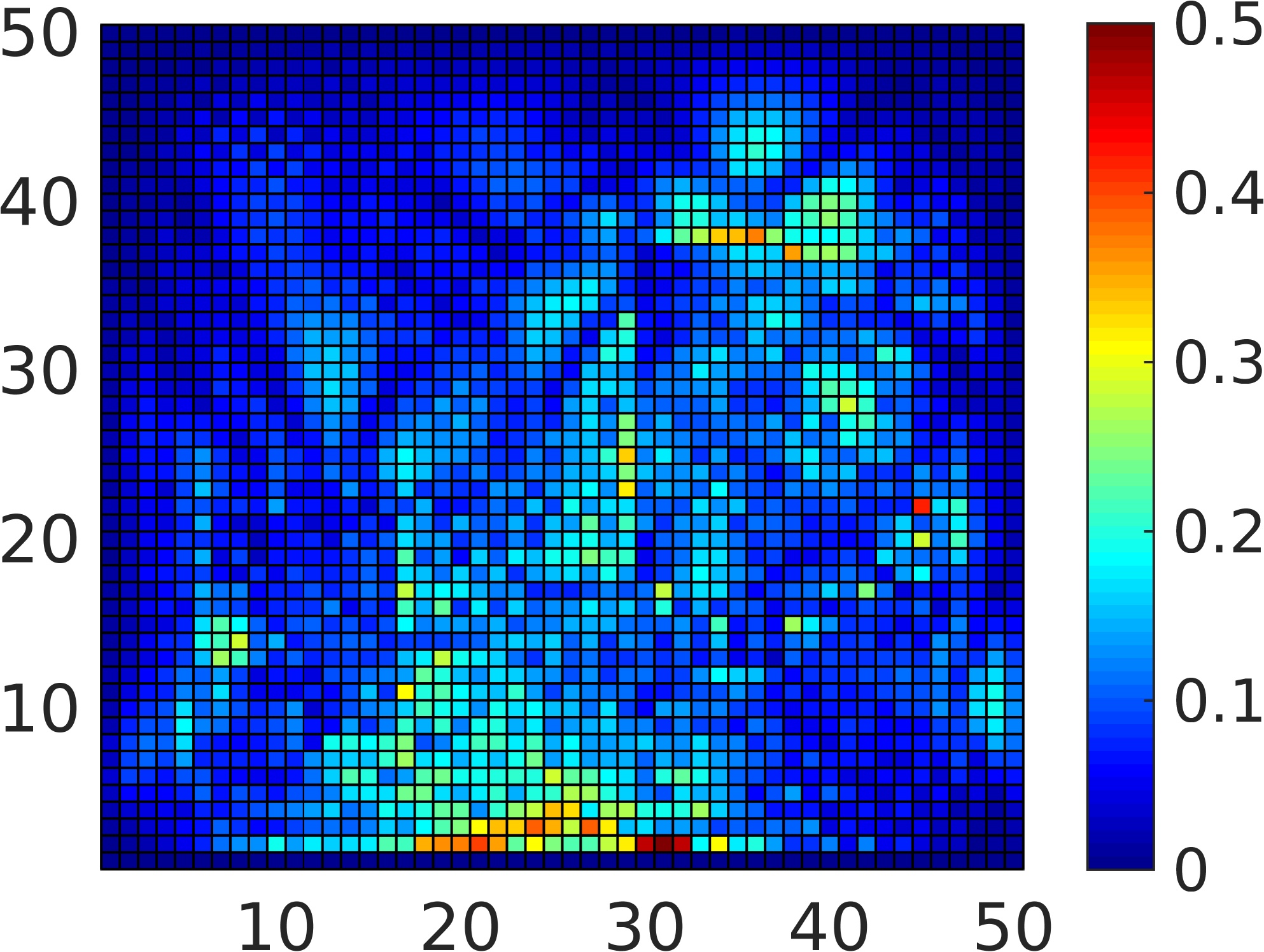}} 
		  \hspace{0.3cm}
	  }
	  \caption{ Posterior standard deviation of $c_1$ in log-scale for Cases A (large SNR, $\dth = 10$), B (medium SNR, $\dth = 12$) and C (small SNR, $\dth = 13$).}
	 \label{fig:StdLarge}
\end{figure}

\begin{figure}[H]{
	\centering
	\captionsetup[subfigure]{labelformat=empty}
% 		\vspace{-0.5cm}
% 		\subfloat[][{With SNR 692}] 
% 		{\includegraphics[width=0.30\textwidth]{FiguresIF/PsiValuesWithVariancesDiagonalSNR3.jpg}} 
% 		\hspace{0.1cm}
		\subfloat[][{ SNR $1.93\times10^3$}] 
		%{\includegraphics[width=0.30\textwidth]{FiguresIF/PsiValuesWithVariancesDiagonalModel.jpg}} 
		{\includegraphics[width=0.30\textwidth]{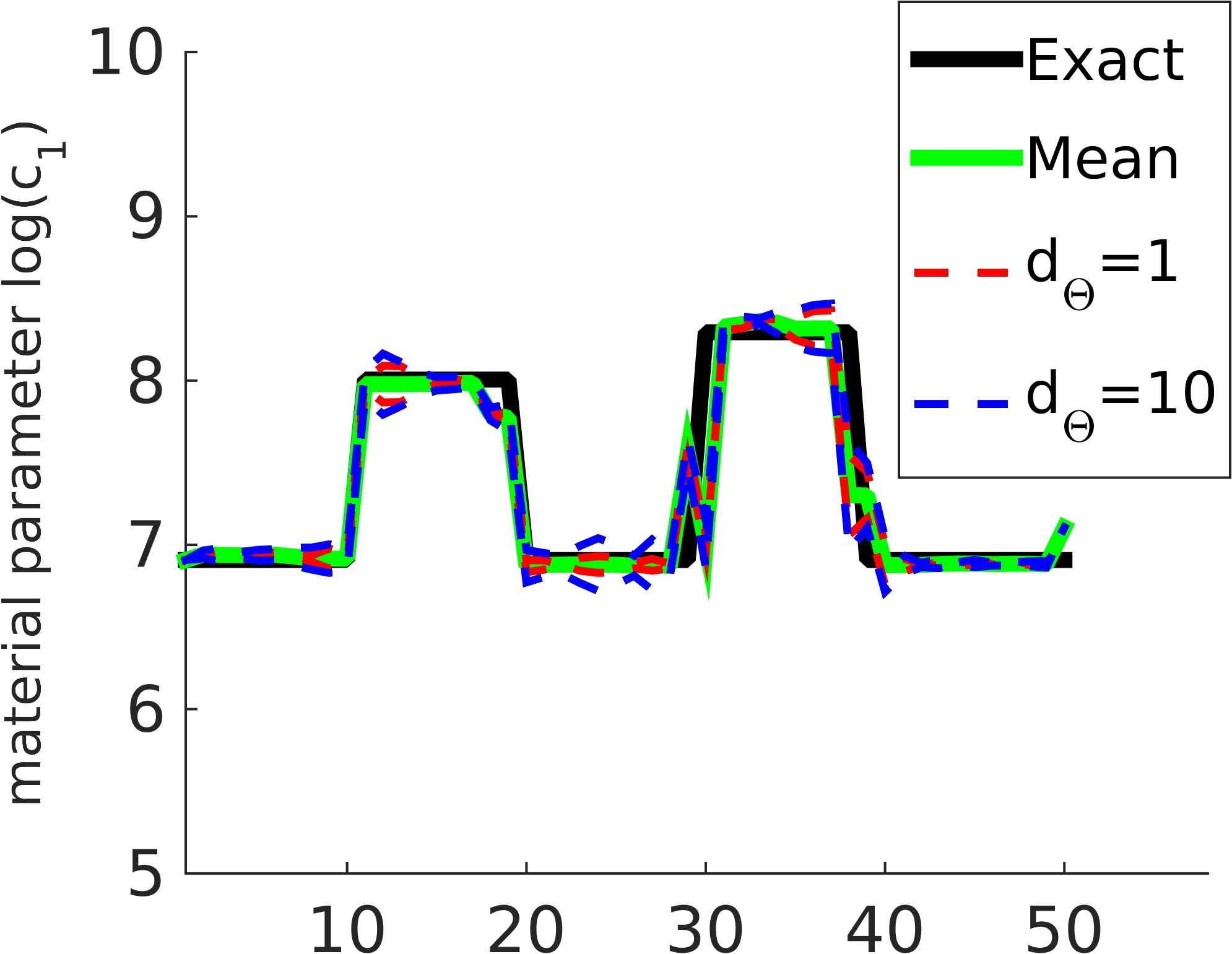}} 
		\hspace{0.1cm}
		\subfloat[][{ SNR $1.89\times10^3$}] 
		%{\includegraphics[width=0.30\textwidth]{FiguresIF/PsiValuesWithVariancesDiagonalSNR5.jpg}} 
		{\includegraphics[width=0.30\textwidth]{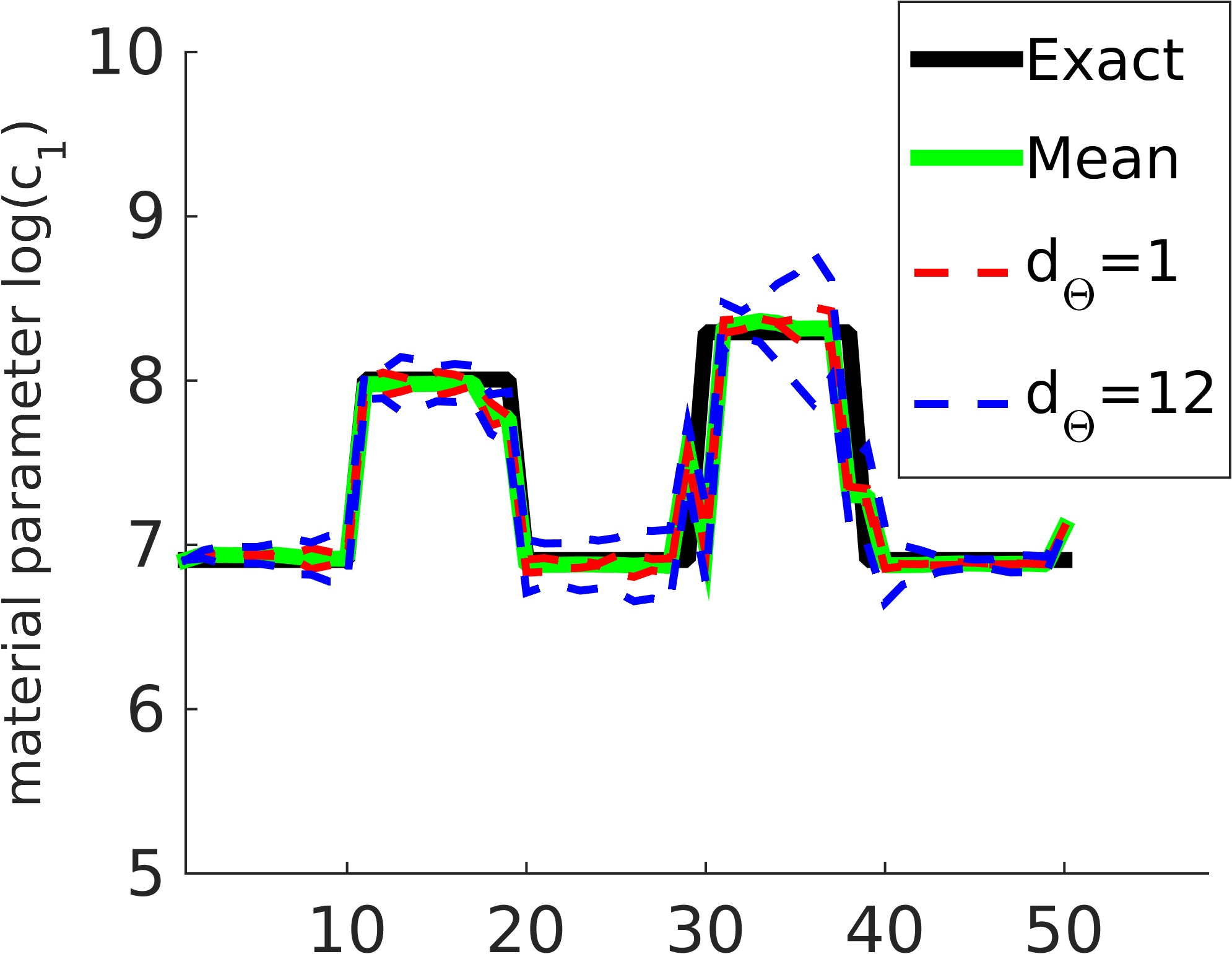}} 
% 		\hspace{0.1cm}
% 		\subfloat[][{ SNR $1.60\times10^3$}] 
% 		{\includegraphics[width=0.30\textwidth]{FiguresIF/PsiValuesWithVariancesDiagonalSNR4.jpg}} 
		\hspace{0.1cm}
		\subfloat[][{ SNR $6.9\times10^2$}] 
		%{\includegraphics[width=0.30\textwidth]{FiguresIF/PsiValuesWithVariancesDiagonalSNR3.jpg}} 
		{\includegraphics[width=0.30\textwidth]{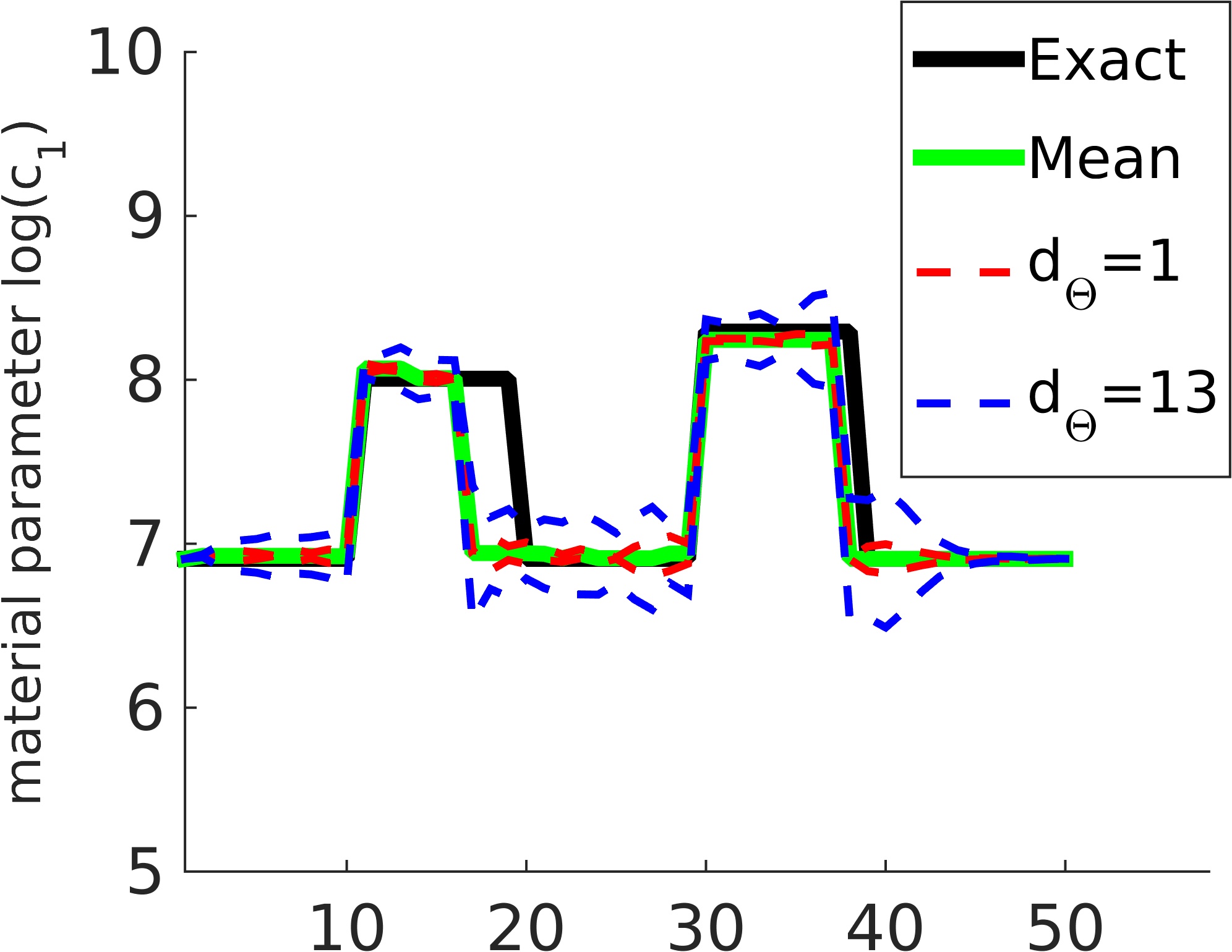}} 
		  \hspace{0.3cm}
	  }
	  \caption{ Posterior mean and credible intervals at $\pm 2$ standard deviations (dashed lines) along
the diagonal from $(0, 0)$ to $(50, 50)$ for various values of $\dth$  and  for Cases A (large SNR), B (medium SNR) and C (small SNR).
 The larger numbers  of $\dth$ correspond to the converged results as determined by \reffig{fig:IG}.}
	 \label{fig:DiagonalCutLarge}
\end{figure}

 \reffig{fig:PosteriorMean} depicts the posterior mean $\bs{\mu}$ for the aforementioned three cases.  \reffig{fig:StdLarge} depicts the spatial distribution of the posterior standard deviation as obtained by using the reduced coordinates. We note that in all cases (low to high SNR), $\bs{\mu}$  provides a reasonable approximation of the ground truth. The advantage of the proposed as well as all Bayesian techniques is that probabilistic estimates can be obtained in the form of the posterior density. This is illustrated in  \reffig{fig:DiagonalCutLarge} which depicts the posterior along the diagonal from $(0,0)$ to $(50,50)$. Firstly, we note that in all cases the posterior quantiles envelop by-and-large the ground truth. Secondly, as expected, these credible intervals are larger in cases where the SNR is smaller (noise is larger). Thirdly and most importantly, we note that these posterior approximations can be obtained by operating on  subspaces of dramatically reduced dimension in relation to the number of 
unknowns ($2500$).


\begin{figure}[H]{
	\centering
	\captionsetup[subfigure]{labelformat=empty}
% 		\vspace{-0.5cm}
% 		\subfloat[][{With SNR 692}] 
% 		{\includegraphics[width=0.30\textwidth]{FiguresIF/InformationGainSNR3.jpg}} 
% 		\hspace{0.1cm}
		\subfloat[][{ SNR $1.93\times10^3$}] 
		{\includegraphics[width=0.31\textwidth]{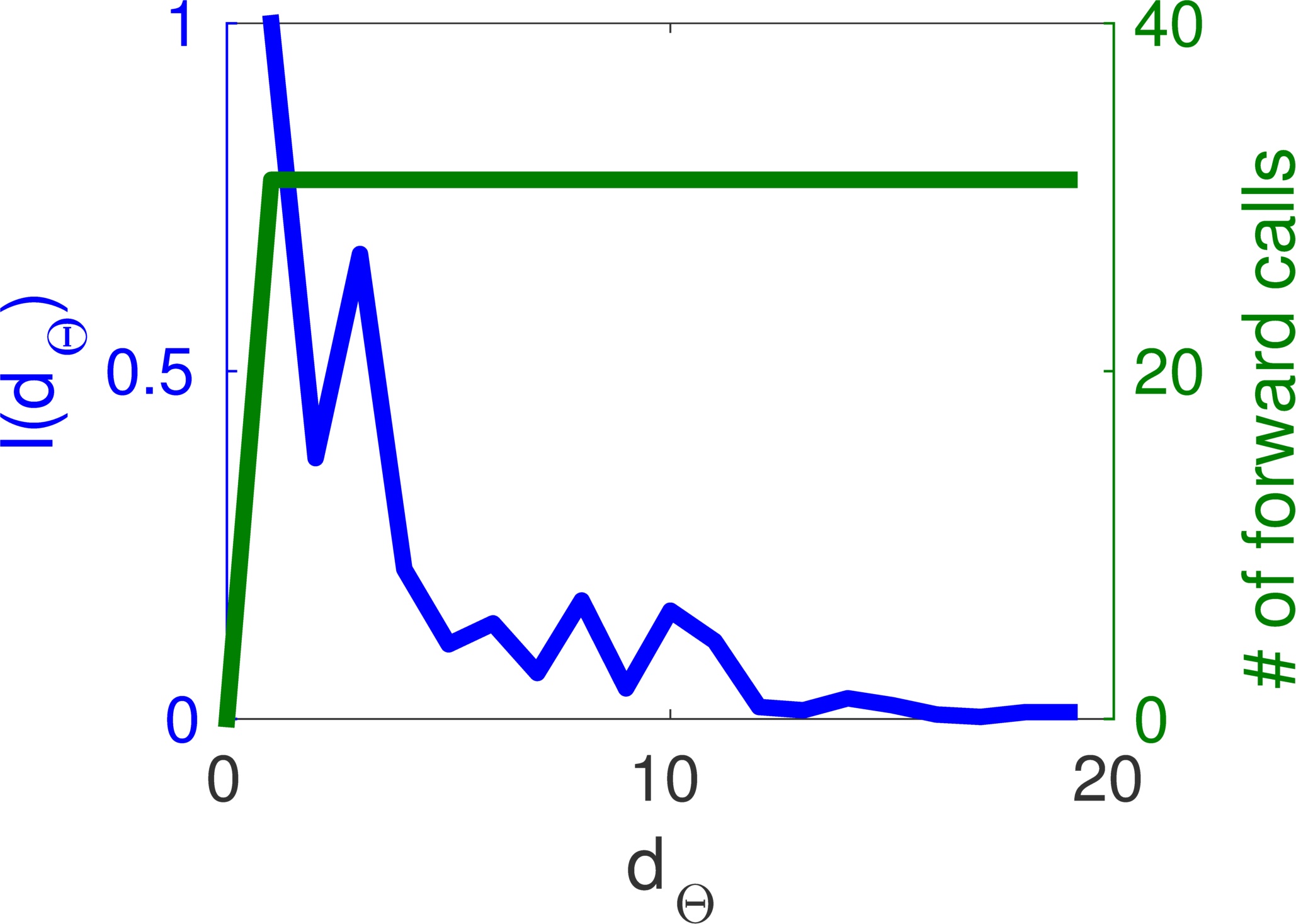}} 
		\hspace{0.1cm}
		\subfloat[][{ SNR $1.89\times10^3$}] 
		{\includegraphics[width=0.31\textwidth]{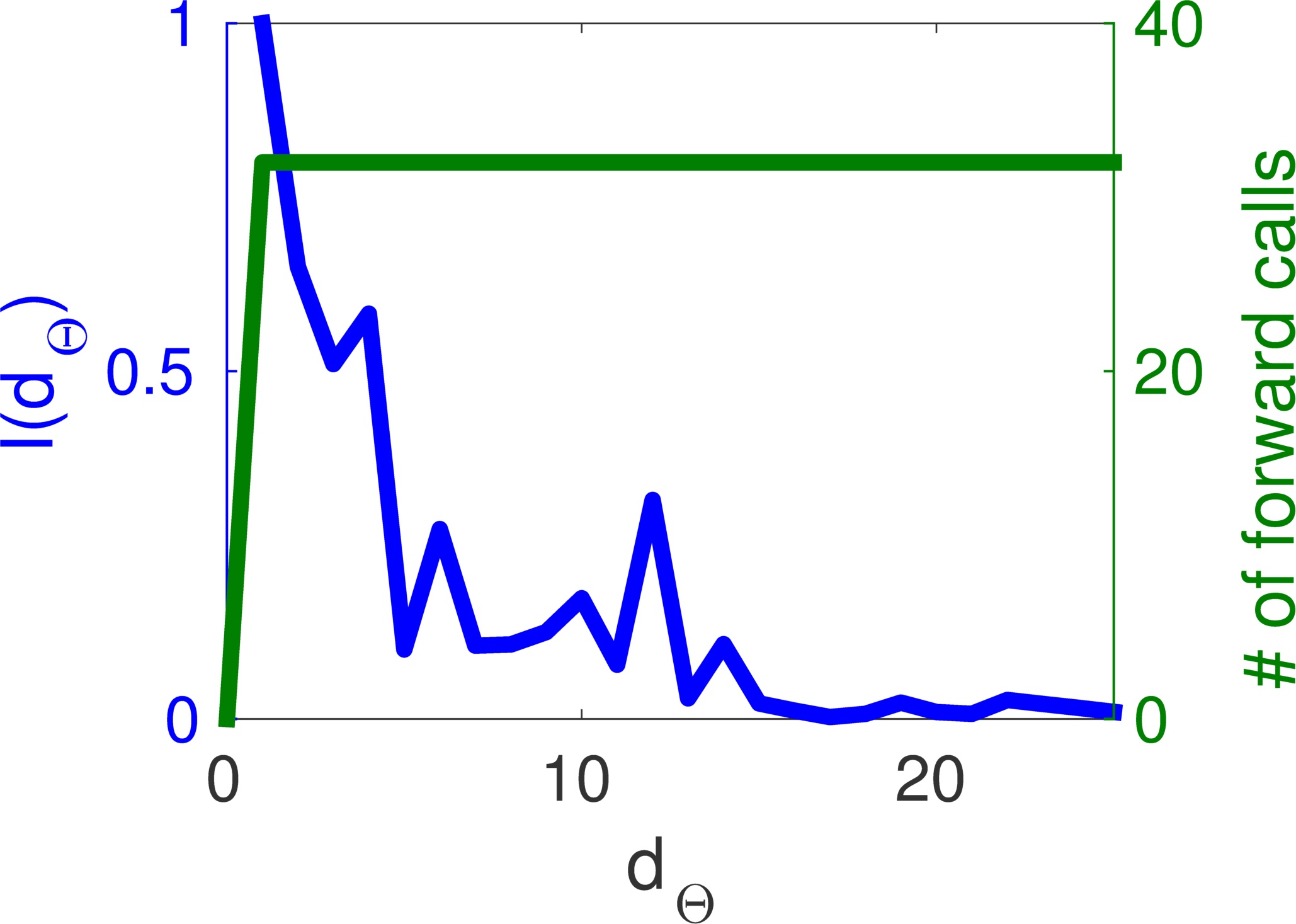}} 
% 		\hspace{0.1cm}
% 		\subfloat[][{ SNR $1.60\times10^3$}] 
% 		{\includegraphics[width=0.31\textwidth]{FiguresIF/InformationGainSNR4.jpg}} 
		\hspace{0.1cm}
		\subfloat[][{ SNR $6.9\times10^2$}] 
		{\includegraphics[width=0.31\textwidth]{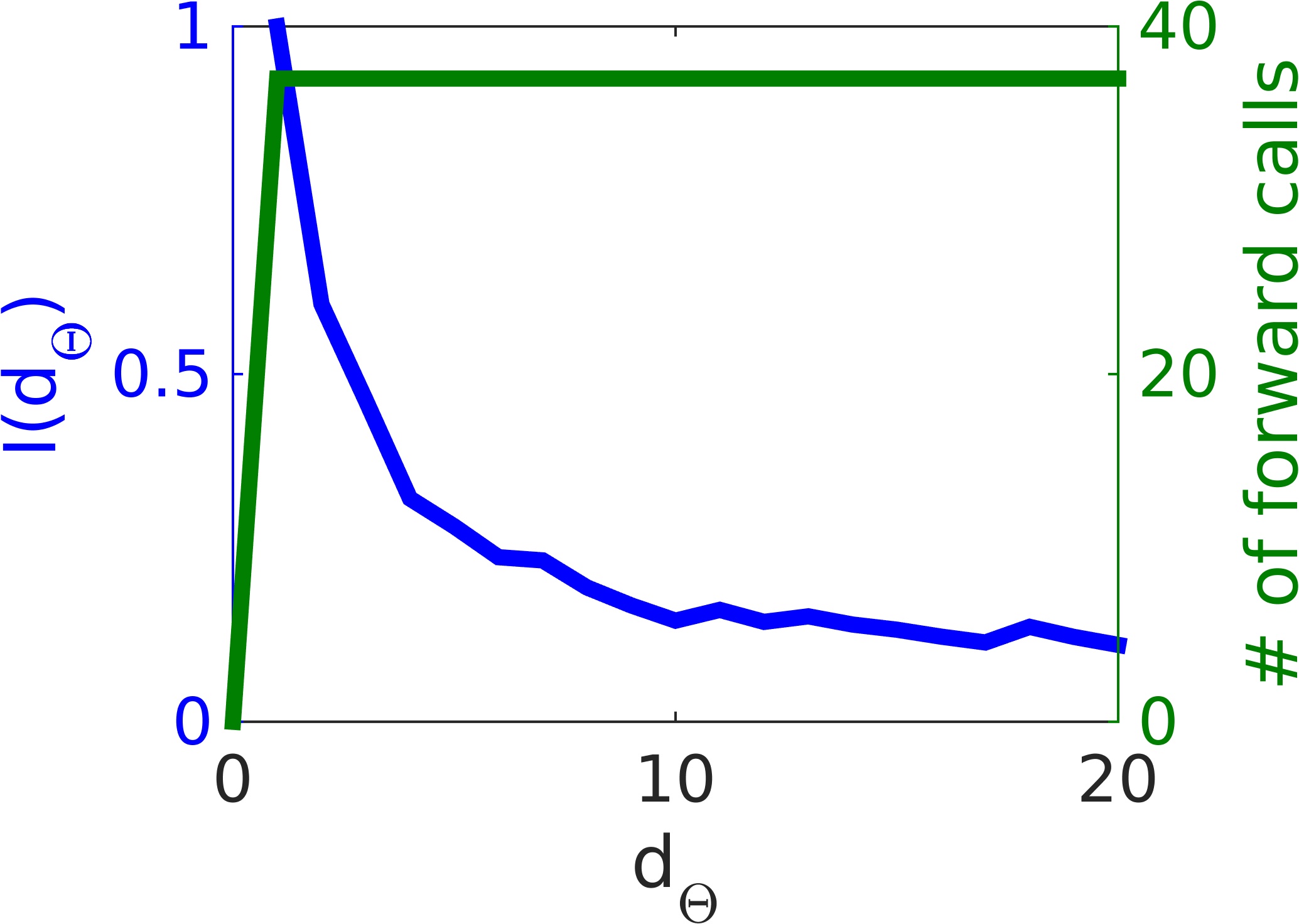}} 
		  \hspace{0.3cm}
	  }
	  \caption{Information gain $I(\dth)$  {\color{blue} ---}  and computational cost {\color[rgb]{0,0.5,0}{---}} as measured by the number of forward calls.}
	 \label{fig:IG}
\end{figure}

 \reffig{fig:IG} depicts the relative information gain $I(\dth)$ (Section \ref{sec:card}) for each SNR. The behavior of the information gain depends on the ratio of the prior $\lambda_{0,i}$ and the posterior of the variance $\lambda_i$.  As with the previous example, it exhibits a relative quick decay after a small number of reduced coordinates have been added. % It shows that the first few bases, here the first $4$ basis vectors, are important to capture the posterior correctly and result in a high information gain when adding these bases. 
 In \reffig{fig:IG} shows also  the number of forward calls as a function of $\dth$. As it was observed previously, the effort is expended in the beginning and in all cases the final result is obtained with less than $40$ forward calls. % Redundant with the first example several forward calls are necessary at the beginning but later, when adding additional bases, no adjustments/improvements of the mean is visible.

\begin{figure}[H]{
	\centering
	\captionsetup[subfigure]{labelformat=empty}
% 		\vspace{-0.5cm}
% 		\subfloat[][{With SNR 692}] 
% 		{\includegraphics[width=0.30\textwidth]{FiguresIF/FMyBasesSNR3.jpg}} 
% 		\hspace{0.1cm}
		\subfloat[][{ SNR $1.93\times10^3$}] 
		{\includegraphics[width=0.30\textwidth]{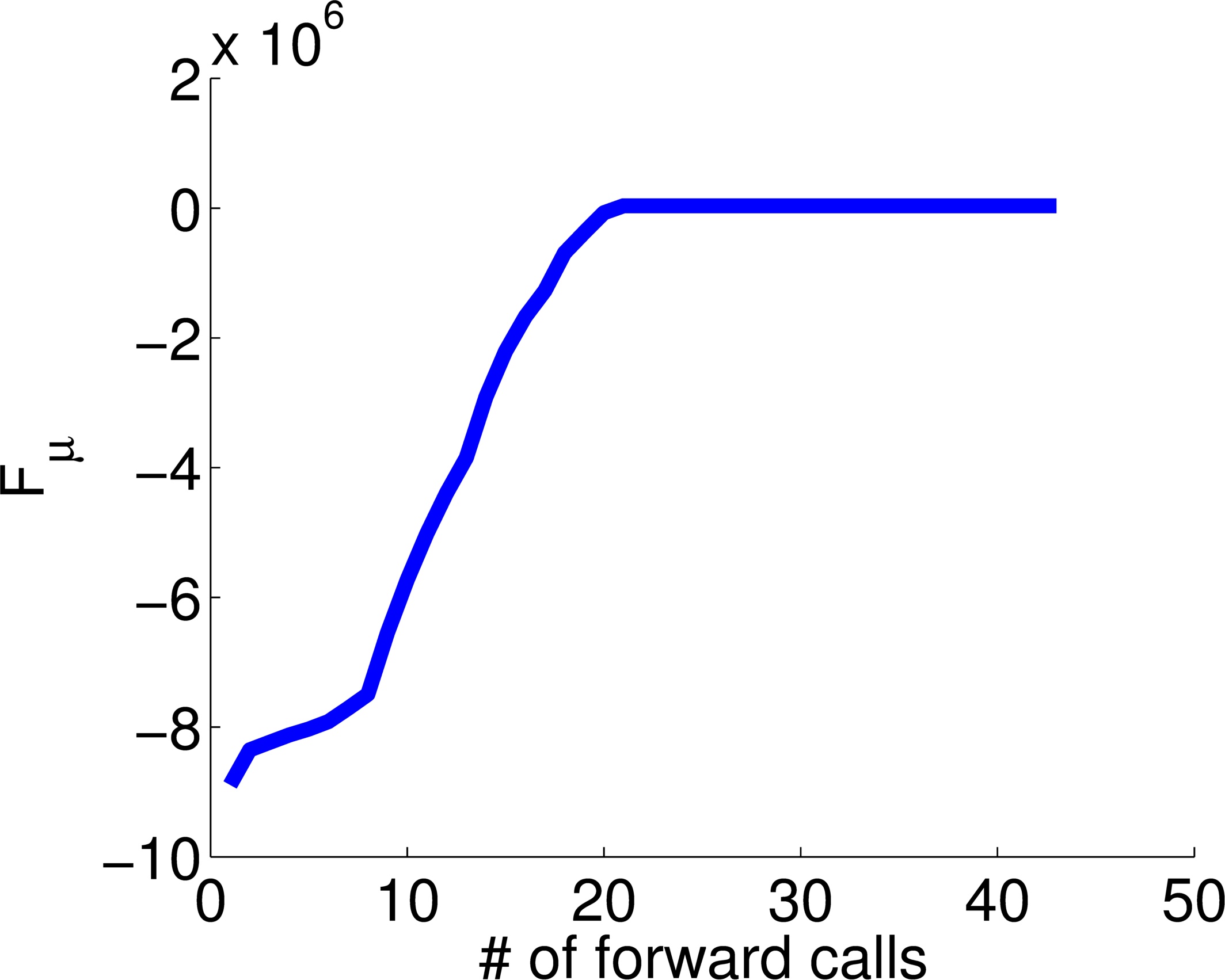}} 
		\hspace{0.1cm}
		\subfloat[][{ SNR $1.89\times10^3$}] 
		{\includegraphics[width=0.30\textwidth]{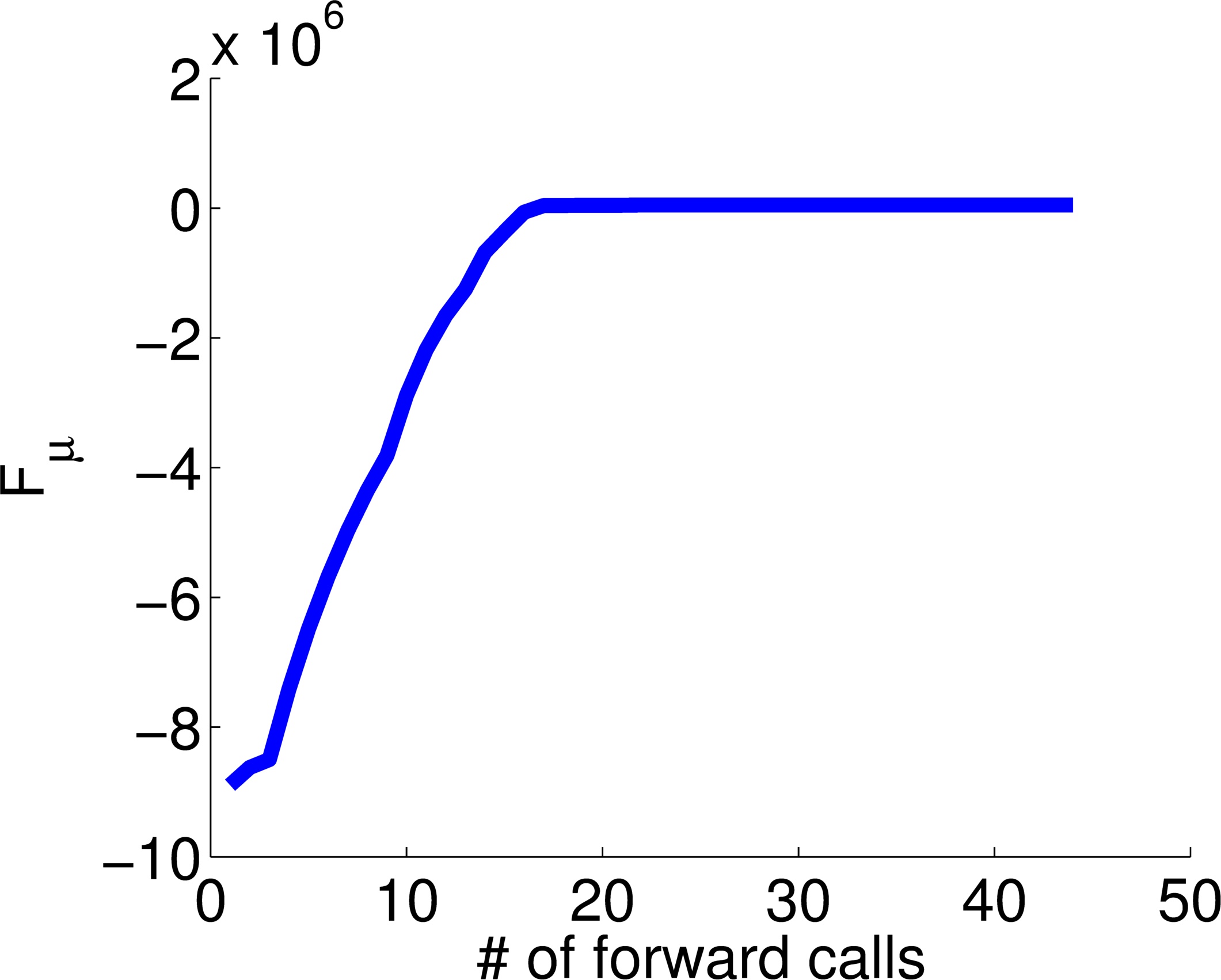}} 
		\hspace{0.1cm}
		\subfloat[][{ SNR $6.9\times10^2$}] 
		{\includegraphics[width=0.30\textwidth]{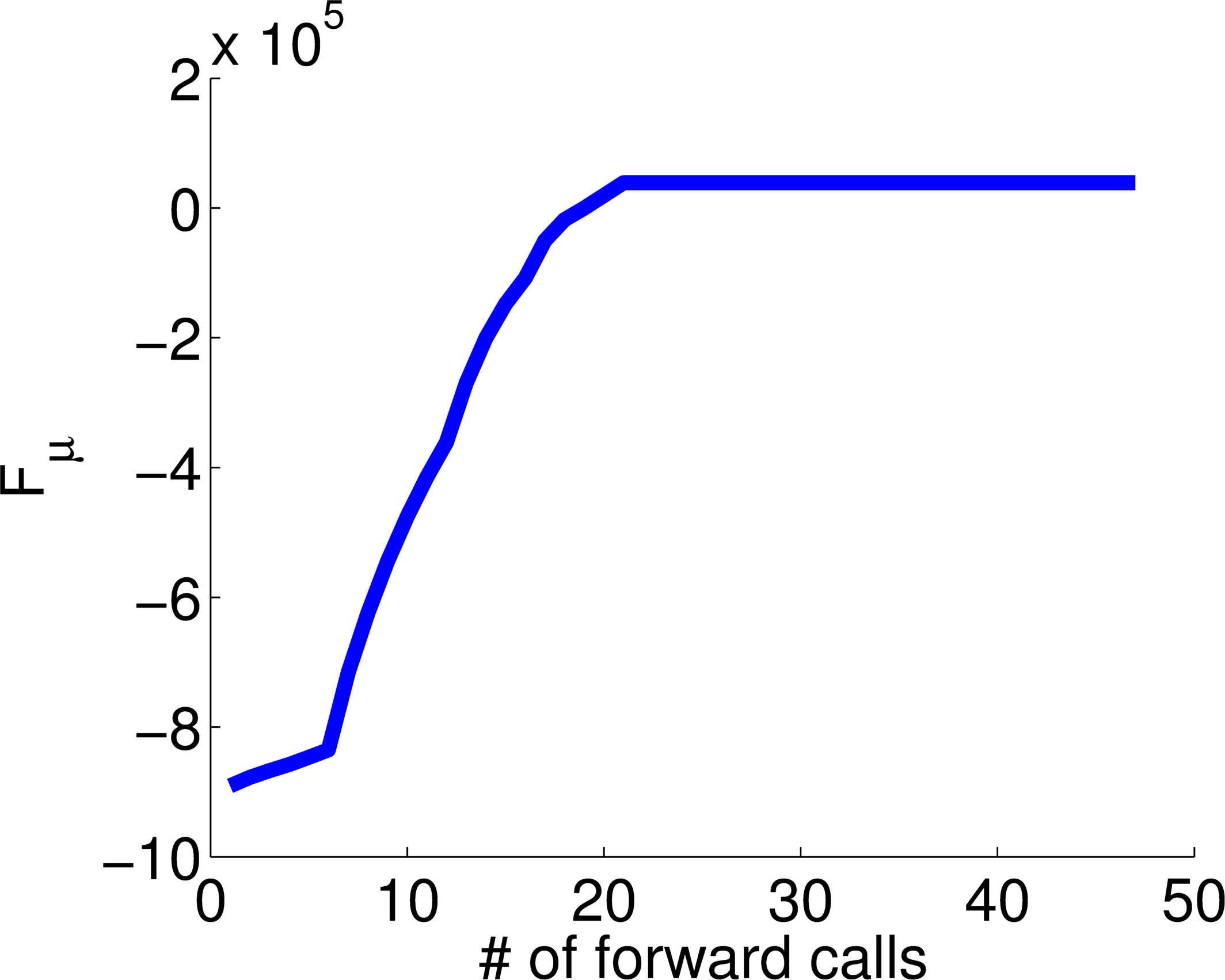}} 
		  \hspace{0.3cm}
	  }
	  \caption{$F_{\mu}$ for the different SNR.}
	 \label{fig:FMyLarge}
\end{figure}

\begin{figure}[H]{
	\centering
	\captionsetup[subfigure]{labelformat=empty}
% 		\vspace{-0.5cm}
% 		\subfloat[][{With SNR 692}] 
% 		{\includegraphics[width=0.30\textwidth]{FiguresIF/FMyBasesSNR3.jpg}} 
% 		\hspace{0.1cm}
		\subfloat[][{SNR $1.93\times10^3$}] 
		{\includegraphics[width=0.30\textwidth]{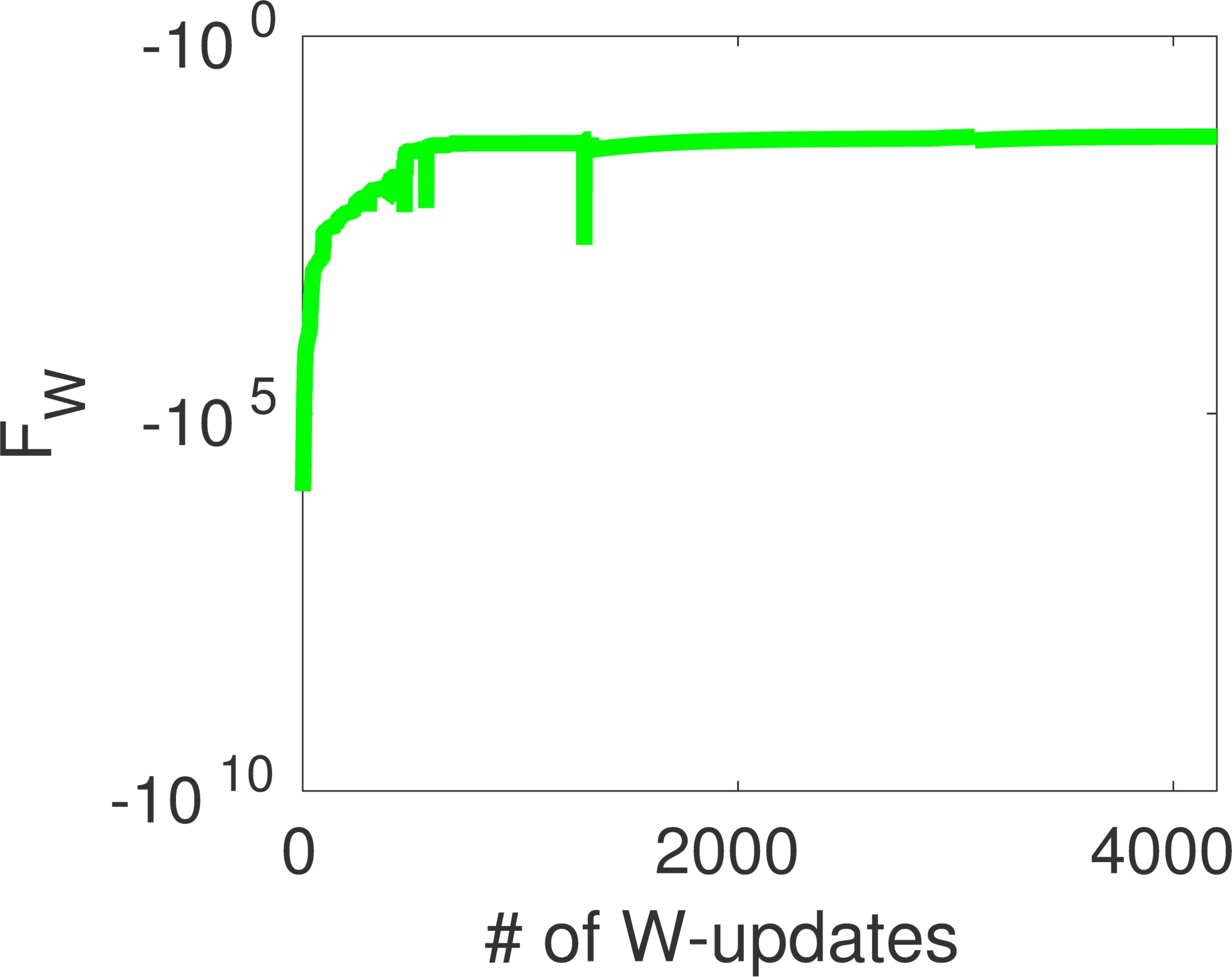}} 
		%{\includegraphics[width=0.30\textwidth]{FiguresIF/FWBases10allModel.jpg}} 
		\hspace{0.1cm}
		\subfloat[][{SNR $1.89\times10^3$}] 
		{\includegraphics[width=0.30\textwidth]{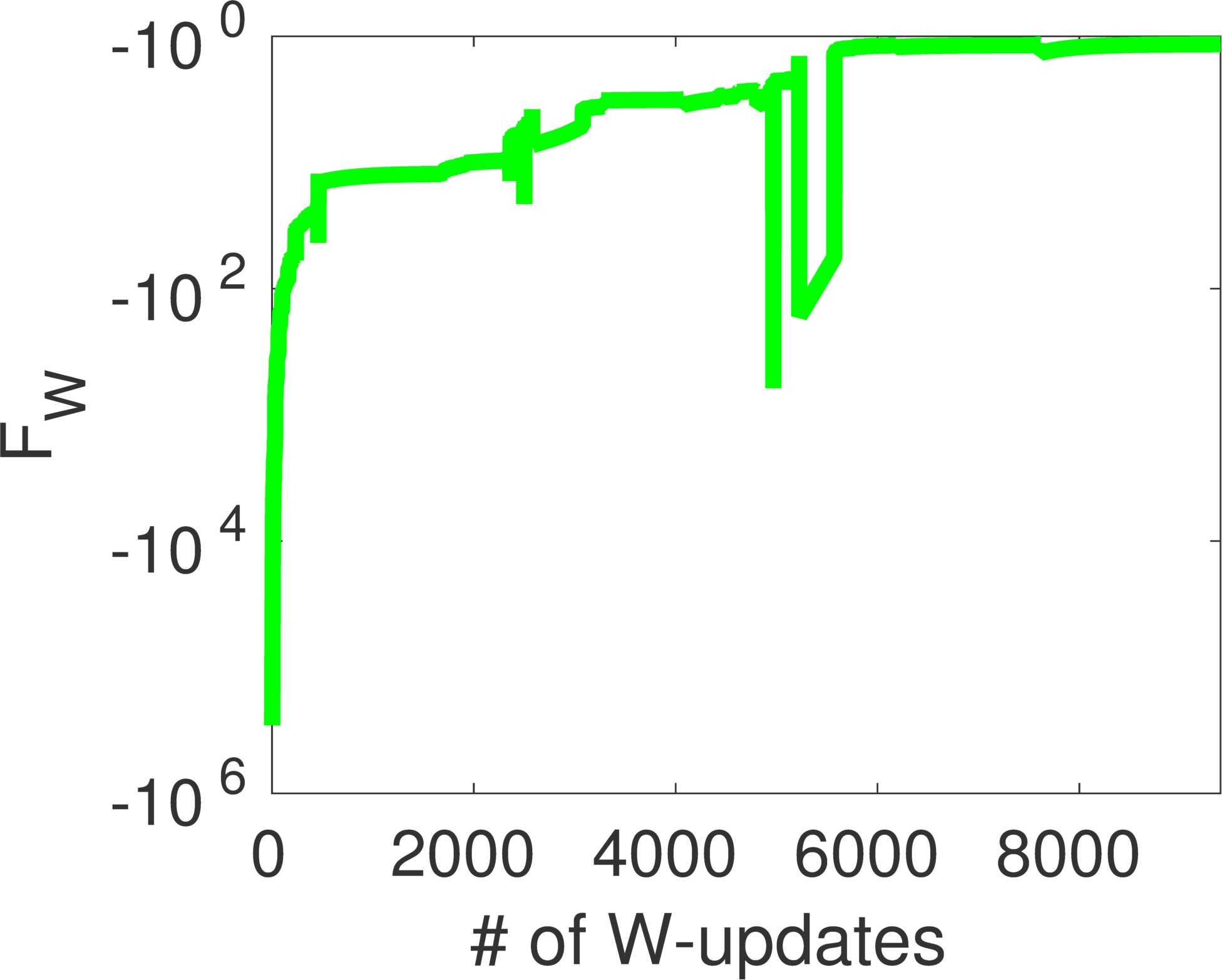}} 
		%{\includegraphics[width=0.30\textwidth]{FiguresIF/FWBases10allSNR5.jpg}}
		\hspace{0.1cm}
		\subfloat[][{SNR $6.9\times10^2$}] 
		{\includegraphics[width=0.30\textwidth]{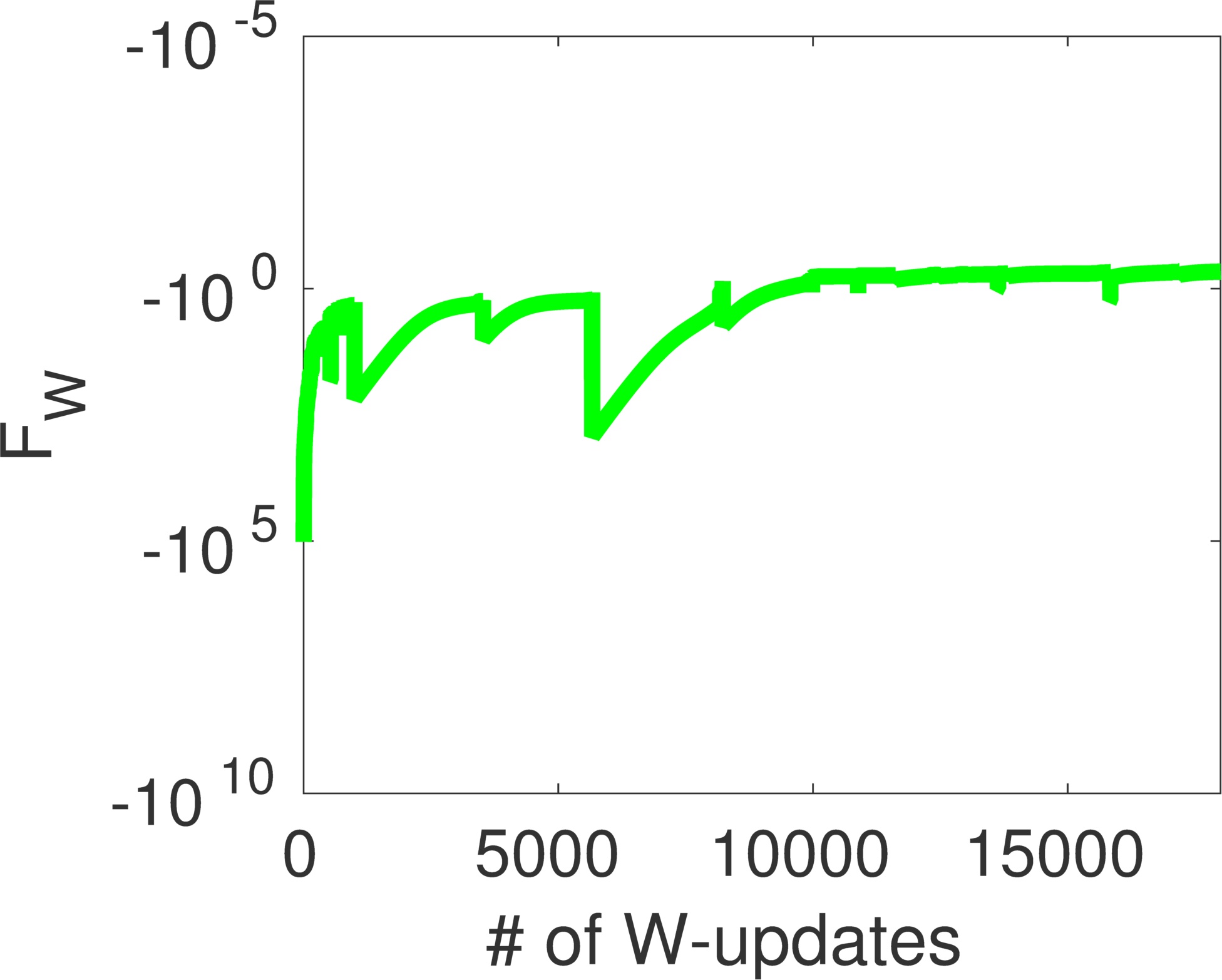}} 
		%{\includegraphics[width=0.30\textwidth]{FiguresIF/FWBases10allSNR4.jpg}} 
		  \hspace{0.3cm}
	  }
	  \caption{$F_W$ for the different SNR and $\dth=2$.}
	 \label{fig:FWLarge}
\end{figure}

\begin{figure}[H]{
	\vspace{-1.5cm}
	\caption*{\hspace{-0.8cm}\textbf{SNR} $\bs{1.93\times10^3}$	\hspace{1.8cm}	\textbf{SNR} $\bs{1.89\times10^3}$	\hspace{1.8cm}	\textbf{SNR} $\bs{6.9\times10^2}$ }
	% \centering
	%\captionsetup[subfigure]{labelformat=empty}
	\captionsetup[subfigure]{position=bottom}
		%\subfloat[][{$\lambda_1^{-1} = 1.0 \times  10^{1}$ }]
		\subfloat[][{$\lambda_1^{-1} = 1.9575 \times  10^{0}$ }]
		{\includegraphics[width=0.30\textwidth]{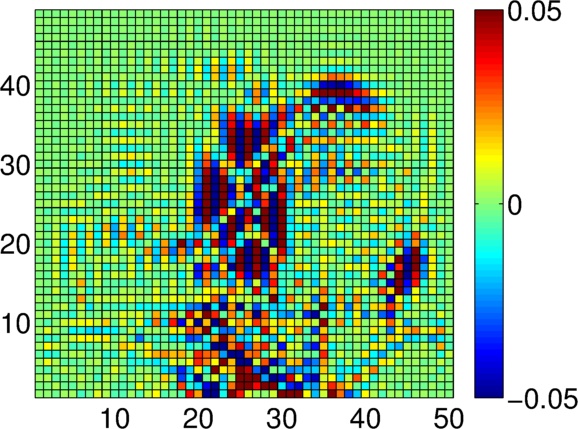} } 
		\hspace{0.1cm}
		%\subfloat[][{$\lambda_1^{-1} = 1.7 \times  10^{1}$ }]
		\subfloat[][{$\lambda_1^{-1} = 1.9825 \times  10^{0}$ }]
		{\includegraphics[width=0.30\textwidth]{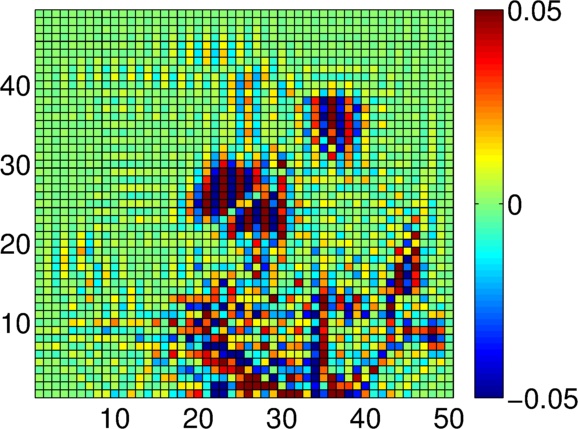} } 
% 		\hspace{0.1cm}
% 		\subfloat[][{$\lambda_1^{-1} = 1.9 \times  10^{1}$ }]
% 		{\includegraphics[width=0.30\textwidth]{FiguresIF/Bestbasis1_SNR4.jpg} } 
		\hspace{0.1cm}
		\subfloat[][{$\lambda_1^{-1} = 1.9996 \times  10^{0}$ }]
		{\includegraphics[width=0.30\textwidth]{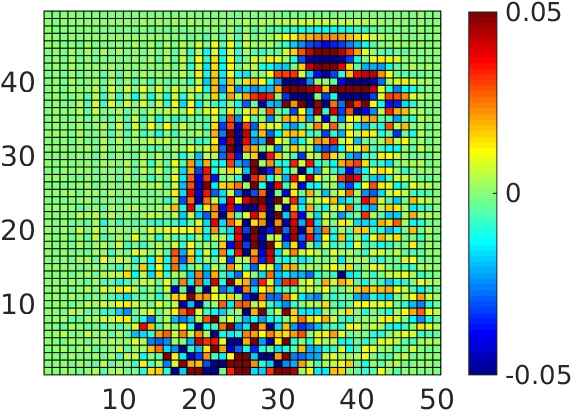} } 
		\\
 		\vspace{-0.2cm}
		%\subfloat[][{$\lambda_2^{-1}= 4.5 \times  10^{0}$ }]
		\subfloat[][{$\lambda_2^{-1}= 1.7933 \times  10^{0}$ }]
		{\includegraphics[width=0.30\textwidth]{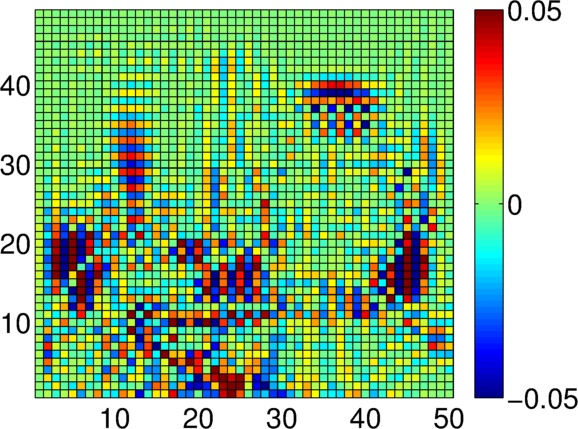} }  		
 		\hspace{0.1cm}
		%\subfloat[][{$\lambda_2^{-1}= 1.5 \times  10^{1}$ }]
		\subfloat[][{$\lambda_2^{-1}= 1.9671 \times  10^{0}$ }]
		{\includegraphics[width=0.30\textwidth]{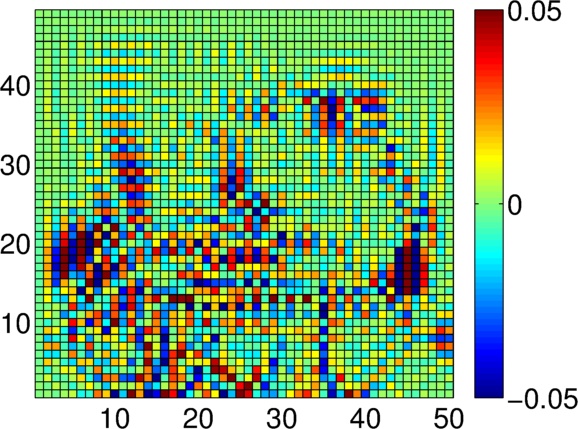} } 
% 		\hspace{0.1cm}		
% 		\subfloat[][{$\lambda_2^{-1} = 1.8 \times  10^{1}$ }]
% 		{\includegraphics[width=0.30\textwidth]{FiguresIF/Bestbasis2_SNR4.jpg} } 
		\hspace{0.1cm}		
		\subfloat[][{$\lambda_2^{-1} = 1.9996 \times  10^{0}$ }]
		{\includegraphics[width=0.30\textwidth]{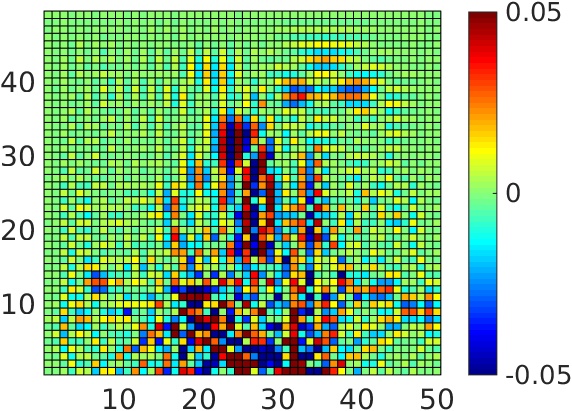} } 
		\\
 		\vspace{-0.2cm}
		%\subfloat[][{$\lambda_3^{-1}  = 1.7 \times  10^{0}$ }]
		\subfloat[][{$\lambda_5^{-1}  = 3.9404 \times  10^{-1}$ }]
		{\includegraphics[width=0.30\textwidth]{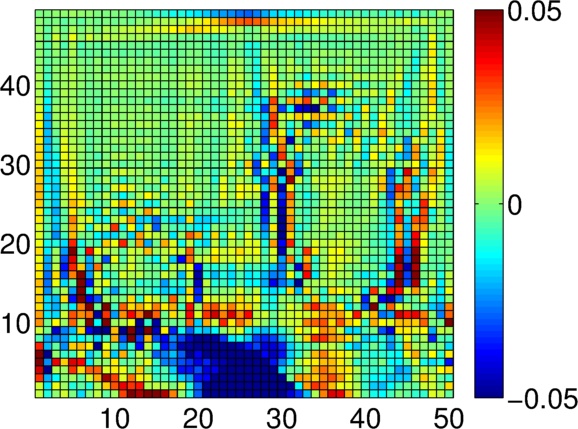} }  		
 		\hspace{0.1cm}
 		%\subfloat[][{$\lambda_3^{-1}  = 1.3 \times  10^{1}$ }]
		\subfloat[][{$\lambda_5^{-1}  = 1.8347 \times  10^{0}$ }]
		{\includegraphics[width=0.30\textwidth]{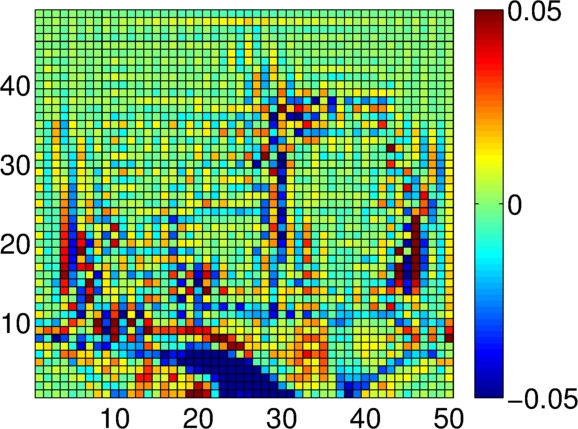} } 
% 		\hspace{0.1cm}
% 		\subfloat[][{$\lambda_3^{-1} = 1.8 \times  10^{1}$ }]
% 		{\includegraphics[width=0.30\textwidth]{FiguresIF/Bestbasis3_SNR4.jpg} } 
		\hspace{0.1cm}
		\subfloat[][{$\lambda_5^{-1} = 1.9994 \times  10^{0}$ }]
		{\includegraphics[width=0.30\textwidth]{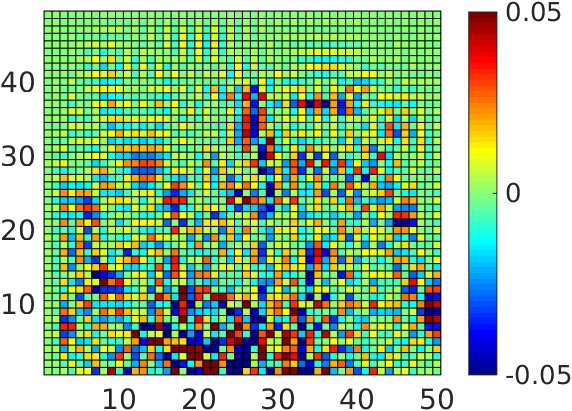} } 
		\\
 		\vspace{-0.2cm}
		%\subfloat[][{$\lambda_4^{-1}= 3.2 \times  10^{-1}$ }]
		\subfloat[][{$\lambda_9^{-1}= 1.3892 \times  10^{-2}$ }]
		{\includegraphics[width=0.30\textwidth]{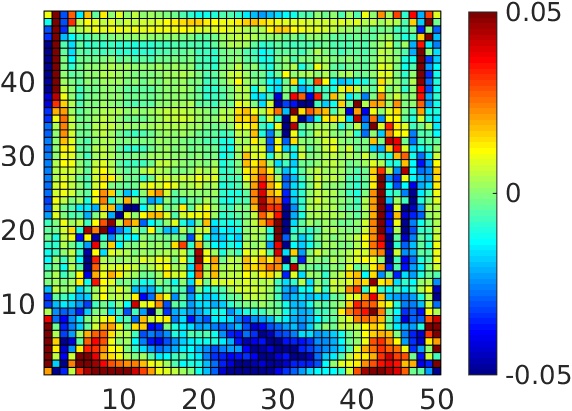} }  		
		\hspace{0.1cm}
		%\subfloat[][{$\lambda_4^{-1}= 5.8 \times  10^{0}$ }]
		\subfloat[][{$\lambda_9^{-1}= 1.0948 \times  10^{0}$ }]
		{\includegraphics[width=0.30\textwidth]{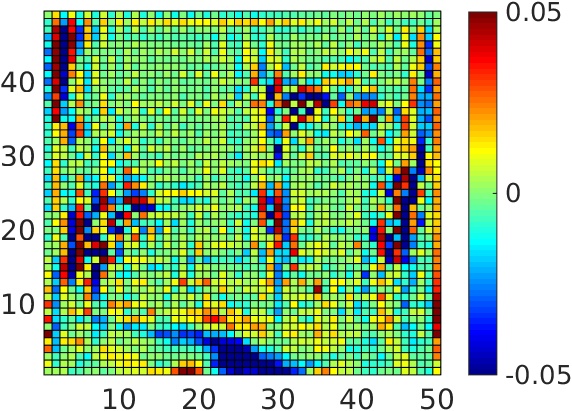} } 
% 		\hspace{0.1cm} 		
% 		\subfloat[][{$\lambda_4^{-1} = 1.4 \times  10^{1}$ }]
% 		{\includegraphics[width=0.30\textwidth]{FiguresIF/Bestbasis4_SNR4.jpg} } 
		\hspace{0.1cm} 		
		\subfloat[][{$\lambda_{9}^{-1} = 1.9993 \times  10^{0}$ }]
		{\includegraphics[width=0.30\textwidth]{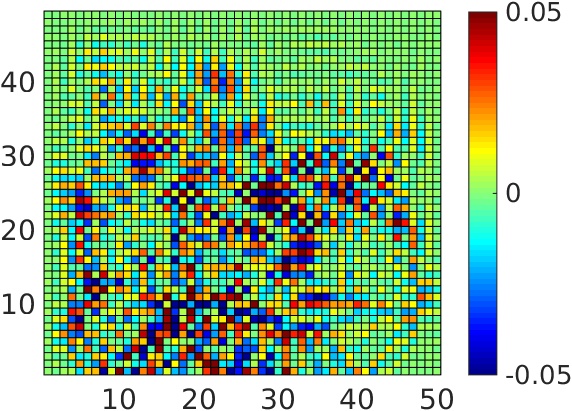} } 
		\\
 		\vspace{-0.2cm}
		%\subfloat[][{$\lambda_5^{-1}  = 1.1 \times  10^{-1}$ }]
		\subfloat[][{$\lambda_{10}^{-1}  = 4.3619 \times  10^{-3}$ }]
		{\includegraphics[width=0.30\textwidth]{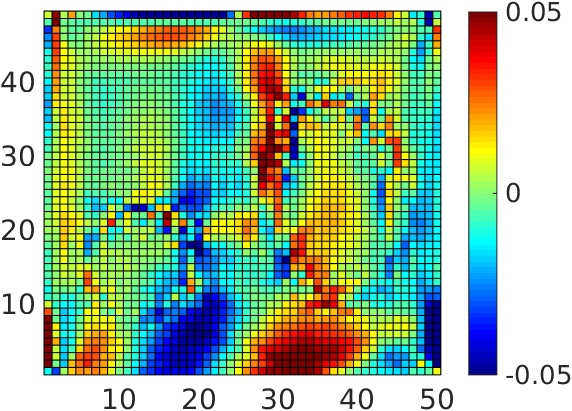} }  		
		\hspace{0.1cm}
		%\subfloat[][{$\lambda_5^{-1}  = 1.7 \times  10^{0}$ }]
		\subfloat[][{$\lambda_{12}^{-1}  = 1.706 \times  10^{-1}$ }]
		{\includegraphics[width=0.30\textwidth]{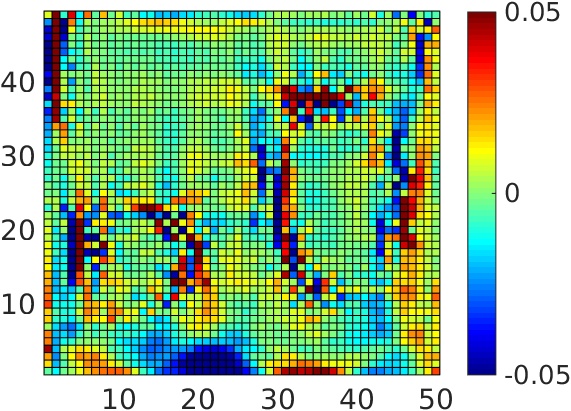} } 
% 		\hspace{0.1cm} 		
% 		\subfloat[][{$\lambda_5^{-1} = 8.8 \times  10^{0}$ }]
% 		{\includegraphics[width=0.30\textwidth]{FiguresIF/Bestbasis5_SNR4.jpg} } 
		\hspace{0.1cm} 		
		\subfloat[][{$\lambda_{13}^{-1} = 1.9989 \times  10^{0}$ }]
		{\includegraphics[width=0.30\textwidth]{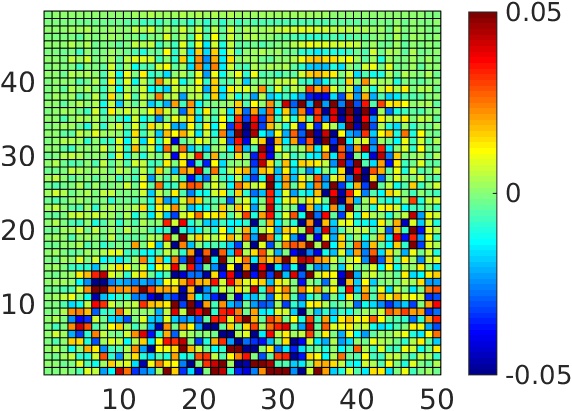} } 
	  }
	  \caption{Some important selected basis vectors for for Cases A (large SNR), B (medium SNR) and C (small SNR) are shown. The basis vectors are ordered based on  decreasing variance $\lambda_i^{-1}$.}\label{fig:BASESLarge}
\end{figure}

  \reffig{fig:FMyLarge} shows the evolution of $F_{\mu}$ as a function of the number of forward calls. %the improvement of the lowerbound during the $\bs{\mu}$-updates within the $\bs{\mu}$-update-algorithms are listed. The improvement occurs when calling the first time the $\bs{\mu}$-update algorithm, which ends if it has been converged. For all three SNR settings around 30 forward calls are necessary to capture $\bs{\mu}$ and to converge. The number of needed iterations changes slightly depending how many iterations one does without regularization. 
\reffig{fig:FWLarge} depicts the corresponding evolution of $F_W$ for $\dth=2$ and for all three SNR cases. %This is shown exemplary for a randomly selected number of basis vectors, here $d_{\bs{\Theta}}10$.
Finally, \reffig{fig:BASESLarge} depicts $5$ basis vectors $\bs{w}_i$ for each SNR in decreasing order based on the corresponding variance $\lambda_{i}^{-1}$. While similarities are observed the basis vectors are not identical as compared across the three different noise levels reflecting the fact that each dataset is informative along different directions in the $\bpsi$ space.  It is clear however that regions in the vicinity of or on  the inclusions exhibit larger posterior variability. As expected the associated variances are larger as the SNR is smaller (i.e. the noise level is higher).

\begin{figure}[H]{
	\centering
	%\captionsetup[subfigure]{labelformat=empty}
		\hspace{2.2cm}
		\subfloat[][{Mean, $d_{\bs{\Theta}}=13$}] 
		{\includegraphics[width=0.32\textwidth]{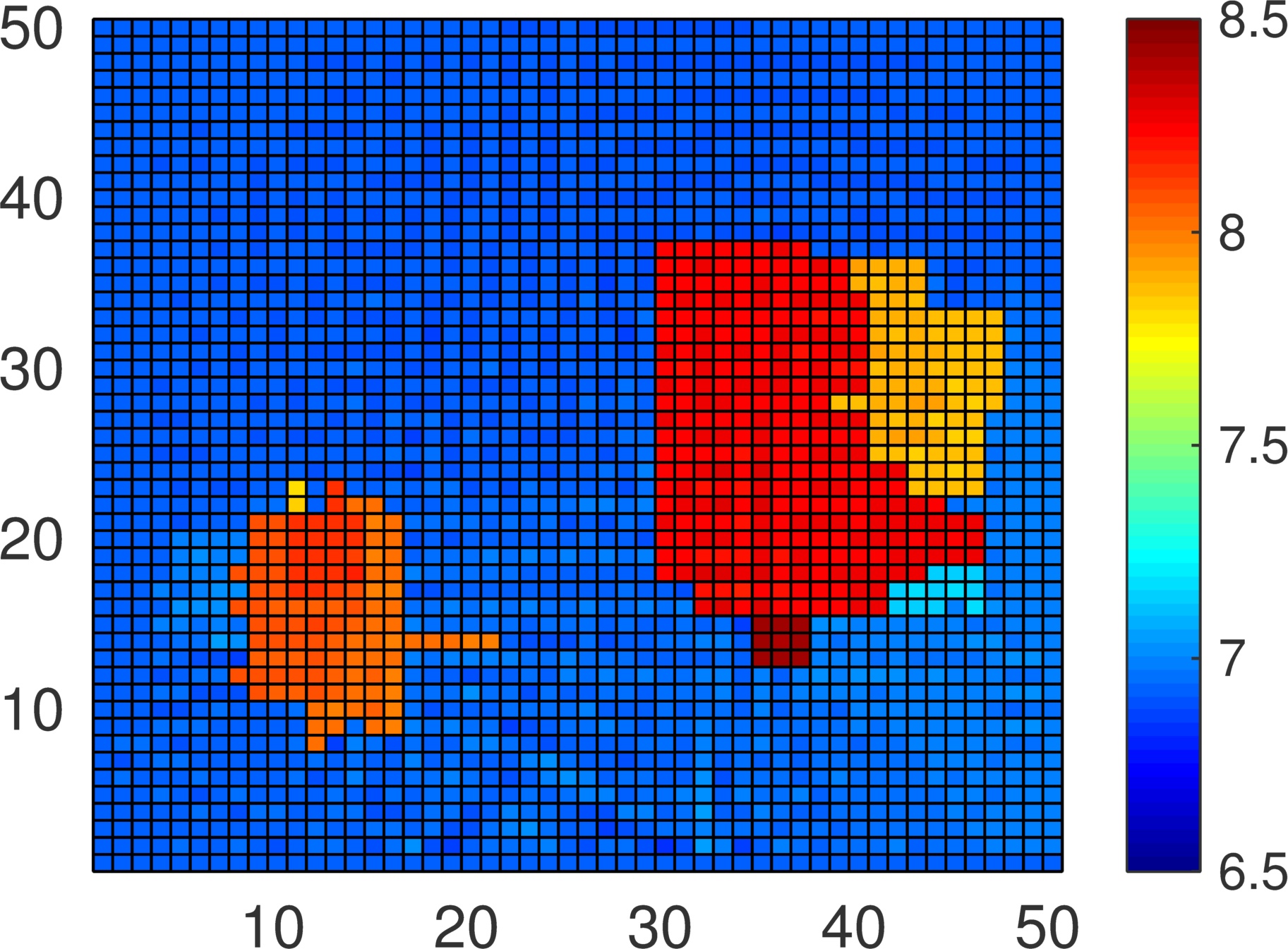}} 
 		\hspace{0.1cm}
% 		\subfloat[][{ Diagonal cut, $d_{\bs{\Theta}}=9$}]
% 		{\includegraphics[width=0.30\textwidth]{FiguresIF/CutThetaISSamplingOnlyLarge.jpg} } 
		\hspace{0.1cm}
		\subfloat[][{ St. dev., $d_{\bs{\Theta}}=13$}]
		  {\includegraphics[width=0.32\textwidth]{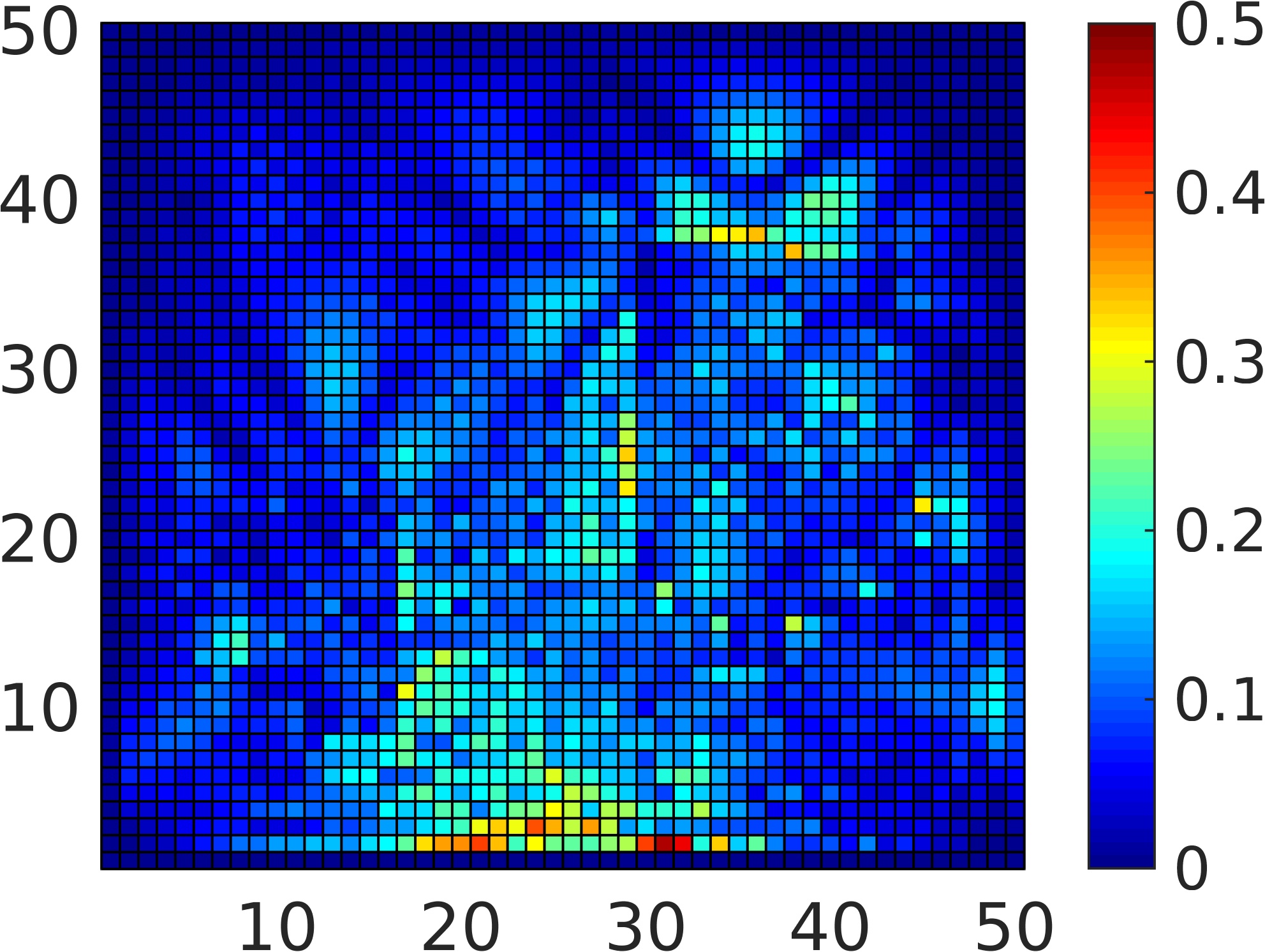} } 	 
	  }
	  \caption{First and second order statistics of the exact posterior ($SNR=6.9\times10^2$) as estimated with Importance Sampling. Figure (a) depicts the posterior mean of $c_1$ in log-scale. Figure (b) depicts the posterior standard deviation.
	  These  should be compared with the VB approximations in \reffig{fig:PosteriorMean} and \reffig{fig:StdLarge}.}
	 \label{fig:PosteriorLargeIS}
\end{figure}

\begin{figure}[H]
	\centering
	\includegraphics[width=0.70\textwidth]{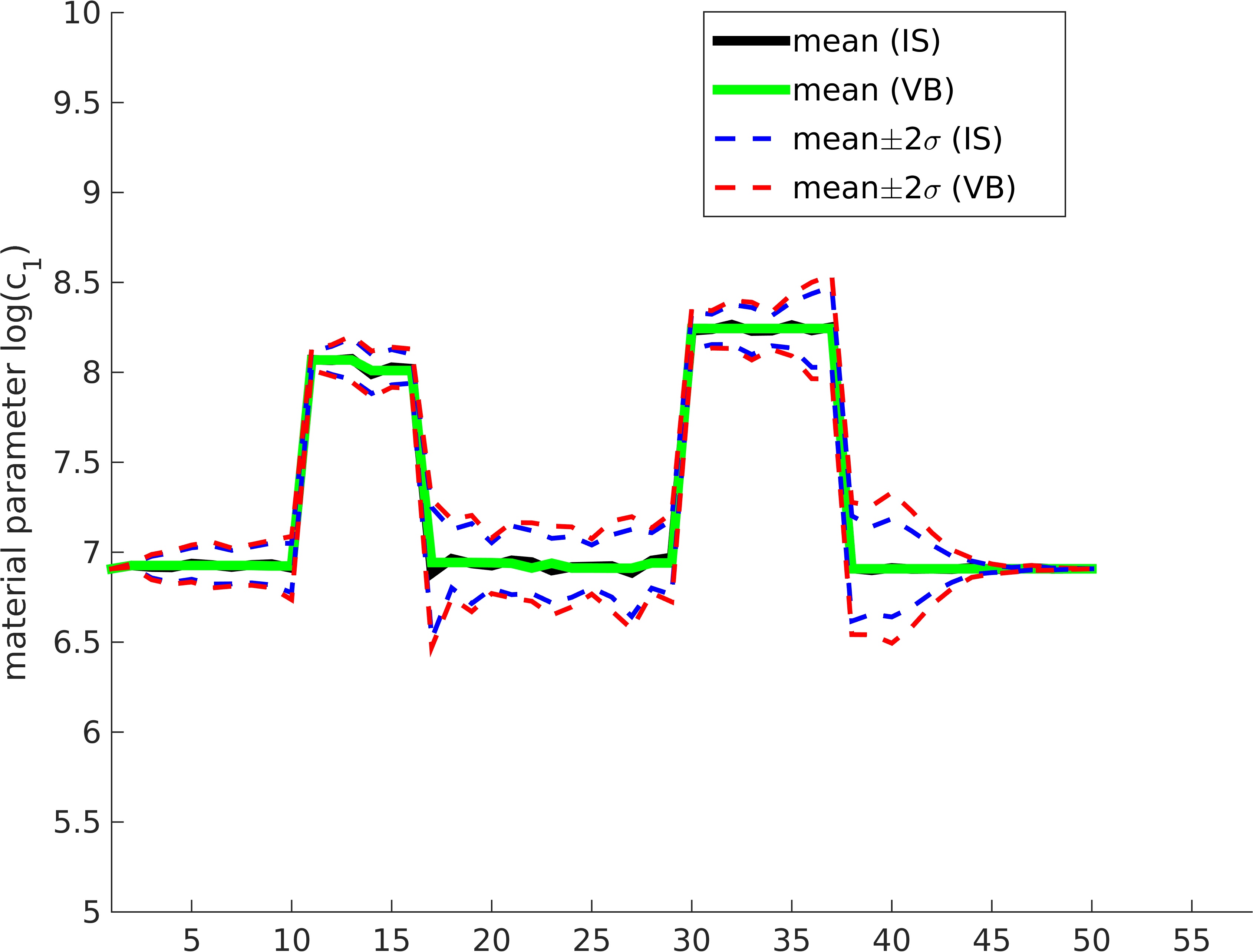}
	  \caption{Posterior mean and posterior quantiles ($\pm2$ standard deviations) along the diagonal from
(0, 0) to (50, 50) for VB and Importance Sampling (IS).}\label{fig:CutThetaVBLargeIS}
\end{figure}

\tcreview{The aforementioned results for the largest noise case ($SNR=6.9\times10^2$) were validated by employing Importance Sampling as discussed in Section \ref{sec:is}. The Effective Sample Size (ESS, \refeq{eq:ess})  was $0.15$ (for $d_{\Theta}=13$) which suggests a good approximation to the actual posterior is provided by the VB result \cite{liu_monte_2001}. More importantly, as it is shown in Figures \ref{fig:PosteriorLargeIS} and \ref{fig:CutThetaVBLargeIS}, the first and second-order statistics of the exact posterior (estimated with Importance Sampling) are very close to the ones computed with the VB approximation.
}

\section{Conclusions}
We introduced  a novel Variational Bayesian framework for the solution of nonlinear inverse problems and demonstrated its capabilities in problems in elastography.
The main advantage of the proposed methodology is the ability to find a much lower-dimensional subspace where a good  approximation to the exact posterior can be obtained. The identification of the reduced basis set is founded on a  fully Bayesian argumentation that employs Variational approximations to the exact posterior.
 Information-theoretic criteria have been proposed in order to adaptively identify  the the cardinality of the reduced coordinates. The posterior approximations are obtained with a limited number of calls to the forward solver for the computation of the response and its derivatives (in all problems considered fewer than $40$ such calls were needed). \tcreview{Furthermore,  with the use of Importance Sampling, the (minute) bias in the posterior estimates can be readily corrected and consistent statistics of the {\em exact posterior} can be readily estimated. }

A possibility that could further reduce the computational effort is the use of forward solvers operating on a hierarchy of resolutions. Starting with the coarsest (and less expensive) model some of the features of the posterior can be obtained with minimal cost and these can be further refined by a smaller number of calls to finer resolution solvers. The resolution of the forward model could also be adaptively altered in regions where the posterior variance appears to be larger. Obtaining efficiently, accurate and fully-Bayesian solutions is a critical step in enabling the use of model-based techniques on a  patient-specific basis for medical diagnosis.

A final extension that is currently under exploration is the use of {\em mixtures} of Gaussian densities in order to provide better approximations to highly non-Gaussian posteriors or even {\em multi-modal} posteriors \cite{choudrey_variational_2003}. Such situations  arise frequently in cases where very sparse and/or very noisy data is  available and represent the most challenging setting for associated inverse problems \cite{koutsourelakis_multi-resolution_2009}. Tools along the aforementioned lines,  offer  appealing possibilities for identifying multiple low-dimensional  subspaces and associated  basis vectors which {\em locally} provide good posterior approximations and when combined, offer an accurate global solution.

 %We introduced in this paper a new algorithm to find a sparse representation with an underdetermined dictionary. With a new dictionary learning algorithm the basis vectors in the selected dictionary are improved. Embedded in a variational Bayesian approach no sampling methods are necessary. We have shown that only a very small number of forward calls are necessary and that with a reduced number of parameters the posterior is captured correctly. The basis vectors which capture most of the variance are selected from the beginning. We also investigated the performance in the presence of different noise levels which could be captured correctly. Therefore, our method can be used successfully to derive unknown material parameters and their uncertainties, for example for medical diagnosis.

 \appendix
 
\section{Expectation-Maximization for the $\bs{\mu}$ prior}
\label{app:mu}
Due to the analytical unavailability of $\log p(\bs{\mu})$ and its derivatives $\frac{ \pa \log p(\bs{\mu})}{\pa \bs{\mu}}$ we employ  an Expectation-Maximization scheme which we describe in here for completeness \cite{dempster_maximum_1977,neal_view_1998}. Proceeding as in \refeq{eq:loglike1} i.e. by making use of Jensen's inequality and an arbitrary distribution $q(\bs{\Phi})$ we can bound $\log p(\bs{\mu})$ as follows:
 \be
 \begin{array}{ll}
   \log p(\bs{\mu}) &= \log \int p(\bs{\mu}|\bs{\Phi}) p(\bs{\Phi})~d\bs{\Phi} \\
   & \log \int \frac{  p(\bs{\mu}|\bs{\Phi}) p(\bs{\Phi})}{q(\bs{\Phi})}  q(\bs{\Phi})~d\bs{\Phi} \\
   & \ge \int q(\bs{\Phi}) \log \frac{  p(\bs{\mu}|\bs{\Phi}) p(\bs{\Phi})}{q(\bs{\Phi})}~d\bs{\Phi}  \\
    & = E_{q(\bs{\Phi})}[\log p(\bs{\mu}|\bs{\Phi})]+E_{q(\bs{\Phi})}[ \log \frac{ p(\bs{\Phi})}{q(\bs{\Phi})}  ].
 \end{array}
 \label{eq:emphi}
\ee
The inequality above becomes an equality only when $q(\bs{\Phi}) \equiv p(\bs{\Phi} | \bs{\mu} )$ i.e. it is the actual posterior on $\bs{\Phi}$ given $\bs{\mu}$.
The latter can be readily established from Equations (\ref{eq:priormuj}) and (\ref{eq:priorphi}) based on which $p(\bs{\Phi} | \bs{\mu} )=\prod_{j=1}^{d_L} Gamma(a_{\phi_j},b_{\phi_j})$ where:
\be
a_{\phi_j}=a_{\phi}+\frac{1}{2}, \quad b_{\phi_j}=b_{\phi}+\frac{1}{2}  (\mu_{k_j}-\mu_{l_j})^2.
\label{eq:phiupd}
\ee
This suggests a two-step procedure for computing $\log p(\bs{\mu})$ and  $\frac{ \pa \log p(\bs{\mu})}{\pa \bs{\mu}}$ for each $\bs{\mu}$:
\bi
\item[(E-step)] Find  $p(\bs{\Phi} | \bs{\mu} )=\prod_{j=1}^{d_L} Gamma(a_{\phi_j},b_{\phi_j})$ from \refeq{eq:phiupd}
\item[(M-step)] Find  $\log p(\bs{\mu})$ and  $\frac{ \pa \log p(\bs{\mu})}{\pa \bs{\mu}}$  from \refeq{eq:emphi} for $q(\bs{\Phi}) \equiv p(\bs{\Phi} | \bs{\mu} )$ as follows:
\be
\begin{array}{ll}
 \log p(\bs{\mu}) & = E_{q(\bs{\Phi})}[\log p(\bs{\mu}|\bs{\Phi})]= -\frac{1}{2} \bs{\mu}^T \bs{L}^T <\bs{\Phi}> \bs{L} \bs{\mu}  \\
 \frac{ \pa \log p(\bs{\mu})}{\pa \bs{\mu}} & = \frac{ \pa }{\pa \bs{\mu}}E_{q(\bs{\Phi})}[\log p(\bs{\mu}|\bs{\Phi})] \\
 & = E_{q(\bs{\Phi})}[\frac{\pa }{\pa \bs{\mu}} \log p(\bs{\mu}|\bs{\Phi})] \\
 & = - \bs{L}^T <\bs{\Phi}> \bs{L} \bs{\mu} 
 \end{array}
 \label{eq:pmu}
\ee
where $<\bs{\Phi}>=E_{q(\bs{\Phi})}[diag(\phi_j)]=diag( \frac{a_{\phi_j}}{b_{\phi_j}})$.
\ei

 % \section{Costs}\label{AppendixCosts}
%\input{appendixCosts}

\newpage
  \bibliographystyle{elsarticle-num} 
  \bibliography{literature_psk}

\begin{thebibliography}{10}
\expandafter\ifx\csname url\endcsname\relax
  \def\url#1{\texttt{#1}}\fi
\expandafter\ifx\csname urlprefix\endcsname\relax\def\urlprefix{URL }\fi
\expandafter\ifx\csname href\endcsname\relax
  \def\href#1#2{#2} \def\path#1{#1}\fi

\bibitem{johannesson_multi-resolution_2005}
G.~Johannesson, R.~E. Glaser, C.~L. Lee, J.~J. Nitao, W.~G. Hanley,
  Multi-resolution markovchain-monte-carlo approach for system identification
  with an application to finite-element models (2005).

\bibitem{torquato_random_2002}
S.~Torquato, Random heterogeneous materials: Microstructure and macroscopic
  properties (2002).

\bibitem{wang_markov_2006}
J.~Wang, N.~Zabaras, A markov random field model of contamination source
  identification in porous media flow, International Journal of Heat and Mass
  Transfer 49~(5) (2006) 939--950.

\bibitem{vogel_computational_2002}
C.~R. Vogel, Computational methods for inverse problems (2002).

\bibitem{kaipio_statistical_2006}
J.~Kaipio, E.~Somersalo, Statistical and computational inverse problems (2006).

\bibitem{glaser_stochastic_2004}
R.~E. Glaser, G.~Johannesson, S.~Sengupta, B.~Kosovic, S.~Carle, G.~A. Franz,
  R.~D. Aines, J.~J. Nitao, W.~G. Hanley, A.~L. Ramirez, {others}, Stochastic
  engine final report: applying markov chain monte carlo methods with
  importance sampling to large-scale data-driven simulation (2004).

\bibitem{weir_fully_1997}
I.~S. Weir, Fully bayesian reconstructions from single-photon emission computed
  tomography data, Journal of the American Statistical Association 92~(437)
  (1997) 49--60.

\bibitem{hegstad_uncertainty_2001}
B.~K.~a. Hegstad, O.~Henning, {others}, Uncertainty in production forecasts
  based on well observations, seismic data, and production history, Spe Journal
  6~(4) (2001) 409--424.

\bibitem{wang_hierarchical_2005}
J.~Wang, N.~Zabaras, Hierarchical bayesian models for inverse problems in heat
  conduction, Inverse Problems 21~(1) (2005) 183.

\bibitem{liu_bayesian_2008}
F.~Liu, M.~J. Bayarri, J.~O. Berger, R.~Paulo, J.~Sacks, A bayesian analysis of
  the thermal challenge problem, Computer Methods in Applied Mechanics and
  Engineering 197~(29) (2008) 2457--2466.

\bibitem{lee_markov_2002}
H.~K. Lee, D.~M. Higdon, Z.~Bi, M.~A. Ferreira, M.~West, Markov random field
  models for high-dimensional parameters in simulations of fluid flow in porous
  media, Technometrics 44~(3) (2002) 230--241.

\bibitem{moral_sequential_2006}
P.~D. Moral, A.~Doucet, A.~Jasra, Sequential Monte Carlo for Bayesian
  Computation, 2006.

\bibitem{koutsourelakis_multi-resolution_2009}
P.-S. Koutsourelakis, A multi-resolution, non-parametric, bayesian framework
  for identification of spatially-varying model parameters, Journal of
  computational physics 228~(17) (2009) 6184--6211.

\bibitem{chopin_free_2012}
N.~Chopin, T.~Lelièvre, G.~Stoltz, Free energy methods for bayesian inference:
  efficient exploration of univariate gaussian mixture posteriors, Statistics
  and Computing 22~(4) (2012) 897--916.

\bibitem{hoang_complexity_2013}
V.~H. Hoang, C.~Schwab, A.~M. Stuart, Complexity analysis of accelerated {MCMC}
  methods for bayesian inversion, Inverse Problems 29~(8) (2013-08) 085010.
\newblock \href {http://dx.doi.org/10.1088/0266-5611/29/8/085010}
  {\path{doi:10.1088/0266-5611/29/8/085010}}.

\bibitem{marzouk_stochastic_2007}
Y.~M. Marzouk, H.~N. Najm, L.~A. Rahn, Stochastic spectral methods for
  efficient bayesian solution of inverse problems, Journal of Computational
  Physics 224~(2) (2007-06-10) 560--586.
\newblock \href {http://dx.doi.org/10.1016/j.jcp.2006.10.010}
  {\path{doi:10.1016/j.jcp.2006.10.010}}.

\bibitem{bilionis_solution_2014}
I.~Bilionis, N.~Zabaras, Solution of inverse problems with limited forward
  solver evaluations: a bayesian perspective, Inverse Problems 30~(1) (2014)
  015004.

\bibitem{oberai_linear_2009}
A.~A. Oberai, N.~H. Gokhale, S.~Goenezen, P.~E. Barbone, T.~J. Hall, A.~M.
  Sommer, J.~Jiang, Linear and nonlinear elasticity imaging of soft tissue in
  vivo: demonstration of feasibility, Physics in medicine and biology 54~(5)
  (2009) 1191.

\bibitem{ganne-carrie_accuracy_2006}
N.~Ganne-Carrié, M.~Ziol, V.~de~Ledinghen, C.~Douvin, P.~Marcellin,
  L.~Castera, D.~Dhumeaux, J.-C. Trinchet, M.~Beaugrand, Accuracy of liver
  stiffness measurement for the diagnosis of cirrhosis in patients with chronic
  liver diseases, Hepatology 44~(6) (2006) 1511--1517.

\bibitem{curtis_genomic_2012}
C.~Curtis, S.~P. Shah, S.-F. Chin, G.~Turashvili, O.~M. Rueda, M.~J. Dunning,
  D.~Speed, A.~G. Lynch, S.~Samarajiwa, Y.~Yuan, {others}, The genomic and
  transcriptomic architecture of 2,000 breast tumours reveals novel subgroups,
  Nature 486~(7403) (2012) 346--352.

\bibitem{muthupillai_magnetic_1995}
R.~Muthupillai, D.~J. Lomas, P.~J. Rossman, J.~F. Greenleaf, A.~Manduca, R.~L.
  Ehman, Magnetic resonance elastography by direct visualization of propagating
  acoustic strain waves, Science 269~(5232) (1995) 1854--1857.

\bibitem{sarvazyan_editorial_2011}
A.~Sarvazyan, T.~J.~Hall, Editorial [{Hot} topic: {Elasticity} {Imaging} {Part}
  {I} \& {II} ({Guest} {Editors}: {Armen} {Sarvazyan} and {Timothy} {J}.
  {Hall})], Current Medical Imaging Reviews 7~(4) (2011) 254--254.
\newblock \href {http://dx.doi.org/10.2174/157340511798038620}
  {\path{doi:10.2174/157340511798038620}}.

\bibitem{doyley_model-based_2012}
M.~M. Doyley, Model-based elastography: a survey of approaches to the inverse
  elasticity problem, Physics in medicine and biology 57~(3) (2012) R35.

\bibitem{ophir_elastography:_1991}
J.~Ophir, I.~Cespedes, H.~Ponnekanti, Y.~Yazdi, X.~Li, Elastography: a
  quantitative method for imaging the elasticity of biological tissues,
  Ultrasonic imaging 13~(2) (1991) 111--134.

\bibitem{bamber_progress_2002}
J.~C. Bamber, P.~E. Barbone, D.~O. Cosgrove, M.~M. Doyley, F.~G. Fuechsel,
  P.~M. Meaney, N.~R. Miller, T.~Shiina, F.~Tranquart, {others}, Progress in
  freehand elastography of the breast, {IEICE} {TRANSACTIONS} on Information
  and Systems 85~(1) (2002) 5--14.

\bibitem{thomas_real-time_2006}
A.~Thomas, T.~Fischer, H.~Frey, R.~Ohlinger, S.~Grunwald, J.-U. Blohmer, K.-J.
  Winzer, S.~Weber, G.~Kristiansen, B.~Ebert, {others}, Real-time
  elastography—an advanced method of ultrasound: first results in 108
  patients with breast lesions, Ultrasound in obstetrics \& gynecology 28~(3)
  (2006) 335--340.

\bibitem{parker_imaging_2011}
K.~J. Parker, M.~M. Doyley, D.~J. Rubens, Imaging the elastic properties of
  tissue: the 20 year perspective, Physics in medicine and biology 56~(1)
  (2011) R1.

\bibitem{barbone_adjoint-weighted_2010}
P.~E. Barbone, C.~E. Rivas, I.~Harari, U.~Albocher, A.~A. Oberai, Y.~Zhang,
  Adjoint-weighted variational formulation for the direct solution of inverse
  problems of general linear elasticity with full interior data, International
  journal for numerical methods in engineering 81~(13) (2010) 1713--1736.

\bibitem{oberai_evaluation_2004}
A.~A. Oberai, N.~H. Gokhale, M.~M. Doyley, J.~C. Bamber, Evaluation of the
  adjoint equation based algorithm for elasticity imaging, Physics in Medicine
  and Biology 49~(13) (2004) 2955.

\bibitem{doyley_enhancing_2006}
M.~M. Doyley, S.~Srinivasan, E.~Dimidenko, N.~Soni, J.~Ophir, Enhancing the
  performance of model-based elastography by incorporating additional a priori
  information in the modulus image reconstruction process, Physics in medicine
  and biology 51~(1) (2006) 95.

\bibitem{arnold_efficient_2010}
A.~Arnold, S.~Reichling, O.~T. Bruhns, J.~Mosler, Efficient computation of the
  elastography inverse problem by combining variational mesh adaption and a
  clustering technique, Physics in Medicine and Biology 55~(7) (2010) 2035.

\bibitem{olson_numerical_2010}
L.~G. Olson, R.~D. Throne, Numerical simulation of an inverse method for tumour
  size and location estimation, Inverse Problems in Science and Engineering
  18~(6) (2010) 813--834.

\bibitem{schleder_diagnostic_2013}
S.~Schleder, L.-M. Dendl, A.~Ernstberger, M.~Nerlich, P.~Hoffstetter, E.-M.
  Jung, P.~Heiss, C.~Stroszczynski, A.~G. Schreyer, Diagnostic value of a
  hand-carried ultrasound device for free intra-abdominal fluid and organ
  lacerations in major trauma patients, Emergency Medicine Journal 30~(3)
  (2013) e20--e20.

\bibitem{beal_variational_2003}
M.~J. Beal, Variational algorithms for approximate Bayesian inference,
  University of London, 2003.

\bibitem{bishop_pattern_2006}
C.~M. Bishop, Pattern recognition and machine learning, springer, 2006.

\bibitem{jordan_introduction_1999}
M.~I. Jordan, Z.~Ghahramani, T.~S. Jaakkola, L.~K. Saul, An introduction to
  variational methods for graphical models, Machine learning 37~(2) (1999)
  183--233.

\bibitem{attias_variational_2000}
H.~Attias, A variational bayesian framework for graphical models, Advances in
  neural information processing systems 12~(1) (2000) 209--215.

\bibitem{wainwright_graphical_2008}
M.~J. Wainwright, M.~I. Jordan, Graphical models, exponential families, and
  variational inference, Foundations and Trends® in Machine Learning 1~(1)
  (2008) 1--305.

\bibitem{chappell_variational_2009}
M.~Chappell, A.~R. Groves, B.~Whitcher, M.~W. Woolrich, {others}, Variational
  bayesian inference for a nonlinear forward model, Signal Processing, {IEEE}
  Transactions on 57~(1) (2009) 223--236.

\bibitem{jin_hierarchical_2010}
B.~Jin, J.~Zou, Hierarchical bayesian inference for ill-posed problems via
  variational method, Journal of Computational Physics 229~(19) (2010)
  7317--7343.

\bibitem{cover_elements_1991}
T.~M. Cover, J.~A. Thomas, Elements of information theory, 1991.

\bibitem{el_moselhy_bayesian_2012}
T.~A. El~Moselhy, Y.~M. Marzouk, Bayesian inference with optimal maps, Journal
  of Computational Physics 231~(23) (2012-10-01) 7815--7850.
\newblock \href {http://dx.doi.org/10.1016/j.jcp.2012.07.022}
  {\path{doi:10.1016/j.jcp.2012.07.022}}.

\bibitem{olshausen_sparse_1997}
B.~A. Olshausen, D.~J. Field, Sparse coding with an overcomplete basis set: {A}
  strategy employed by {V}1?, Vision Research 37~(23) (1997) 3311--3325.
\newblock \href {http://dx.doi.org/10.1016/S0042-6989(97)00169-7}
  {\path{doi:10.1016/S0042-6989(97)00169-7}}.

\bibitem{lewicki_learning_2000}
M.~S. Lewicki, T.~J. Sejnowski, Learning overcomplete representations, Neural
  computation 12~(2) (2000) 337--365.

\bibitem{holzapfel_nonlinear_2000}
G.~A. Holzapfel, Nonlinear solid mechanics (2000).

\bibitem{mase_continuum_2009}
G.~T. Mase, R.~E. Smelser, G.~E. Mase, Continuum mechanics for engineers, {CRC}
  press, 2009.

\bibitem{bonet_worked_2012}
J.~Bonet, A.~J. Gil, R.~D. Wood, Worked Examples in Nonlinear Continuum
  Mechanics for Finite Element Analysis, 2012.

\bibitem{gokhale_solution_2008}
N.~H. Gokhale, P.~E. Barbone, A.~A. Oberai, Solution of the nonlinear
  elasticity imaging inverse problem: the compressible case, Inverse Problems
  24~(4) (2008-08) 045010.
\newblock \href {http://dx.doi.org/10.1088/0266-5611/24/4/045010}
  {\path{doi:10.1088/0266-5611/24/4/045010}}.

\bibitem{adams_sobolev_2003}
R.~A. Adams, J.~J.~F. Fournier, Sobolev Spaces, Academic Press.

\bibitem{goenezen_solution_2011}
S.~Goenezen, P.~Barbone, A.~A. Oberai,
  \href{http://www.sciencedirect.com/science/article/pii/S0045782510003749}{Solution
  of the nonlinear elasticity imaging inverse problem: {The} incompressible
  case}, Computer Methods in Applied Mechanics and Engineering 200~(13–16)
  (2011) 1406--1420.
\newblock \href {http://dx.doi.org/10.1016/j.cma.2010.12.018}
  {\path{doi:10.1016/j.cma.2010.12.018}}.
\newline\urlprefix\url{http://www.sciencedirect.com/science/article/pii/S0045782510003749}

\bibitem{zienkiewicz_finite_1977}
O.~C. Zienkiewicz, R.~L. Taylor, The finite element method (1977).

\bibitem{hughes_finite_2000}
T.~J.~R. Hughes, T.~Hughes, The Finite Element Method: Linear Static and
  Dynamic Finite Element Analysis, Dover Pubn Inc, 2000-08-16.

\bibitem{e_principles_2011}
W.~E, Principles of {Multiscale} {Modeling}, 1st Edition, Cambridge University
  Press, Cambridge, UK ; New York, 2011.

\bibitem{schillings_sparse_2013}
C.~Schillings, C.~Schwab, Sparse, adaptive smolyak quadratures for bayesian
  inverse problems, Inverse Problems 29~(6) (2013-06) 065011.
\newblock \href {http://dx.doi.org/10.1088/0266-5611/29/6/065011}
  {\path{doi:10.1088/0266-5611/29/6/065011}}.

\bibitem{giles_introduction_2000}
M.~B. Giles, N.~A. Pierce, An introduction to the adjoint approach to design,
  Flow, turbulence and combustion 65~(3) (2000) 393--415.

\bibitem{hinze_optimization_2009}
M.~Hinze, R.~Pinnau, M.~Ulbrich, S.~Ulbrich, Optimization with {PDE}
  constraints, in: Optimization with Pde Constraints, Vol.~23, 2009, pp.
  1--270.

\bibitem{papadimitriou_direct_2008}
D.~I. Papadimitriou, K.~C. Giannakoglou, Direct, adjoint and mixed approaches
  for the computation of hessian in airfoil design problems, International
  Journal for Numerical Methods in Fluids 56~(10) (2008) 1929--1943.

\bibitem{Orginos:1118470}
K.~Orginos, \href{https://cds.cern.ch/record/1118470}{{A solver for multiple
  right hand sides.}}, PoS LATTICE 2007 (2007) 042.
\newline\urlprefix\url{https://cds.cern.ch/record/1118470}

\bibitem{gutknecht_block_2009}
M.~H. Gutknecht, T.~Schmelzer,
  \href{http://www.sciencedirect.com/science/article/pii/S0024379508003479}{The
  block grade of a block {Krylov} space}, Linear Algebra and its Applications
  430~(1) (2009) 174--185.
\newblock \href {http://dx.doi.org/10.1016/j.laa.2008.07.008}
  {\path{doi:10.1016/j.laa.2008.07.008}}.
\newline\urlprefix\url{http://www.sciencedirect.com/science/article/pii/S0024379508003479}

\bibitem{calvetti_tikhonov_2003}
D.~Calvetti, L.~Reichel, Tikhonov regularization of large linear problems,
  {BIT} Numerical Mathematics 43~(2) (2003) 263--283.

\bibitem{bardsley_gaussian_2013}
J.~M. Bardsley, Gaussian markov random field priors for inverse problems,
  Inverse Probl. Imaging 7~(2) (2013) 397--416.

\bibitem{schwab_sparse_2012}
C.~Schwab, A.~M. Stuart, Sparse deterministic approximation of bayesian inverse
  problems, Inverse Problems 28~(4) (2012-04) 045003.
\newblock \href {http://dx.doi.org/10.1088/0266-5611/28/4/045003}
  {\path{doi:10.1088/0266-5611/28/4/045003}}.

\bibitem{richards_quantitative_2007}
M.~S. Richards, Quantitative three dimensional elasticity imaging (2007).

\bibitem{rivaz_real-time_2011}
H.~Rivaz, E.~M. Boctor, M.~Choti, G.~D. Hager, {others}, Real-time regularized
  ultrasound elastography, Medical Imaging, {IEEE} Transactions on 30~(4)
  (2011) 928--945.

\bibitem{jin_variational_2012}
B.~Jin, A variational bayesian method to inverse problems with impulsive noise,
  Journal of Computational Physics 231~(2) (2012) 423--435.

\bibitem{arridge_approximation_2006}
S.~R. Arridge, J.~P. Kaipio, V.~Kolehmainen, M.~Schweiger, E.~Somersalo,
  T.~Tarvainen, M.~Vauhkonen, Approximation errors and model reduction with an
  application in optical diffusion tomography, Inverse Problems 22~(1) (2006)
  175.

\bibitem{kaipio_statistical_2007}
J.~Kaipio, E.~Somersalo, Statistical inverse problems: Discretization, model
  reduction and inverse crimes, Journal of Computational and Applied
  Mathematics 198~(2) (2007) 493--504.

\bibitem{koutsourelakis_novel_2012}
P.-S. Koutsourelakis, A novel bayesian strategy for the identification of
  spatially varying material properties and model validation: an application to
  static elastography, International Journal for Numerical Methods in
  Engineering 91~(3) (2012) 249--268.

\bibitem{gelman_bayesian_2003}
A.~Gelman, J.~Carlin, H.~Stern, D.~Rubin, Bayesian {Data} {Analysis}, 2nd
  Edition, Chapman \& Hall/CRC, 2003.

\bibitem{honarvar_sparsity_2012}
M.~Honarvar, R.~S. Sahebjavaher, S.~E. Salcudean, R.~Rohling, Sparsity
  regularization in dynamic elastography, Physics in medicine and biology
  57~(19) (2012) 5909.

\bibitem{mackay_choice_1998}
D.~J.~C. {MacKay}, Choice of basis for laplace approximation, Machine Learning
  33~(1) (1998-10) 77--86.
\newblock \href {http://dx.doi.org/10.1023/A:1007558615313}
  {\path{doi:10.1023/A:1007558615313}}.

\bibitem{bui-thanh_extreme-scale_2012}
T.~Bui-Thanh, C.~Burstedde, O.~Ghattas, J.~Martin, G.~Stadler, L.~C. Wilcox,
  Extreme-scale {UQ} for bayesian inverse problems governed by {PDEs}, in:
  Proceedings of the International Conference on High Performance Computing,
  Networking, Storage and Analysis, {IEEE} Computer Society Press, 2012, p.~3.

\bibitem{tiangang_cui_likelihood-informed_2014}
J.~M. Tiangang~Cui, Likelihood-informed dimension reduction for nonlinear
  inverse problems, Inverse Problems 30~(11).
\newblock \href {http://dx.doi.org/10.1088/0266-5611/30/11/114015}
  {\path{doi:10.1088/0266-5611/30/11/114015}}.

\bibitem{tipping_probabilistic_1999}
M.~E. Tipping, C.~M. Bishop, Probabilistic principal component analysis,
  Journal of the Royal Statistical Society: Series B (Statistical Methodology)
  61~(3) (1999) 611--622.

\bibitem{dobigeon_bayesian_2010}
N.~Dobigeon, J.-Y. Tourneret, Bayesian orthogonal component analysis for sparse
  representation, Signal Processing, {IEEE} Transactions on 58~(5) (2010)
  2675--2685.

\bibitem{candes_robust_2006}
E.~J. Candes, J.~Romberg, T.~Tao, Robust uncertainty principles: Exact signal
  reconstruction from highly incomplete frequency information, Ieee
  Transactions on Information Theory 52~(2) (2006-02) 489--509.
\newblock \href {http://dx.doi.org/10.1109/TIT.2005.862083}
  {\path{doi:10.1109/TIT.2005.862083}}.

\bibitem{wipf_sparse_2004}
D.~P. Wipf, B.~D. Rao, Sparse bayesian learning for basis selection, Ieee
  Transactions on Signal Processing 52~(8) (2004-08) 2153--2164.
\newblock \href {http://dx.doi.org/10.1109/TSP.2004.831016}
  {\path{doi:10.1109/TSP.2004.831016}}.

\bibitem{lee_efficient_2006}
H.~Lee, A.~Battle, R.~Raina, A.~Y. Ng, Efficient sparse coding algorithms, in:
  Advances in neural information processing systems, 2006, pp. 801--808.

\bibitem{seeger_variational_2010}
M.~W. Seeger, D.~P. Wipf, Variational bayesian inference techniques, Ieee
  Signal Processing Magazine 27~(6) (2010-11) 81--91.
\newblock \href {http://dx.doi.org/10.1109/MSP.2010.938082}
  {\path{doi:10.1109/MSP.2010.938082}}.

\bibitem{bui-thanh_adaptive_2012}
T.~Bui-Thanh, O.~Ghattas, D.~Higdon, Adaptive hessian-based nonstationary
  gaussian process response surface method for probability density
  approximation with application to bayesian solution of large-scale inverse
  problems, Siam Journal on Scientific Computing 34~(6) (2012) A2837--A2871.
\newblock \href {http://dx.doi.org/10.1137/110851419}
  {\path{doi:10.1137/110851419}}.

\bibitem{peierls_minimum_1938}
R.~Peierls, On a {Minimum} {Property} of the {Free} {Energy}, Physical Review
  54~(11) (1938) 918--919.
\newblock \href {http://dx.doi.org/10.1103/PhysRev.54.918}
  {\path{doi:10.1103/PhysRev.54.918}}.

\bibitem{muirhead_aspects_1982}
R.~Muirhead, Aspects of Multivariate Statistical Theory, 1982.

\bibitem{bardsley_hierarchical_2010}
J.~M. Bardsley, D.~Calvetti, E.~Somersalo, Hierarchical regularization for
  edge-preserving reconstruction of {PET} images, Inverse Problems 26~(3)
  (2010-03) 035010.
\newblock \href {http://dx.doi.org/10.1088/0266-5611/26/3/035010}
  {\path{doi:10.1088/0266-5611/26/3/035010}}.

\bibitem{choudrey_variational_2003}
R.~A. Choudrey, S.~J. Roberts, Variational mixture of bayesian independent
  component analyzers, Neural Computation 15~(1) (2003-01) 213--252.
\newblock \href {http://dx.doi.org/10.1162/089976603321043766}
  {\path{doi:10.1162/089976603321043766}}.

\bibitem{wen_feasible_2013}
Z.~Wen, W.~Yin, A feasible method for optimization with orthogonality
  constraints, Mathematical Programming 142~(1) (2013-12) 397--434.
\newblock \href {http://dx.doi.org/10.1007/s10107-012-0584-1}
  {\path{doi:10.1007/s10107-012-0584-1}}.

\bibitem{barzilai_2-point_1988}
J.~Barzilai, J.~Borwein, 2-point step size gradient methods, Ima Journal of
  Numerical Analysis 8~(1) (1988-01) 141--148.
\newblock \href {http://dx.doi.org/10.1093/imanum/8.1.141}
  {\path{doi:10.1093/imanum/8.1.141}}.

\bibitem{dempster_maximum_1977}
A.~Dempster, N.~Laird, D.~Rubin, Maximum likelihood from incomplete data via em
  algorithm, Journal of the Royal Statistical Society Series B-Methodological
  39~(1) (1977) 1--38.

\bibitem{neal_view_1998}
R.~M. Neal, G.~E. Hinton, A view of the {EM} algorithm that justifies
  incremental, sparse, and other variants, in: M.~I. Jordan (Ed.), Learning in
  Graphical Models, Vol.~89, 1998, pp. 355--368.

\bibitem{itti_bayesian_2009}
L.~Itti, P.~Baldi, Bayesian surprise attracts human attention, Vision Research
  49~(10) (2009-06-02) 1295--1306.
\newblock \href {http://dx.doi.org/10.1016/j.visres.2008.09.007}
  {\path{doi:10.1016/j.visres.2008.09.007}}.

\bibitem{kong_sequential_1994}
A.~Kong, J.~S. Liu, W.~H. Wong,
  \href{http://www.tandfonline.com/doi/abs/10.1080/01621459.1994.10476469}{Sequential
  {Imputations} and {Bayesian} {Missing} {Data} {Problems}}, Journal of the
  American Statistical Association 89~(425) (1994) 278--288.
\newblock \href {http://dx.doi.org/10.1080/01621459.1994.10476469}
  {\path{doi:10.1080/01621459.1994.10476469}}.
\newline\urlprefix\url{http://www.tandfonline.com/doi/abs/10.1080/01621459.1994.10476469}

\bibitem{robert_monte_2004}
C.~P. Robert, G.~Casella, Monte {Carlo} {Statistical} {Methods}, 2nd Edition,
  Springer New York, 2004.

\bibitem{wellman_breast_1999}
P.~Wellman, R.~D. Howe, E.~Dalton, K.~A. Kern, Breast tissue stiffness in
  compression is correlated to histological diagnosis, Harvard {BioRobotics}
  Laboratory Technical Report.

\bibitem{liu_monte_2001}
J.~Liu, Monte {Carlo} {Strategies} in {Scientific} {Computing}, Springer
  {Series} in {Statistics}, Springer, 2001.

\bibitem{samani_method_2004}
A.~Samani, D.~Plewes, A method to measure the hyperelastic parameters of ex
  vivo breast tissue samples, Physics in Medicine and Biology 49~(18)
  (2004-09-21) 4395--4405.
\newblock \href {http://dx.doi.org/10.1088/0031-9155/49/18/014}
  {\path{doi:10.1088/0031-9155/49/18/014}}.

\bibitem{ohagan_measurement_2009}
J.~J. O'Hagan, A.~Samani, Measurement of the hyperelastic properties of 44
  pathological ex vivo breast tissue samples, Physics in Medicine and Biology
  54~(8) (2009-04-21) 2557--2569.
\newblock \href {http://dx.doi.org/10.1088/0031-9155/54/8/020}
  {\path{doi:10.1088/0031-9155/54/8/020}}.

\bibitem{mooney_theory_1940}
M.~Mooney, Theory of large elastic deformation, Journal of Applied Physics 11
  (1940) 582--592.
\newblock \href {http://dx.doi.org/10.1063/1.1712836}
  {\path{doi:10.1063/1.1712836}}.

\bibitem{rivlin_large_1948}
R.~Rivlin, Large elastic deformations of isotropic materials. further
  developments of the general theory, Philosophical Transactions of the Royal
  Society of London Series a-Mathematical and Physical Sciences 241~(835)
  (1948) 379--397.
\newblock \href {http://dx.doi.org/10.1098/rsta.1948.0024}
  {\path{doi:10.1098/rsta.1948.0024}}.

\bibitem{simo_quasi-incompressible_1991}
J.~Simo, R.~Taylor, Quasi-incompressible finite elasticity in principal
  stretches - continuum basis and numerical algorithms, Computer Methods in
  Applied Mechanics and Engineering 85~(3) (1991-02) 273--310.
\newblock \href {http://dx.doi.org/10.1016/0045-7825(91)90100-K}
  {\path{doi:10.1016/0045-7825(91)90100-K}}.

\bibitem{schoeberl_comparison_2013}
M.~Sch\"oberl, Comparison of different optimization algorithms for nonlinear
  inverse problems in biomechanics (2013).

\end{thebibliography}

% %% else use the following coding to input the bibitems directly in the
% %% TeX file.
% \end{thebibliography}
\end{document}